\documentclass[
amsmath,
prd,nofootinbib,floatfix,11pt,
]{revtex4}

\newcommand\Zdisc{Z_{\rm disc}}
\newcommand\Zexc{Z_{\rm exc}}
\newcommand\beq{\begin{eqnarray}}
\newcommand\eeq{\end{eqnarray}}
\newcommand\meff{m_{\rm eff}}
\newcommand\missET{E_T^{\rm miss}}
\newcommand\ETmiss{E_T^{\rm miss}}
\newcommand\sigmabar{\overline\sigma}
\newcommand\sigmab{\Delta_b}
\def\lsim{\mathrel{\rlap{\lower4pt\hbox{$\sim$}}
    \raise1pt\hbox{$<$}}}                
\def\gsim{\mathrel{\rlap{\lower4pt\hbox{$\sim$}}
    \raise1pt\hbox{$>$}}}            

\allowdisplaybreaks
\usepackage{graphicx}
\usepackage{setspace}

\usepackage{dcolumn}
\usepackage{bm}

\begin{document}

\renewcommand{\theequation}{\arabic{section}.\arabic{equation}}
\renewcommand{\thefigure}{\arabic{section}.\arabic{figure}}
\renewcommand{\thetable}{\arabic{section}.\arabic{table}}

\title{\large \baselineskip=20pt 
Vectorlike leptons at the Large Hadron Collider}

\author{Nilanjana Kumar$^1$ and Stephen P.~Martin$^{1,2}$}
\affiliation{
{\it $^1$Department of Physics, Northern Illinois University, DeKalb IL 60115} \\
{\it $^2$Fermi National Accelerator Laboratory, P.O. Box 500, Batavia IL 60510}
}

\begin{abstract}\normalsize \baselineskip=14pt
We study the prospects for excluding or discovering vectorlike leptons 
using multilepton events at the LHC. We consider models in which the 
vectorlike leptons decay to tau leptons. If the vectorlike leptons are 
weak isosinglets, then discovery in multilepton states is found to be 
extremely challenging. For the case that the vectorlike leptons are weak 
isodoublet, we argue that there may be an opportunity for exclusion for 
masses up to about 275 GeV by direct searches with existing LHC data at 
$\sqrt{s}=8$ TeV. We also discuss prospects for exclusion or discovery 
at the LHC with future $\sqrt{s}=13$ TeV data.
\end{abstract}

\maketitle

\vspace{-0.3in}

\tableofcontents
\baselineskip=15.4pt
\baselineskip=15.2pt

\section{Introduction \label{sec:intro}}
\setcounter{equation}{0}
\setcounter{figure}{0}
\setcounter{table}{0}
\setcounter{footnote}{1}

Vectorlike quarks and leptons are hypothetical new fermions that transform
in non-chiral representations of the unbroken Standard
Model (SM) gauge group. They are among the simplest viable SM
extensions near the electroweak scale. Vectorlike fermions can have
electroweak singlet masses that dominate over the contributions to
their masses from Yukawa couplings to the Higgs boson. This means that their
loop-induced contributions to precision
electroweak observables and radiative Higgs decays and production
obey decoupling with large masses. Therefore, vectorlike fermions are less
constrained than extra chiral families, which are now ruled out by 
a combination of direct searches and the
observations of the 125 GeV Higgs boson production and decay. 
In the absence of large lepton flavor violation, 
general mass limits on vectorlike fermions with non-exotic electric charges therefore
follow only from direct searches.

Besides the mere fact that they are possible, there are a variety of 
motivations to consider vectorlike fermions. From a top-down 
perspective, phenomenological models motivated by string theory or large 
extra dimensions are well-known to be often replete with such particles.  
In weak-scale supersymmetry, the mass of the lightest Higgs scalar boson 
can be raised by introducing new vectorlike heavy chiral supermultiplets 
with large Yukawa couplings 
\cite{Moroi:1991mg,Moroi:1992zk,Babu:2004xg,Babu:2008ge,Martin:2009bg, 
Graham:2009gy,Martin:2010dc,Endo:2011mc,Evans:2011uq,Li:2011ab, 
Moroi:2011aa,Endo:2011xq,Endo:2012rd,Nakayama:2012zc,Martin:2012dg}. The 
correction to $M_h$ is positive if the vectorlike fermions are lighter 
than their scalar partners, which implies that the former could be the 
first physics beyond the SM to be detected at the Large Hadron Collider 
(LHC). Some other interesting discussions of the possible role of 
vectorlike leptons in physics beyond the SM are given in 
refs.~\cite{Dimopoulos:1990kc,Sher:1995tc,Thomas:1998wy,Frampton:1999xi,delAguila:2010vg,
Dermisek:2012as,Joglekar:2012vc,Kearney:2012zi,Halverson:2014nwa,
Arina:2012aj,Batell:2012zw,Dermisek:2012ke,Schwaller:2013hqa,
Dermisek:2013gta,Fairbairn:2013xaa,Ishiwata:2013gma,Altmannshofer:2013zba,
Falkowski:2013jya,Dobrescu:2014fca,Ellis:2014dza,Falkowski:2014ffa,
Holdom:2014rsa,Dermisek:2014cia,Dobrescu:2014esa,Dermisek:2014qca,
Holdom:2014boa,Ishiwata:2015cga,Bizot:2015zaa,Dermisek:2015oja,Bhattacharya:2015qpa}.
For vector-like leptons with large exotic charges, there is also a possibility of
indirect searches from the loop-induced process
$pp \rightarrow pp \gamma\gamma$ \cite{Fichet:2013gsa}.

In this paper we consider the LHC exclusion and discovery reach for
vectorlike leptons in two scenarios.
First, we consider $SU(2)_L$-singlet charged 
vectorlike leptons $\tau^{\prime\pm}$
which under $SU(3)_C\times SU(2)_L\times
U(1)_Y$ transform as 2-component left-handed fermions\footnote{For reviews
of the 2-component fermion notation followed here, 
see refs.~\cite{Dreiner:2008tw,Martin:2012us}.}
\beq
\tau' + \overline \tau' = ({\bf1},{\bf1},+1)+({\bf1},{\bf1},-1).
\eeq
The second scenario consists of pure $SU(2)_L$-doublet particles
$L' = (\nu', \tau^{\prime -})$ and their antiparticles
$\overline L' = (\tau^{\prime +}, \overline \nu')$,
which transform under $SU(3)_C\times SU(2)_L\times
U(1)_Y$ as 2-component left-handed fermions
\beq
L' + \overline L' = ({\bf1},{\bf2},-1/2)+({\bf1},{\bf2},+1/2).
\eeq
In the following, we will refer to these as the Singlet VLL and Doublet VLL
models, respectively.
For simplicity, we consider these two possibilities separately,
although models in which they are combined or replicated
are certainly feasible,
and would have a richer phenomenology.

The main source of vectorlike lepton masses are weak singlet terms. If 
these were the only sources of vectorlike lepton mass, then in the 
Singlet VLL model the $\tau'$ would be absolutely stable, causing 
possible problems due to its presence as a charged exotic stable relic 
in the universe. In the Doublet VLL model the $\nu'$ would be stable. We 
will therefore assume that the vectorlike leptons mix through Yukawa 
interactions with the ordinary known leptons of the SM, allowing them to 
have 2-body decays to SM leptons and bosons $W,Z,h$, as described in 
more detail in the next section. Our premise, motivated by the relative 
weakness of lepton flavor-violation constraints involving the $\tau$ 
lepton compared to the electron and muon, is that the vector-like lepton 
coupling to SM leptons is mostly with the third family, and therefore 
the $\tau'$ and $\nu'$ decay mostly to final states involving the $\tau$ 
lepton and Standard Model neutrinos. This is the most pessimistic 
possibility for LHC reach, due to the relatively lower detection 
efficiency and higher fake rates for $\tau$ candidates in the LHC 
detectors. An earlier study \cite{Dermisek:2014qca} instead considered 
the more optimistic possibility that vectorlike leptons produced at the 
LHC will decay mostly to muons. At this writing, there do not appear to 
be any official limits on vectorlike leptons from the LHC detector 
collaborations, so that the only constraint comes from the non-discovery 
by the LEP $e^+e^-$ collider experiments, of about $M_{\tau'} > 100$ 
GeV. In the following, we will consider the possibilities of setting 
limits on vectorlike leptons that decay to $\tau$ and Standard Model 
neutrinos using existing ATLAS multilepton searches, and using our own 
alternative search criteria, for $pp$ collisions at $\sqrt{s}=8$ TeV 
with 20 fb$^{-1}$ of integrated luminosity. We will then consider the 
projected exclusion and discovery reaches of the LHC in multilepton 
searches in future runs with $\sqrt{s}=13$ TeV. In these 3-, 4-, and 
5-lepton searches, we rely both on $\tau$ leptons that are detected as 
hadronic taus, $\tau_h$, and on $\tau$ leptons that decay leptonically 
to electrons and muons.

In considerations of exclusion and discovery prospects, we will employ 
the following criteria. For a given hypothesis $H_0$, we estimate the 
median expected p-value $p$, which is the probability, 
for data generated under a hypothesis $H$, of observing 
a result of equal or greater incompatibility with $H_0$. 
As is conventional in high-energy physics,
the $p$-value is converted 
to a significance $Z$ according to
\beq
Z &=& \sqrt{2}\, {\rm erfc}^{-1}(2 p).
\eeq
In the case of a 
standard Gaussian distribution, $Z$ corresponds to the number of 
standard deviations. In the case of discovery,
$H_0$ is the background-only hypothesis of a 
Poisson distribution of events with mean $b$, while $H$ is the hypothesis
of a Poisson distribution of signal and background events with mean $s+b$.
However, the background levels are not, and will not be, 
known with perfect accuracy. Therefore we 
include the effects of 
a variance $\Delta_b$ in the expected number of background events. 
An approximation (see ref.~\cite{Cowan}, and also refs.~\cite{Cowan:2010js} and 
\cite{LiMa,Cousins:2008zz}) 
for the median expected discovery significance is then:
\beq
\Zdisc = 
\left [ 2\left ((s+b) \ln \left [
\frac{(s+b)(b+\sigmab^2)}{b^2+(s+b)\sigmab^2}
\right ]-
\frac {b^2}{\sigmab^2} \ln\left [1+ \frac {\sigmab^2 s}{b(b+\sigmab^2)}\right ]
\right)\right]^{1/2} .
\label{eq:Zdisc}
\eeq
In the special case that the background expectation is known perfectly, so that
$\Delta_b = 0$, this reduces to 
\beq
\Zdisc = \sqrt{2 [(s+b) \ln(1+s/b) - s]},
\eeq
which would further reduce to $s/\sqrt{b}$ in the limit of large $b$.
However, when $b$ is small, $s/\sqrt{b}$ greatly overestimates the expected
significance. For a discovery criterion, we use $\Zdisc > 5$, 
corresponding to a $p$-value range $p< 2.86 \times 10^{-7}$,
and we will use eq.~(\ref{eq:Zdisc}) with the somewhat arbitrary 
choices $\Delta_b = 0.1b$, $0.2b$, and $0.5b$, corresponding to a 10\%, 20\%,
and 50\% uncertainty in the background.

For exclusion, the role of $H_0$ is played by the signal plus background 
hypothesis, and $H$ is the background-only hypothesis.
We then find, based on methods in refs.~\cite{Cowan,Cowan:2010js}, an 
estimate for the median expected exclusion significance:
\beq
\Zexc = 
\left [2 \left \{ s-b \ln \left (\frac{b+s+x}{2b} \right ) 
- \frac{b^2}{\Delta_b^2} \ln \left (\frac{b-s+x}{2b} \right ) \right \} -
(b + s - x) (1 + b/\Delta_b^2) \right ]^{1/2},
\label{eq:Zexc}
\eeq
where
\beq
x = \sqrt{(s+b)^2 - 4 s b \Delta_b^2/(b + \Delta_b^2)} .
\eeq
In the special case $\Delta_b = 0$, eq.~(\ref{eq:Zexc}) reduces to
\beq
\Zexc = \sqrt{2 (s - b \ln(1 + s/b))},
\eeq
which further reduces to $s/\sqrt{b}$ in the limit of large $b$. Again,
for small $b$, the latter overestimates the expected exclusion significance.
For a median expected 95\% confidence level (CL) exclusion
($p = 0.05$), we will use
eq.~(\ref{eq:Zexc}) with 
$\Zexc > 1.645$, and again consider
$\Delta_b = 0.1b$, $0.2b$, and $0.5b$. 
In the case of the Singlet VLL model only, where the signal is quite small, 
we will also consider the very optimistic case $\Delta_b=0$.

\section{Production and decay of vectorlike leptons\label{sec:productiondecay}}
\setcounter{equation}{0}
\setcounter{figure}{0}
\setcounter{table}{0}
\setcounter{footnote}{1}

In the Singlet VLL model, the fermion mass terms and
$\tau'$ mixing with the Standard Model lepton
can be obtained from the Lagrangian
written in 2-component fermion form as
\beq
-{\cal L} &=& m_{\tau'} \tau' \overline \tau' + \epsilon H L \overline \tau'
           + y_\tau H L \overline \tau  + {\rm c.c.}
\label{eq:masses_singlet}
\eeq
where $H$ is the SM Higgs complex doublet scalar field,
$L = (\tau, \nu_\tau)$ is the SM third family lepton doublet
in the gauge eigenstate basis, $y_\tau$ is the SM $\tau$ Yukawa coupling,
and $\epsilon$ is the mixing Yukawa coupling. The charged fermion mass matrix
in the gauge eigenstate basis is
\beq
-{\cal L} &=& \begin{pmatrix} \tau & \tau' \end{pmatrix} {\cal M}
\begin{pmatrix} \overline \tau \\ \overline \tau' \end{pmatrix}
+ {\rm c.c.} ,
\eeq
where
\beq
{\cal M} &=& \begin{pmatrix}
y_\tau v & \epsilon v \\
0 & m_{\tau'} \end{pmatrix} ,
\eeq
and $v= \langle H \rangle = 174$ GeV is the Higgs vacuum expectation
value (VEV). For the Doublet VLL model, the Lagrangian is
\beq
-{\cal L} &=&
m_{\tau'} L' \overline L' +
\epsilon H L' \overline \tau + y_\tau H L \overline \tau + {\rm c.c.},
\label{eq:masses_doublet}
\eeq
so that the charged mass matrix is instead
\beq
{\cal M} = \begin{pmatrix}
y_\tau v & 0 \\
\epsilon v & m_{\tau'} \end{pmatrix} .
\eeq
In both cases, we will take the Yukawa coupling $\epsilon$ to be small.
Then, neglecting effects suppressed by $\epsilon$,
the charged lepton mass eigenstates include just a $\tau'$
with mass $M_{\tau'} \approx m_{\tau'}$,
and the SM tau lepton with mass
$M_{\tau} \approx y_\tau v$.

In the Doublet VLL model,
there is also a $\nu'$ state, with mass degenerate with the
$\tau'$ at tree level.
Taking into account 1-loop
radiative corrections \cite{Thomas:1998wy} while
still neglecting the effects quadratic in $\epsilon$, there is
a small mass splitting
\beq
M_{\nu'} = M_{\tau'} - \frac{\alpha}{2} M_Z f(M^2_{\tau'}/M_Z^2),
\eeq
where \cite{Thomas:1998wy}
\beq
f(r) = \frac{\sqrt{r}}{\pi} \int_0^1 dx (2-x) \ln(1 + x/r(1-x)^2).
\label{eq:deff}
\eeq
is positive and approaches 1  from below as $r$ becomes very large.
For $M_{\tau'} = (100, 200, 300$ GeV, and $\infty$), this mass splitting
is respectively about (258, 297, 313, 355) MeV, and will be 
only very slightly increased
by mixing, by approximately $\epsilon^2 v^2/2 M_{\tau'}$.
For kinematic purposes, we will therefore simply
take $M_{\nu'} = M_{\tau'}$.

\begin{figure}[!tb]
  \begin{minipage}[]{0.5\linewidth}
    \includegraphics[width=7.0cm,angle=0]{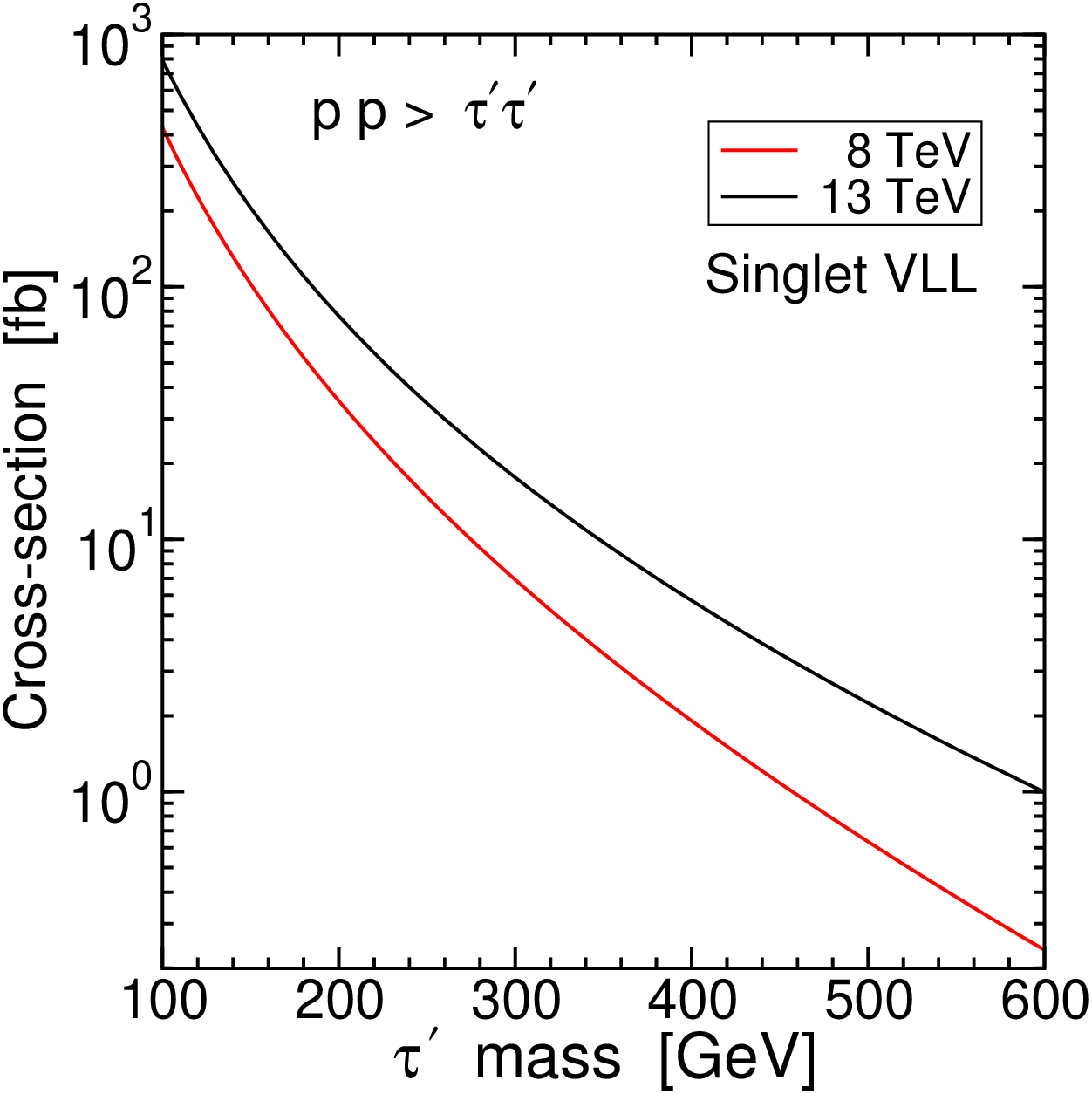}
  \end{minipage}
\begin{minipage}[]{0.49\linewidth}
  \caption{\label{fig:cross_sections_singlet} The production cross section for
  $p p \rightarrow \tau^{\prime +} \tau^{\prime -}$ as a function of
  the mass $M_{\tau^\prime}$, for the LHC at $\sqrt{s} = 8$ and 13 TeV, 
  in the Singlet VLL model, mediated by the interactions in 
  eq.~(\ref{eq:production_couplings_singlet}). 
}
\end{minipage}
\end{figure}

\begin{figure}[!tb]
  \begin{minipage}[]{0.495\linewidth}
    \includegraphics[width=7.5cm,angle=0]{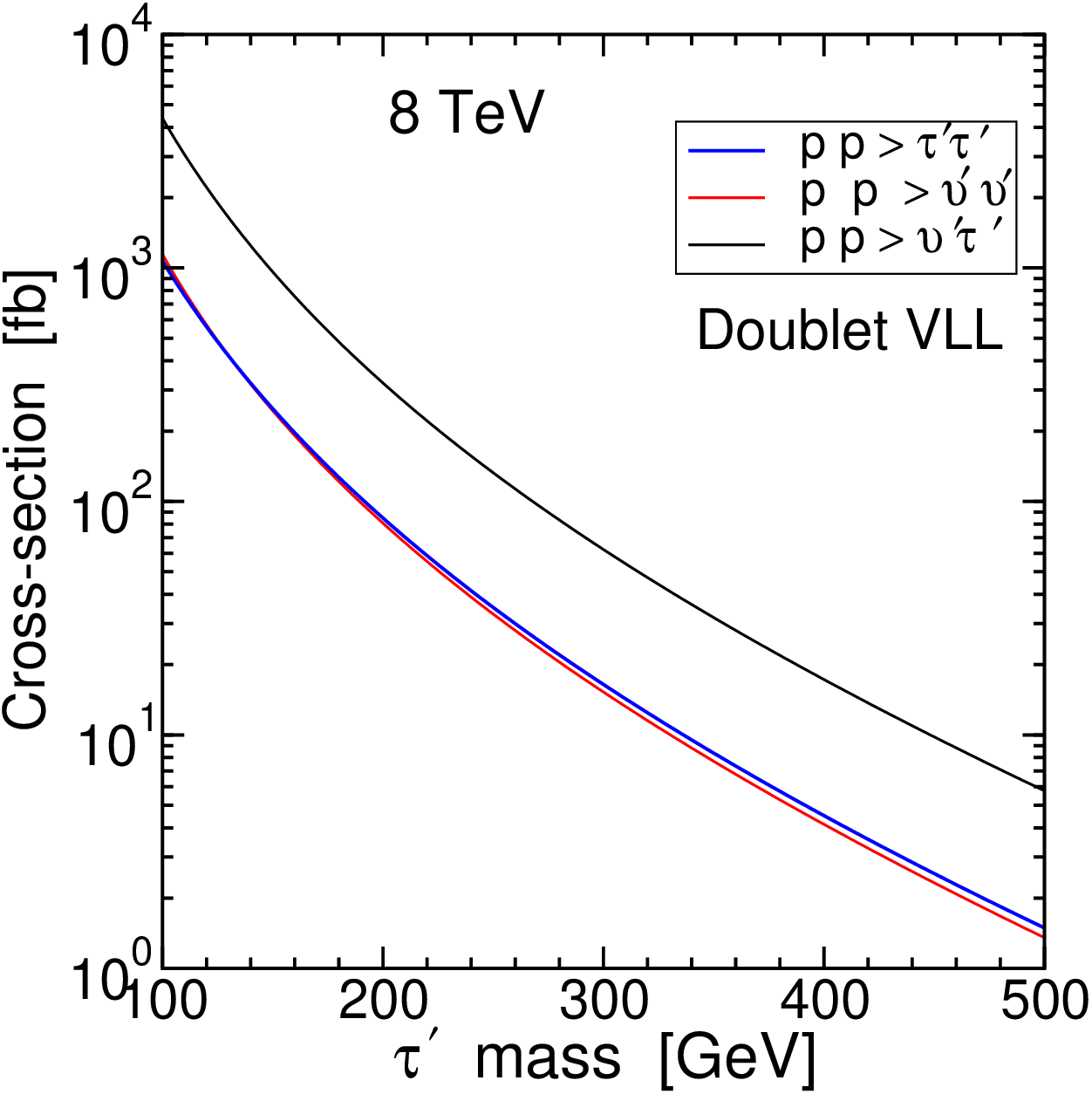}
  \end{minipage}
    \begin{minipage}[]{0.495\linewidth}
    \includegraphics[width=7.5cm,angle=0]{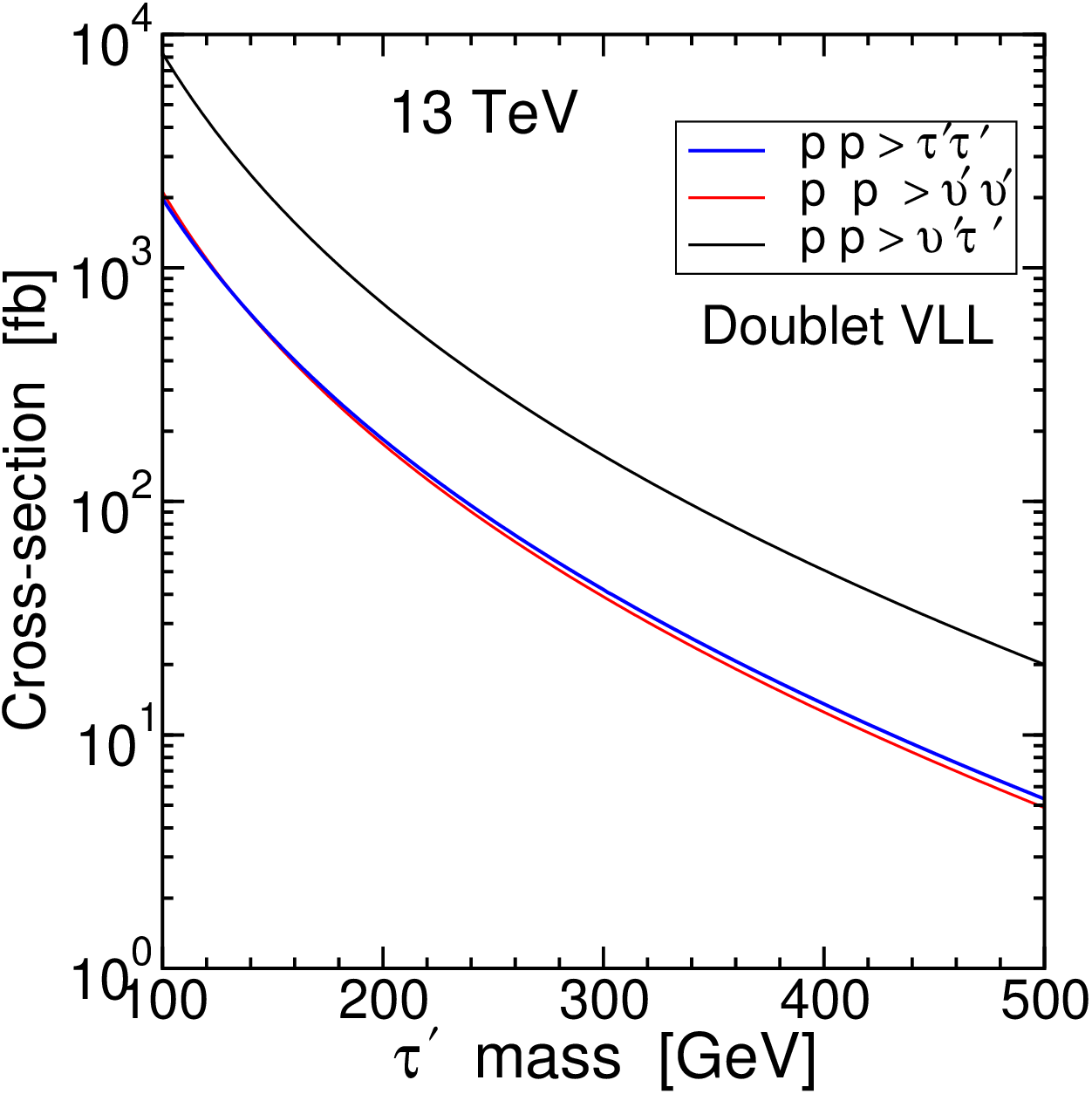}
  \end{minipage}
\caption{\label{fig:cross_sections_doublet} The production cross sections for,
  from bottom to top,
  $\nu' \overline \nu'$ and
  $\tau^{\prime +} \tau^{\prime -}$ and 
  the combined cross-section for $\nu' \tau^{\prime +}$ and
  $\overline \nu' \tau^{\prime -}$, as a function of
  the common mass $M_{\tau^\prime} = M_{\nu'}$, 
  for the LHC at $\sqrt{s} = 8$ (left panel) and 13 TeV (right panel),
  in the Doublet VLL model, mediated by the interactions in
  eq.~(\ref{eq:production_couplings_doublet}).   
}
\end{figure}

The production rates for vectorlike leptons are governed to a very good
approximation by their lepton flavor-preserving interactions with the
electroweak vector bosons. In 2-component fermion
notation [with a metric signature ($-$,$+$,$+$,$+$)]
in the mass eigenstate basis, the Singlet
VLL model has, neglecting terms quadratic in $\epsilon$:
\beq
{\cal L}_{\rm int} &=&
\frac{g s_W^2}{c_W} Z_\mu \left (
  \tau^{\prime\dagger} \sigmabar^\mu \tau^\prime
  - \overline\tau^{\prime\dagger} \sigmabar^\mu \overline\tau^\prime
\right )
-e A_\mu \left (
  \tau^{\prime\dagger} \sigmabar^\mu \tau^\prime
  - \overline\tau^{\prime\dagger} \sigmabar^\mu \overline\tau^\prime
\right )
,
\label{eq:production_couplings_singlet}
\eeq
where $e$ is the QED coupling, $g$ is the $SU(2)_L$ coupling, and $s_W$,
$c_W$ are the sine and cosine of the weak mixing angle, with
$e = g s_W$. For the Doublet VLL model, the Lagrangian governing
$\tau'$ and $\nu'$ production is similarly:
\beq
{\cal L}_{\rm int} &=&
\frac{g}{\sqrt{2}} W_\mu^+ (
  \overline\tau^{\prime\dagger}\sigmabar^\mu \overline\nu' +
  \nu^{\prime\dagger}\sigmabar^\mu \tau'
)
+ \frac{g}{\sqrt{2}} W_\mu^- (
  \tau^{\prime\dagger}\sigmabar^\mu \nu' +
  \overline\nu^{\prime\dagger}\sigmabar^\mu \overline\tau'
)
-e A_\mu \left (
  \tau^{\prime\dagger} \sigmabar^\mu \tau^\prime
  - \overline\tau^{\prime\dagger} \sigmabar^\mu \overline\tau^\prime
\right )
\nonumber \\ &&
+ \frac{g}{c_W}\left (s_W^2 - \frac{1}{2}\right ) Z_\mu \left (
\tau^{\prime\dagger} \sigmabar^\mu \tau^\prime
-
\overline\tau^{\prime\dagger} \sigmabar^\mu \overline\tau^\prime
\right )
+
\frac{g}{2 c_W} Z_\mu \left (
\nu^{\prime\dagger} \sigmabar^\mu \nu^\prime
-
\overline\nu^{\prime\dagger} \sigmabar^\mu \overline\nu^\prime
\right )
.
\label{eq:production_couplings_doublet}
\eeq
In the Singlet VLL model, the production channel at the LHC is:
\beq
pp &\rightarrow& \tau^{\prime +} \tau^{\prime -},
\label{eq:pptauptaup}
\eeq
through $s$-channel $Z,\gamma$,
while in the Doublet VLL model one has in addition the processes involving
the heavy vectorlike Dirac neutrino:
\beq
pp &\rightarrow& \nu' \overline \nu',
\label{eq:ppnupnup}
\\
pp &\rightarrow& \nu' \tau^{\prime +} ,
\label{eq:ppnuptaup1}
\\
pp &\rightarrow& \overline \nu' \tau^{\prime -} .
\label{eq:ppnuptaup2}
\eeq
In both cases, the production rates are a function of only one free parameter,
the mass $M_{\tau'}$. They are shown for $\sqrt{s} = 8$ and 13 TeV
in Figure \ref{fig:cross_sections_singlet}
for the Singlet VLL model and in Figure \ref{fig:cross_sections_doublet}
for the Doublet VLL model.
It is evident that the production cross sections are much larger in the
Doublet VLL model than in the Singlet VLL model.
This is partly because of the larger couplings, 
but also because $\tau^{\prime}$ is accompanied by $\nu^{\prime}$
in the doublet case. 
Indeed, the largest rate is
from the $\nu'\tau^{\prime\pm}$ modes mediated by $s$-channel 
$W^\pm$ bosons. The
LHC prospects for exclusion or discovery of the Doublet VLL model are
therefore much brighter than for the Singlet VLL model, as we will see below.

We now turn to the interactions that mediate vectorlike lepton decays,
which arise due to the mixing parameter $\epsilon$. 
Working to linear order in $\epsilon$, we have
for the Singlet VLL model:
\beq
{\cal L}_{\rm int} &=&
g^{W^+}_{\nu^\dagger \tau'} \left [W^+_\mu (\nu^\dagger \sigmabar^\mu \tau') +
W_\mu^- (\tau^{\prime \dagger} \sigmabar^\mu \nu) \right ]
+
g^Z_{\tau^\dagger \tau'} Z_\mu \left (\tau^\dagger \sigmabar^\mu \tau' +
\tau^{\prime \dagger} \sigmabar^\mu \tau \right )
\nonumber \\ &&
+ (y^h_{\tau\overline \tau'} h \tau \overline \tau' + {\rm c.c.})
\eeq
where $h$ is the real scalar field for the 125 GeV Higgs boson, and 
\beq
g^{W^+}_{\nu^\dagger \tau'} &=& \epsilon M_W/M_{\tau'} ,
\\
g^Z_{\tau^\dagger \tau'} &=& -\epsilon M_Z/\sqrt{2} M_{\tau'} ,
\\
y^h_{\tau\overline \tau'} &=& -\epsilon/\sqrt{2}.
\eeq
The resulting decay widths for $\tau'$ to SM states are:
\beq
\Gamma (\tau' \rightarrow W \nu) &=&
\frac{M_{\tau'}}{32\pi} (1-r_W)^2 (2 + 1/r_W)
|g^{W^+}_{\nu^\dagger \tau'}|^2 ,
\\
\Gamma (\tau' \rightarrow Z \tau) &=&
\frac{M_{\tau'}}{32\pi} (1-r_Z)^2 (2 + 1/r_Z)
|g^Z_{\tau^\dagger \tau'}|^2 ,
\\
\Gamma (\tau' \rightarrow h \tau) &=&
\frac{M_{\tau'}}{32\pi} (1-r_h)^2 |y^h_{\tau\overline \tau'}|^2 .
\eeq
where $r_X = M_X^2/M_{\tau'}^2$ for $X=W,Z,h$.

For the Doublet VLL model, the interactions that mediate decays of $\tau'$ and
$\nu'$ are:
\beq
{\cal L}_{\rm int} &=&
g^{W^+}_{\overline \tau^{\dagger} \overline \nu'}
\left [W^+_\mu (\overline \tau^{\dagger} \sigmabar^\mu \overline \nu') +
W_\mu^- (\overline \nu^{\prime \dagger} \sigmabar^\mu \overline \tau) \right ]
+g^Z_{\overline\tau^{\dagger} \overline\tau'} Z_\mu
\left (\overline\tau^\dagger \sigmabar^\mu \overline \tau' +
\overline \tau^{\prime \dagger} \sigmabar^\mu \overline \tau \right )
\nonumber \\ &&
+ (y^h_{\tau'\overline \tau} h \tau' \overline \tau + {\rm c.c.})
\eeq
where, again working to linear order in $\epsilon$,
\beq
g^{W^+}_{\overline \tau^{\dagger} \overline \nu'}
&=& -\epsilon M_W/M_{\tau'} ,
\\
g^Z_{\overline\tau^{\dagger} \overline\tau'} &=&
-\epsilon M_Z/\sqrt{2} M_{\tau'} ,
\\
y^h_{\tau'\overline \tau} &=& -\epsilon/\sqrt{2}.
\eeq
The resulting decay widths for $\tau'$ and $\nu'$ to SM states within
this approximation are:
\beq
\Gamma (\tau' \rightarrow W \nu) &=& 0 ,
\\
\Gamma (\tau' \rightarrow Z \tau) &=&
\frac{M_{\tau'}}{32\pi} (1-r_Z)^2 (2 + 1/r_Z)
|g^Z_{\overline\tau^{\dagger} \overline\tau'}|^2 ,
\\
\Gamma (\tau' \rightarrow h \tau) &=&
\frac{M_{\tau'}}{32\pi} (1-r_h)^2 |y^h_{\tau'\overline \tau}|^2 ,
\\
\Gamma (\nu' \rightarrow W \tau) &=&
\frac{M_{\nu'}}{32\pi} (1-r_W)^2 (2 + 1/r_W)
|g^{W^+}_{\overline \tau^{\dagger} \overline \nu'}|^2 ,
\\
\Gamma (\nu' \rightarrow Z \nu) &=& \Gamma (\nu' \rightarrow h \nu)\,=\, 0.
\eeq

The resulting branching ratios only depend on
the single parameter $M_{\tau'}$, as all of the widths are proportional
to $\epsilon^2$. These
are shown in Figure \ref{fig:BRs} for $\tau'$ in the
Singlet VLL model (left panel) and the Doublet VLL model (right panel).
Note that for $M_{\tau'} \gg M_h, M_Z, M_W$, the results asymptotically
approach:%
\begin{figure}[!tb]
  \begin{minipage}[]{0.495\linewidth}
    \includegraphics[width=7.5cm,angle=0]{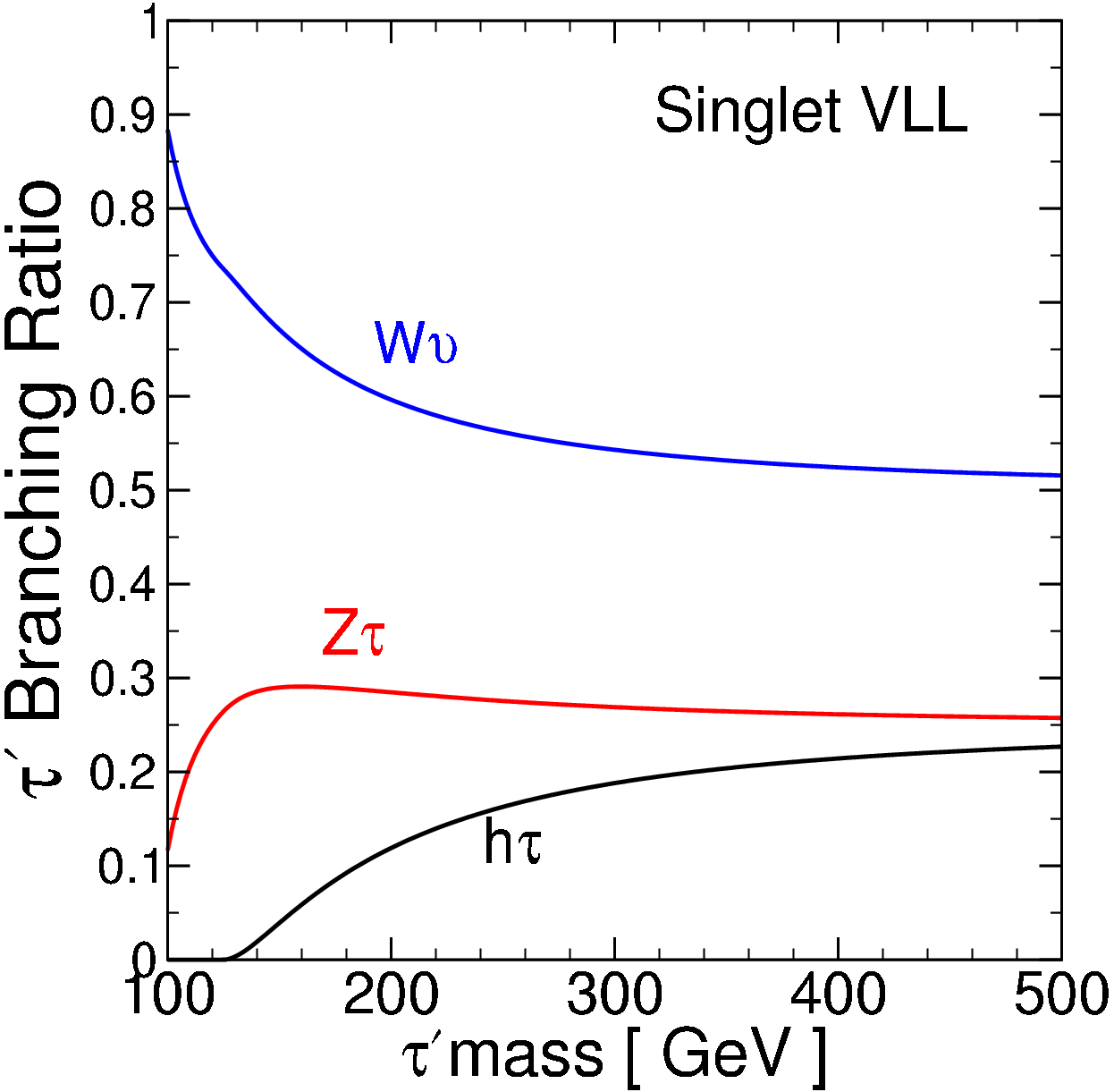}
  \end{minipage}
    \begin{minipage}[]{0.495\linewidth}
    \includegraphics[width=7.5cm,angle=0]{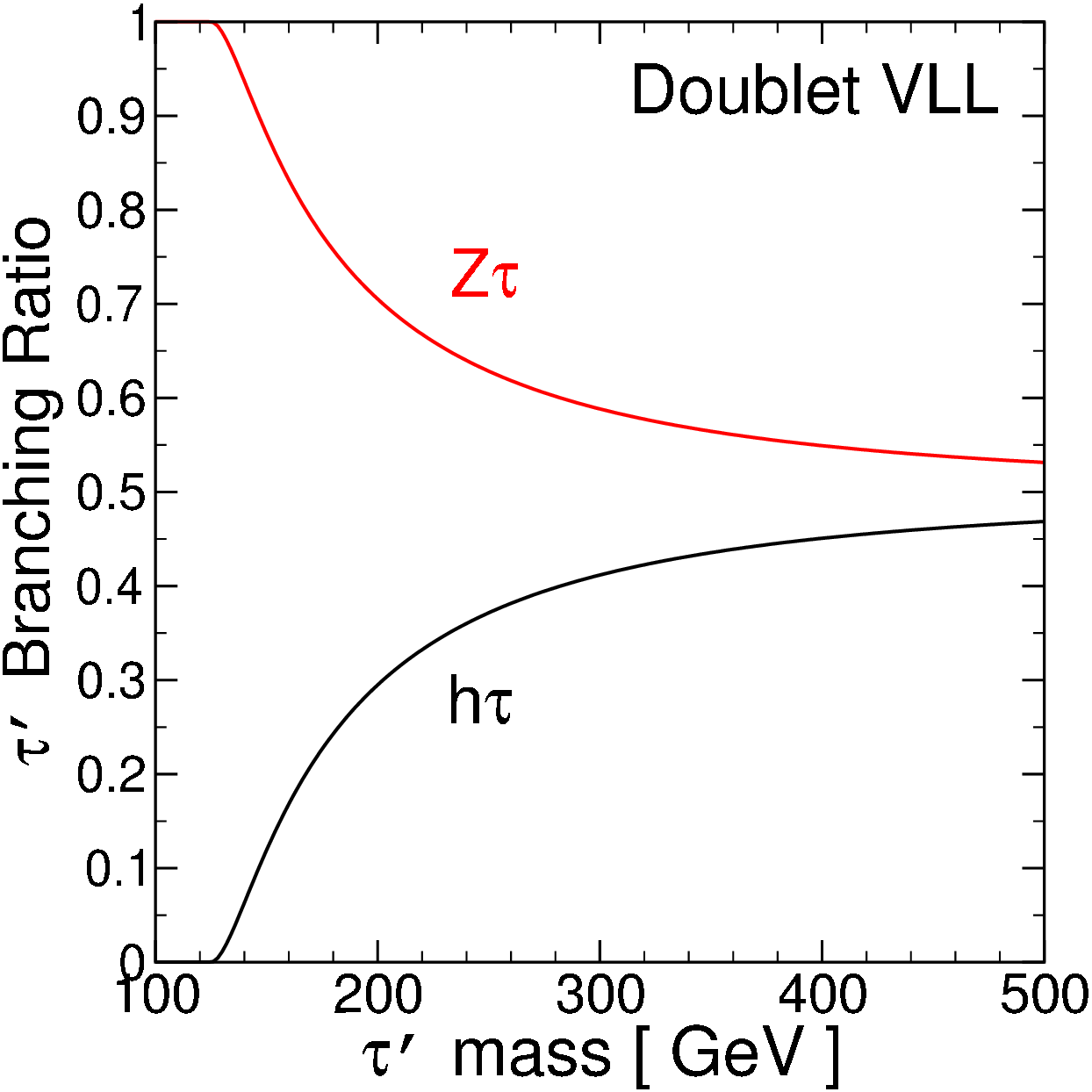}
  \end{minipage}
\caption{\label{fig:BRs} The branching ratios for $\tau' \rightarrow W\nu$ and
$Z\tau$ and $h\tau$, as a function of $M_{\tau'}$,
for the Singlet VLL model (left panel) and
the Doublet VLL model (right panel).}
\end{figure}
\beq
\mbox{BR}(\tau' \rightarrow W \nu) : \mbox{BR}(\tau' \rightarrow Z\tau) :
\mbox{BR}(\tau' \rightarrow h\tau) &=& 
\left \{ \begin{array}{ll}
2:1:1\qquad\!& \mbox{(Singlet VLL model)},\\
0:1:1\qquad\!& \mbox{(Doublet VLL model)} .
\end{array} \right.
\phantom{xx}
\eeq
Here, we have assumed that the highly kinematically suppressed
decay of $\tau'$ to $\nu'$ is negligible.
To justify this, note that from ref.~\cite{Thomas:1998wy}:
\beq
\Gamma ( \tau' \rightarrow \nu' \pi^-) =
(3.1 \times 10^{-14}\> \mbox{GeV}) f^3 \sqrt{1 - 0.155/f^2},
\eeq
where $f = f(M_{\tau'}^2/M_Z^2)$ from eq.~(\ref{eq:deff}).
This decay width will
be smaller provided that the dimensionless mixing Yukawa coupling
satisfies $\epsilon \gsim 2 \times 10^{-7}$. This is also
very roughly the condition needed for the decays of $\tau'$ and $\nu'$ to
have a decay length $c\tau$ less than the centimeter scale, with
some dependence of course on the mass.
The Doublet VLL model has 
\beq
\mbox{BR}(\nu' \rightarrow W^+ \tau^-) &=& 
\mbox{BR}(\overline \nu' \rightarrow W^- \tau^+) \>=\> 1.
\eeq
This reflects our assumption of no mass mixing between $\nu'$ and the
SM neutrinos. The large branching ratio of $\nu'$ into states with taus and
possible leptons from the $W$ decay helps the Doublet VLL model 
exclusion and discovery prospects.

\section{Event simulation\label{sec:simulation}}
\setcounter{equation}{0}
\setcounter{figure}{0}
\setcounter{table}{0}
\setcounter{footnote}{1}

From the results of the preceding section, 
we find the following signals for the Singlet VLL model
from $\tau'$ pair production eq.~(\ref{eq:pptauptaup}):
\beq
&& ZZ\tau^+\tau^-,
\qquad\quad 
Zh\tau^+\tau^-,
\qquad\quad
hh\tau^+\tau^-,
\\
&& ZW^\pm\tau^\mp + \missET,
\qquad\quad
hW^\pm\tau^\mp + \missET,
\qquad\quad
W^+W^- + \missET,
\eeq
while only the first 3 are produced by $\tau'$ pair production in 
the Doublet VLL model. However, the Doublet VLL model also has signals
\beq
W^+W^-\tau^+\tau^-
\eeq
from $\nu'$ pair production eq.~(\ref{eq:ppnupnup}), and
\beq
&& ZW^\pm\tau^+\tau^-
\\
&& hW^\pm\tau^+\tau^-
\eeq
from $\tau'\nu'$ production eq.~(\ref{eq:ppnuptaup1}) and 
eq.~(\ref{eq:ppnuptaup2}). In this paper, we consider final states
with three or more leptons (including electrons 
and muons from leptonic tau decays, 
as well as hadronic taus, $\tau_h$) 
that arise from these. This includes leptons
coming from Higgs decays directly to taus and to $W$ boson pairs,
for which we use $M_h = 125$ GeV and the branching ratios 
\beq
\mbox{BR}(h \rightarrow \tau^+\tau^-) &=& 0.0605,
\\
\mbox{BR}(h \rightarrow W^+W^-) &=& 0.21.
\eeq
In our signals, we often distinguish events depending on whether 
two opposite sign same flavor (OSSF) leptons reconstruct a $Z$ boson.

For both the Singlet and Doublet
VLL models, we have implemented the production and decay of 
$\tau'$ and $\nu'$
and their antiparticles in Madgraph 5 \cite{Alwall:2011uj},
which was used to generate both signal and background events.
The couplings of the vectorlike leptons were 
discussed above. These couplings are included in the model files 
of FeynRules \cite{Alloul:2013bka} 
to calculate the Feynman rules for the implementation into Madgraph. 
PYTHIA \cite{Sjostrand:2006za} was used for showering and hadronization.  
In order to do the detector simulation we used 
Delphes 3 \cite{deFavereau:2013fsa}. 
In some cases below we found it useful to veto $b$ jets 
in order to reduce backgrounds including 
$t \overline t Z$ and $t \overline t W$ and $t \overline t h$.
We chose the $b$-tagging efficiency for true $b$ jets to be 0.7, 
the efficiency of mistagging a charm jet as 
$b$ jet was 0.1, while for up, down and strange the 
mistagging efficiency was chosen to be 0.001.
We have used the default Delphes tau tagging efficiency of 0.4, 
and tau misidentification rate for QCD jets is 0.001. 

The main physics backgrounds for multilepton channels are $WZ$, $ZZ$, 
$t\overline{t}Z$, $t\overline {t}W$, $hZ$, $t\overline{t}h$, 
$WWZ$, $ZZW$, $ZZZ$, $hh$. They have also been simulated by Madgraph. 
We used $K$ factors found from 
NLO and NNLO cross sections at 8 TeV from
\cite{Campbell:1999ah,Campanario:2008yg,Garzelli:2012bn,Heinemeyer:2013tqa}. 
At 8 TeV the $K$ factors are 1.58, 1.47, 1.38, 1.58, 1.315, 1.44, 1.8, 
1.59, 1.591 following the same order of backgrounds as above excluding $hh$. 
Production of $hh$ background includes 
the triangular and the box diagram. But the box diagram is not include 
in the Madgraph ``heft'' model package, so we generated 
events for $hh$ using ``heft'' but we took the cross-section 
for $hh$ production from \cite{Chen:2013emb}.
We took the same $K$ factors to approximate the cross-sections at 13 TeV.
Except for the SM Higgs boson ($h$), every other particle in the background processes 
was forced to decay leptonically (including to tau leptons) 
in order to increase the yield in the simulation. 
In the cases of $h$ decays to $WW$, $ZZ$, $b\bar b$, $\tau^+\tau^-$,
and $gg$, we modified the Madgraph couplings of $h$ to ensure agreement
of the branching ratios with the theoretical predictions from HDECAY 
\cite{Djouadi:1997yw}
for $M_h = 125$ GeV.
At each of $\sqrt{s} = 8$ and 13 TeV, we generated 100,000 events for 
each of the backgrounds except for $hh$ where we generated 500,000 events. 
To be conservative, we did not include $K$ factors for the signal processes.
The $K$-factors for $SU(2)_L$ triplets was recently found \cite{Ruiz:2015zca} 
to be in the range of about
1.17 to 1.2 for $\sqrt{s}=13$ TeV, and the results for singlets and doublets 
should be about the same.

In the following sections,
our study is divided in the following manner. First we looked
at existing $\sqrt{s} = 8$ TeV multilepton searches by ATLAS, which 
were originally aimed at 
supersymmetric models, but are re-purposed here for vectorlike leptons. Unfortunately,
we find no sensitivity to the Singlet VLL model here, so our analysis is confined
to the Doublet VLL model. We then propose more inclusive 4-lepton
searches, which are studied for the Doublet VLL model with
$\sqrt{s}=8$ TeV.
Finally, we consider the prospects for 
4-lepton and 5-lepton signals at $\sqrt{s}=13$ TeV 
for both the Doublet and Singlet VLL models, as well as an optimistic variant of the
Singlet VLL model in which one arbitrarily takes BR$(\tau' \rightarrow Z\tau)=1$.


\section{ATLAS multilepton searches at $\sqrt{s} = 8$ TeV\label{sec:ATLAS8TeV}}
\setcounter{equation}{0}
\setcounter{figure}{0}
\setcounter{table}{0}
\setcounter{footnote}{1}

As discussed above, vectorlike lepton models have a good possibility to 
provide a beyond Standard Model signature when pair-produced at LHC and 
multilepton final-state channels are considered. In the following 
section we study the Doublet VLL Model first at $\sqrt{s} = 8$ TeV. We 
find that there is an opportunity to set limits on this model by using 
existing ATLAS searches at LHC \cite{ATLASmultilepton,ATLAS4l} on 
3-lepton and 4-lepton channels at $\sqrt{s} = 8$ TeV with $\int L$ dt = 
20.3 fb$^{-1}$. We look at 3-lepton channels first,
and compared the visible signal cross sections after cuts 
with the limits from \cite{ATLASmultilepton}. Similarly, for the 4-lepton 
analysis we compared with limits from \cite{ATLAS4l}. While studying the 
VLL Doublet model we always refer to 
$M_{\tau^{\prime}}=M_{\nu^{\prime}}$ as $M_{\tau^{\prime}}$. 
For our study, we generated
100,000 signal events at $\sqrt{s}=8$ TeV, for each of $M_{\tau'} =$ 
110, 130, 150, 180, 200, 250, 300, 400, and 500 GeV.

\subsection{Three-lepton searches for the Doublet VLL model}
\setcounter{equation}{0}
\setcounter{footnote}{1}
In this section we consider a search strategy based on requiring at least 
three leptons, following the selection criteria used by the ATLAS 
search at $\sqrt{s}=8$ TeV 
and $\int L$ dt = 20.3 fb$^{-1}$, 
described in \cite{ATLASmultilepton}. 
Lepton candidates $(e,\mu,\tau_h)$ are required to satisfy:
\beq
p_{T}  &>& \mbox{15 GeV},
\\
|\eta| &<& 2.4,
\label{eq:etacut}
\\
\Delta R_{\ell,\ell'} &>& 0.1\>\> (\mbox{for each $\ell,\ell' = e,\mu,\tau_{h}$}). 
\\
\Delta R_{\ell j} &>& 0.3\>\> (\mbox{for each jet and $\ell = e,\mu,\tau_{h}$}). 
\label{eq:DeltaRcut}
\eeq
Events are then selected with at least three leptons, with at least one electron or muon satisfying a $p_T$ trigger requirement:
\beq
N(e,\mu,\tau_h) &\geq& 3,
\\
p_{T} (e_1/\mu_1) &>& \mbox{26 GeV}.
\eeq
After this selection, events are classified into two channels. One is 
events with at least three electron or muon candidates, and the other is 
events with exactly two electrons or muons and at least one hadronic tau 
($2e/\mu+\geq 1\tau_h$). Events are then further classified into three 
categories. The first category is events with at least one Opposite Sign 
Same Flavor (OSSF) pair of leptons with 2-body invariant mass within 20 
GeV of the $Z$ boson mass. This category is referred to as 
on-$Z$. The second category is events with an OSSF pair that does not 
satisfy the on-$Z$ requirement, and this category is called off-$Z$ 
OSSF. All the remaining events contribute to the last category which is 
off-$Z$ no-OSSF. Tables \ref{tab:table1} and \ref{tab:table2} list the 
visible cross sections we find after cuts for the signals in each 
category, for the Doublet VLL model with various $M_{\tau'}$, along with 
the corresponding ATLAS limits from \cite{ATLASmultilepton}.
\begin{table}
\caption{Visible cross sections $\sigma_v$
in the $\geq 3e/\mu$ channel that pass 
the on-$Z$, off-$Z$, and off-$Z$, no-OSSF selections, in the Doublet VLL model
at $\sqrt{s} = 8$ TeV.
The last line shows the ATLAS limit obtained  
at $\sqrt{s} = 8$ TeV with $\int L$ dt = 20.3 fb$^{-1}$, 
from \cite{ATLASmultilepton}.}
\label{tab:table1}
\centering
\begin{center}
\begin{tabular}{|c|c|c|c|}
\hline
$M_{\tau^\prime}$ (GeV)	& $\sigma_{v}$ on-$Z$ (fb) & $\sigma_{v}$ off-$Z$ OSSF (fb) & $\sigma_{v}$ off-$Z$ no-OSSF (fb) \\
\hline
110&	18.54  &2.25  &1.18\\
130&	15.40  &3.27  &1.62\\
150&	11.50  &2.76  &1.67\\
180&	6.49	&1.92  &1.27\\
200&	4.50	&1.65	&1.10\\
250&	2.00	&1.02   &0.56\\
300&	0.96	&0.57   &0.32\\
400&	0.26	&0.21   &0.11\\
500&	0.08	&0.08	&0.04\\
\hline
ATLAS limit & 31 & 2.5 & 0.89\\
\hline
\end{tabular}
\vspace{-0.3cm}
\end{center}
\caption{Visible cross sections $\sigma_v$
in the $2e/\mu + \geq 1 \tau_h$ channel that pass 
the on-$Z$, off-$Z$, and off-$Z$, no-OSSF selections, in the Doublet VLL model
at $\sqrt{s} = 8$ TeV.
The last line shows the ATLAS limit obtained at $\sqrt{s} = 8$ TeV  
with $\int L$ dt = 20.3 fb$^{-1}$ 
from \cite{ATLASmultilepton}.}
\label{tab:table2}
\centering
\begin{center}
\begin{tabular}{|c|c|c|c|}
\hline
$M_{\tau^\prime}$ (GeV)	& $\sigma_{v}$ on-$Z$ (fb) & $\sigma_{v}$ off-$Z$ OSSF (fb) & $\sigma_{v}$ off-$Z$ no-OSSF (fb) \\
\hline
110&	12.96	&3.24	&6.19\\
130&	11.58	&2.78	&8.08\\
150&	7.57	&2.25	&7.08\\
180&	3.92	&1.43	&5.18\\
200&	2.68	&1.16	&4.18\\
250&	1.04	&0.57	&2.07\\
300&	0.47	&0.33	&1.10\\
400&	0.11	&0.11	&0.34\\
500&	0.03	&0.04	&0.12\\
\hline
ATLAS limit &  207 & 14.0 &4.3 \\
\hline
\end{tabular}
\end{center}
\end{table}

\begin{figure}[!tb]
\begin{minipage}[]{0.495\linewidth}
\includegraphics[width=7.5cm,angle=0]{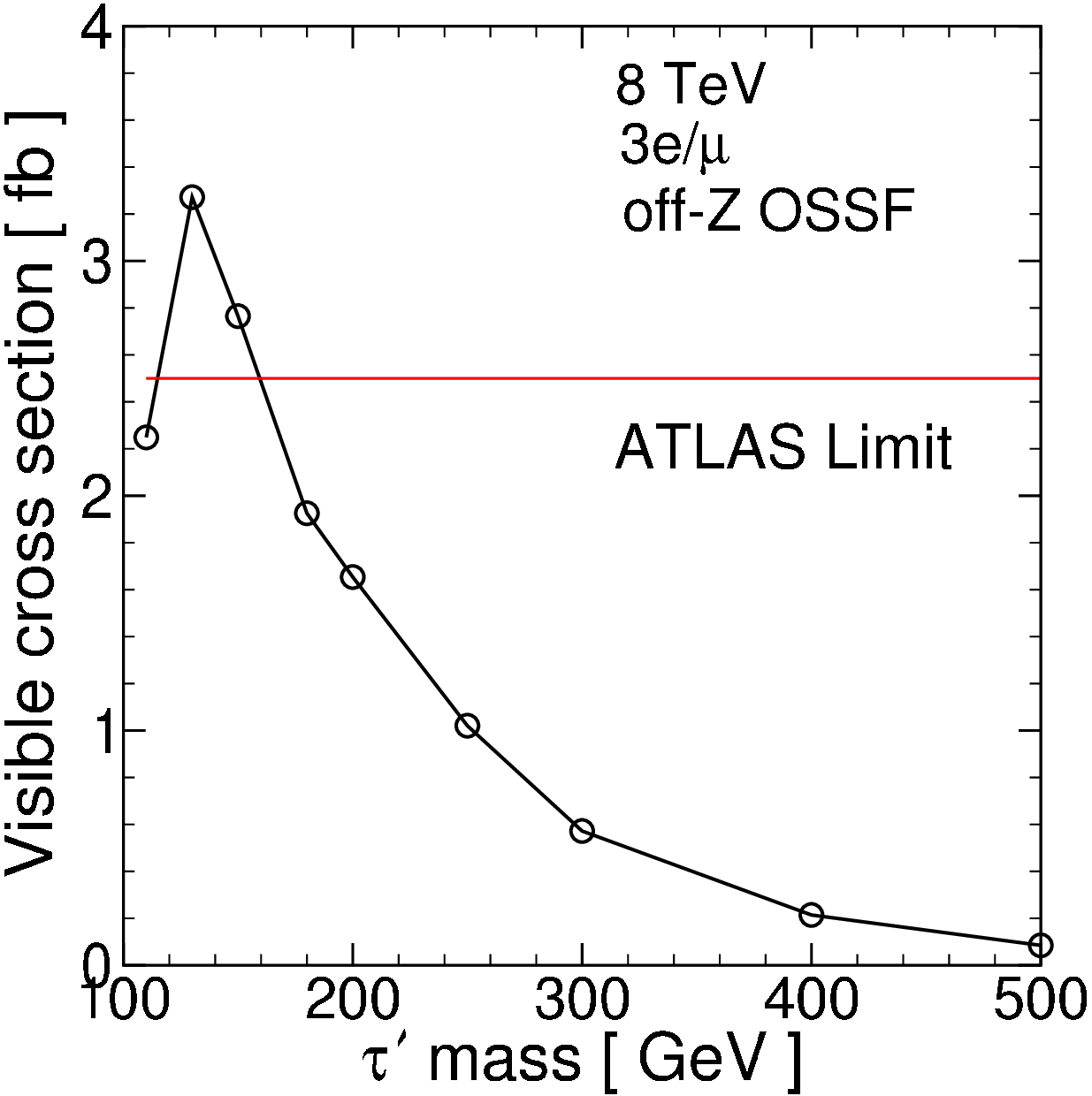}
\end{minipage}
\begin{minipage}[]{0.495\linewidth}
\begin{flushright}
\includegraphics[width=7.5cm,angle=0]{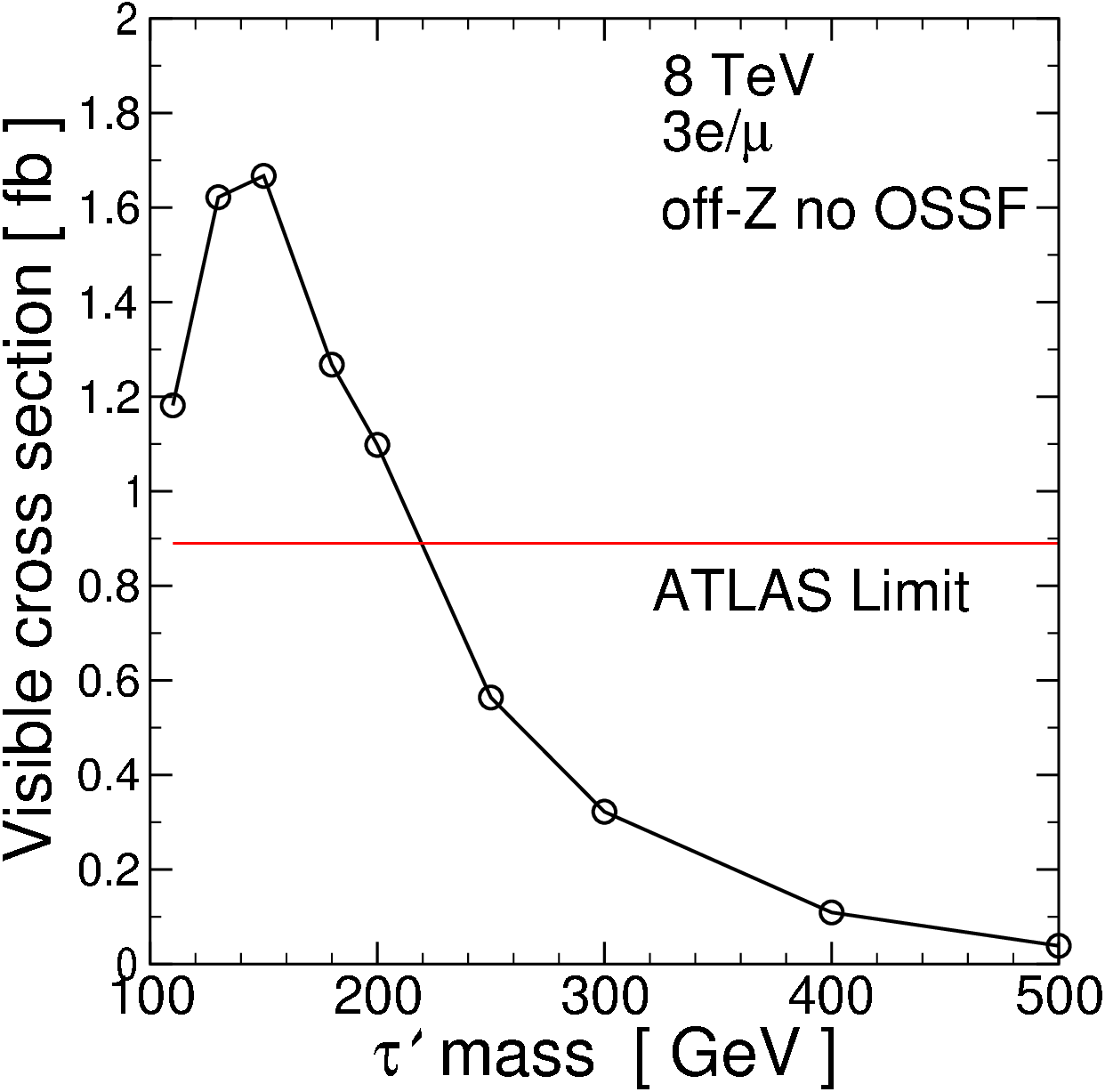}
\end{flushright}
\end{minipage}
\vspace{0.4cm}\\
\begin{minipage}[]{0.50\linewidth}
\includegraphics[width=7.5cm,angle=0]{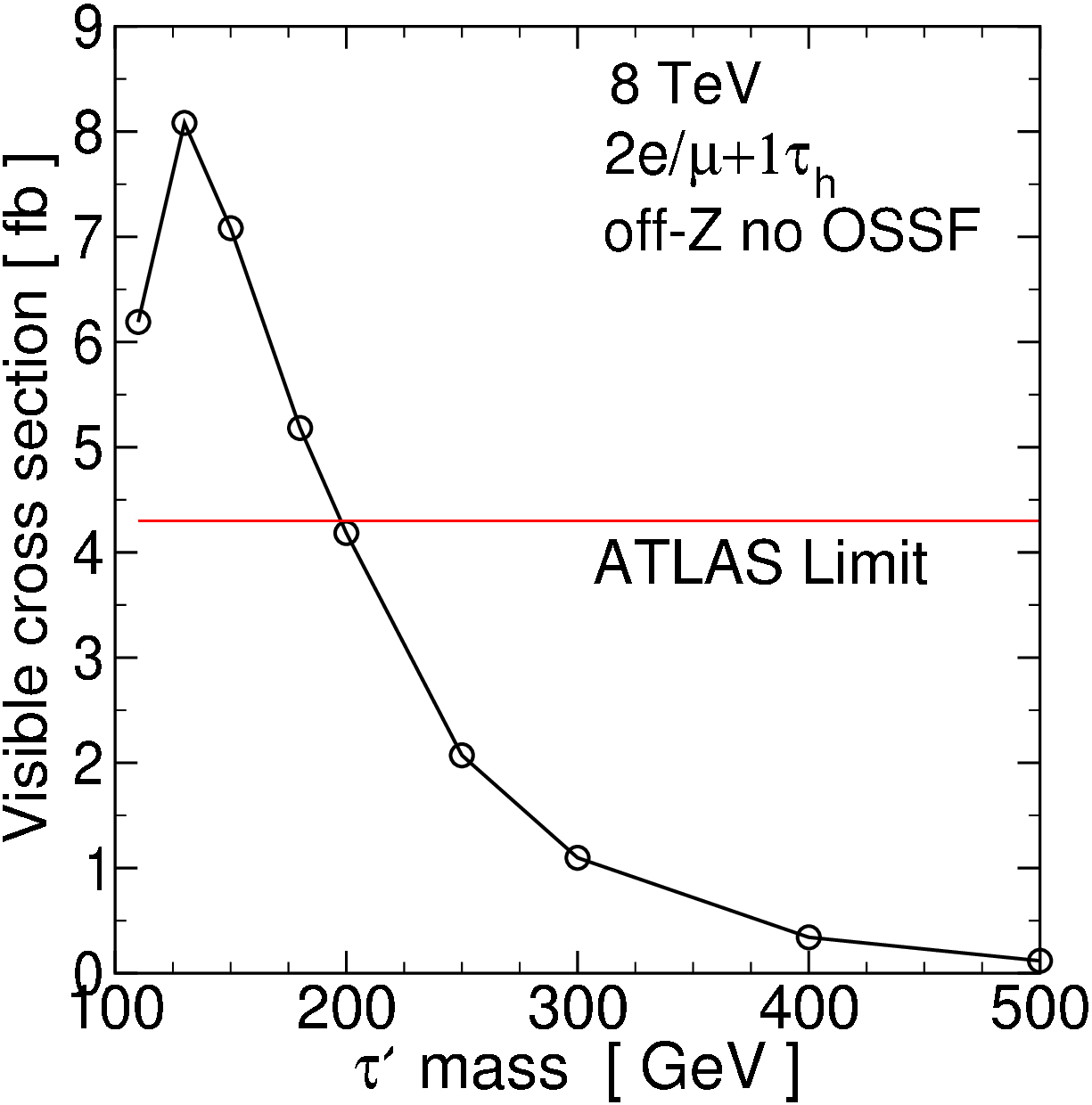}
\end{minipage}
\begin{minipage}[]{0.4\linewidth}
\phantom{x}
\end{minipage}
\begin{minipage}[]{0.45\linewidth}
\caption{\label{fig:3l}
Total visible cross sections $\sigma_v$, for the Doublet VLL model,
that pass selections:
$\geq 3e/\mu$ off-$Z$ OSSF (top left),
$\geq 3e/\mu$ off-$Z$ no-OSSF (top right), and 
$2e/\mu+ \geq 1 \tau_h$ off-$Z$ no-OSSF (bottom left).
The results are given
as functions of $M_{\tau^{\prime}} = M_{\nu'}$, at $\sqrt{s} = 8 $ TeV,
and are as given in Tables \ref{tab:table1} and \ref{tab:table2}. 
For comparison, the ATLAS limits at $\sqrt{s} = 8 $ TeV and 
$\int L$ dt = 20.3 fb$^{-1}$ 
from ref.~\cite{ATLASmultilepton} are represented by the horizontal red lines.
}
\end{minipage}
\end{figure}

From Tables \ref{tab:table1} and \ref{tab:table2}, we see that there are 
three categories in which the signal cross-sections after cuts can 
exceed the ATLAS bounds for a range of $M_{\tau'}$. Those cases are 
$\geq 3e/\mu$ off-$Z$ OSSF, and $\geq 3e/\mu$ off-$Z$ no-OSSF, and 
$2e/\mu + \geq 1 \tau_h$ off-$Z$ no-OSSF. The estimated visible 
cross-sections and corresponding ATLAS limits for these cases are 
depicted graphically in Figure \ref{fig:3l}. The best reaches seem to be 
in the off-$Z$ no-OSSF channels, where our estimates for the visible 
signal cross-section exceeds the ATLAS limit for all masses up to about 
$M_{\tau'}$ = 200 GeV for both $\geq 3e/\mu$ and $2e/\mu + \geq 1 
\tau_h$, where we expect more than about 22 and 84 signal events, 
respectively. It must be kept in mind that our studies based on 
Madgraph, Pythia and Delphes are certainly only an approximation to the 
real ATLAS (or CMS) experimental responses to signal events. We have not 
attempted to perform a detailed validation of our estimates in this 
case, as a true exclusion can only be established by the experimental 
collaborations in any case. Nevertheless, we can conclude from this 
study that using the 3-lepton channels, there is at least a possibility 
to set a limit on vectorlike lepton production in the Doublet VLL model 
using existing LHC Run 1 data at $\sqrt{s}=8$ TeV data.

\subsection{Four-lepton searches for the Doublet VLL model}
\setcounter{equation}{0}
\setcounter{footnote}{1}
In this section we consider 4-lepton signals for the Doublet VLL model 
at $\sqrt{s}=8$ TeV, this time using 
the selection criteria of the ATLAS search reported in ref.~\cite{ATLAS4l}.
Lepton candidates are required to satisfy:
\beq
\mbox{$p_{T} (e,\mu) > 10$ GeV},
\\
\mbox{$p_{T} (\tau_{h}) > 20$ GeV},
\eeq
along with the same pseudo-rapidity and isolation requirements of 
eqs.~(\ref{eq:etacut}) and (\ref{eq:DeltaRcut}) above. 
Events are then required to have at least 4 leptons, of which at least 2 must be $e,\mu$:
\beq
N(e,\mu,\tau_h) &\geq& 4,
\\
N(e,\mu) &\geq& 2,
\eeq
and to pass through at least 
one of the following four trigger criteria:
\begin{itemize}
\item $p_{T} (e_1/\mu_1) > 25$ GeV for a single isolated $e$ or $\mu$.
\item $p_{T} (e_1) > 14$ GeV, $p_{T} (e_2) > 14$ GeV or $p_{T} (e_1) > 25$ GeV, $p_{T} (e_2) > 10$ GeV for double $e$.
\item $p_{T} (\mu_1) > 14$ GeV, $p_{T} (\mu_2) > 14$ GeV or $p_{T} (\mu_1) > 18$ GeV, $p_{T} (\mu_2) > 10$ GeV for double $\mu$.
\item $p_{T} (e_1/\mu_1) > 14$ GeV, $p_{T} (e_1/\mu_1) > 10$ GeV or $p_{T} (e_1/\mu_1) > 18$ GeV, $p_{T} (e_2/\mu_2) > 10$ GeV for $e+\mu$ events.
\end{itemize}
After these selections, events are classified in three signal regions, 
which are called SR0, SR1, SR2, following \cite{ATLAS4l}. These have, 
respectively: at least 4 $e/\mu$, exactly 3 $e/\mu$ and at least 1 
$\tau_h$, and exactly 2 $e/\mu$ and at least 2 $\tau_h$. Events are then 
further classified into two categories, called no-$Z$ and on-$Z$, which 
respectively veto against the presence of a $Z$ boson or require the 
presence of $Z$ boson. This is done by looking for an OSSF pair of 
leptons ($e$ or $\mu$) that yield invariant mass values in the $M_Z \pm 
10$ GeV interval. The no-$Z$ class is further divided into two regions, 
a and b, classified by $\ETmiss$ and $\meff$ as defined in Table 5 of 
\cite{ATLAS4l}. Hence the signal is studied in nine signal regions in 
all. Tables \ref{tab:table3}, \ref{tab:table4}, and \ref{tab:table5} 
show results from our simulation 
for the visible signal cross sections for the Doublet 
VLL model in each category, as well as the corresponding ATLAS limits 
from ref.~\cite{ATLAS4l}.
\begin{table}[!p]
\caption{Total visible cross sections $\sigma_v$  
in the $\geq \!4e/\mu$ channels 
that pass three different selection requirements
for the signal regions, for the Doublet VLL model, with $\sqrt{s} = 8$ TeV. 
The last line shows 
the ATLAS limit obtained
at $\sqrt{s} = 8$ TeV with $\int L$ dt = 20.3 fb$^{-1}$ 
from \cite{ATLAS4l}.}
\label{tab:table3}
\centering
\begin{center}
\begin{tabular}{|c|c|c|c|}
\hline
$M_{\tau^\prime}$ (GeV)	& SR0noZa $\sigma_{v}$ (fb) & SR0noZb $\sigma_{v}$ (fb) & SR0Z  $\sigma_{v}$ (fb) \\
\hline
110&	0.195&	0.133	&0.589\\
130&	0.112&	0.047	&0.488\\
150&	0.262&	0.144	&0.502\\
180&	0.184&	0.100	&0.429\\
200&	0.132&	0.080	&0.399\\
250&	0.097&	0.073	&0.221\\
300&	0.047&	0.036	&0.131\\
400&	0.019&	0.017	&0.046\\
500&	0.006&	0.006	&0.016\\
\hline
ATLAS limit & 0.29 & 0.18 & 0.40 \\
\hline
\end{tabular}
\vspace{-0.1cm}
\end{center}
\caption{Total visible cross sections $\sigma_v$  
in the $3e/\mu\, +\! \geq\! 1 \tau_h$ channels 
that pass three different selection requirements
for the signal regions, for the Doublet VLL model, at $\sqrt{s} = 8$ TeV. 
The last line shows the ATLAS limit obtained
at $\sqrt{s} = 8$ TeV with $\int L$ dt = 20.3 fb$^{-1}$ 
from \cite{ATLAS4l}.}
\label{tab:table4}
\centering
\begin{center}
\begin{tabular}{|c|c|c|c|}
\hline
$M_{\tau^\prime}$ (GeV)	& SR1noZa $\sigma_{v}$ (fb) & SR1noZb $\sigma_{v}$ (fb) & SR1Z  $\sigma_{v}$ (fb) \\
\hline
110&	0.312   &0.070	&0.191\\
130&	0.421	&0.104	&0.208\\
150&	0.421	&0.113	&0.159\\
180&	0.346	&0.137	&0.217\\
200&	0.270	&0.156	&0.178\\
250&	0.178	&0.131	&0.127\\
300&	0.098	&0.089	&0.078\\
400&	0.032	&0.032	&0.029\\
500&	0.012	&0.012	&0.011\\
\hline
ATLAS Limit  &  0.28  & 0.17 &  0.26 \\
\hline
\end{tabular}
\vspace{-0.1cm}
\end{center}
\caption{Total visible cross sections $\sigma_v$  
in the $2e/\mu\, +\! \geq\! 2 \tau_h$ channel 
that pass three different selection requirements
for the signal regions, in the Doublet VLL model, at $\sqrt{s} = 8 $ TeV. 
The last line shows the ATLAS limit obtained
at $\sqrt{s} = 8$ TeV with $\int L$ dt = 20.3 fb$^{-1}$ 
from \cite{ATLAS4l}.
\label{tab:table5}}
\begin{center}
\begin{tabular}{|c|c|c|c|}
\hline
$M_{\tau^\prime}$ (GeV)	& SR2noZa $\sigma_{v}$ (fb) & SR2noZb $\sigma_{v}$ (fb) & SR2Z  $\sigma_{v}$ (fb) \\
\hline
110&	0.078&	0.054&	0.061\\
130&	0.109&	0.042&	0.106\\
150&	0.111&	0.065&	0.080\\
180&	0.111&	0.059&	0.046\\
200&	0.109&	0.065&	0.072\\
250&	0.083&	0.056&	0.034\\
300&	0.051&	0.043&	0.021\\
400&	0.017&	0.017&	0.008\\
500&	0.006&  0.006&	0.002\\
\hline
ATLAS Limit & 0.45 & 0.43 & 0.17 \\ 
\hline
\end{tabular}
\end{center}
\end{table}

\begin{figure}[]
\begin{minipage}[]{0.495\linewidth}
\includegraphics[width=7.5cm,angle=0]{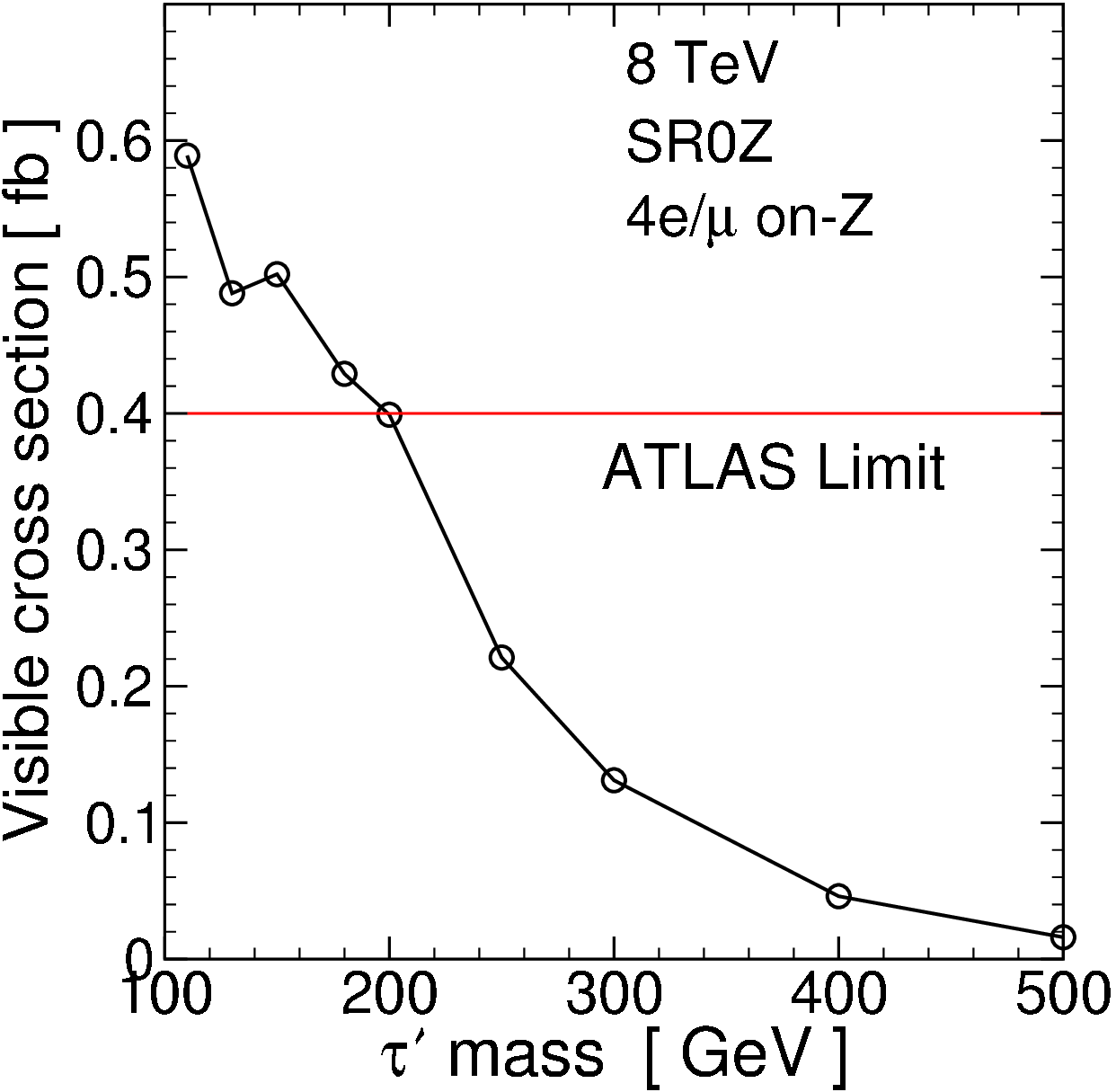}
\end{minipage}
\begin{minipage}[]{0.495\linewidth}
\begin{flushright}
\includegraphics[width=7.5cm,angle=0]{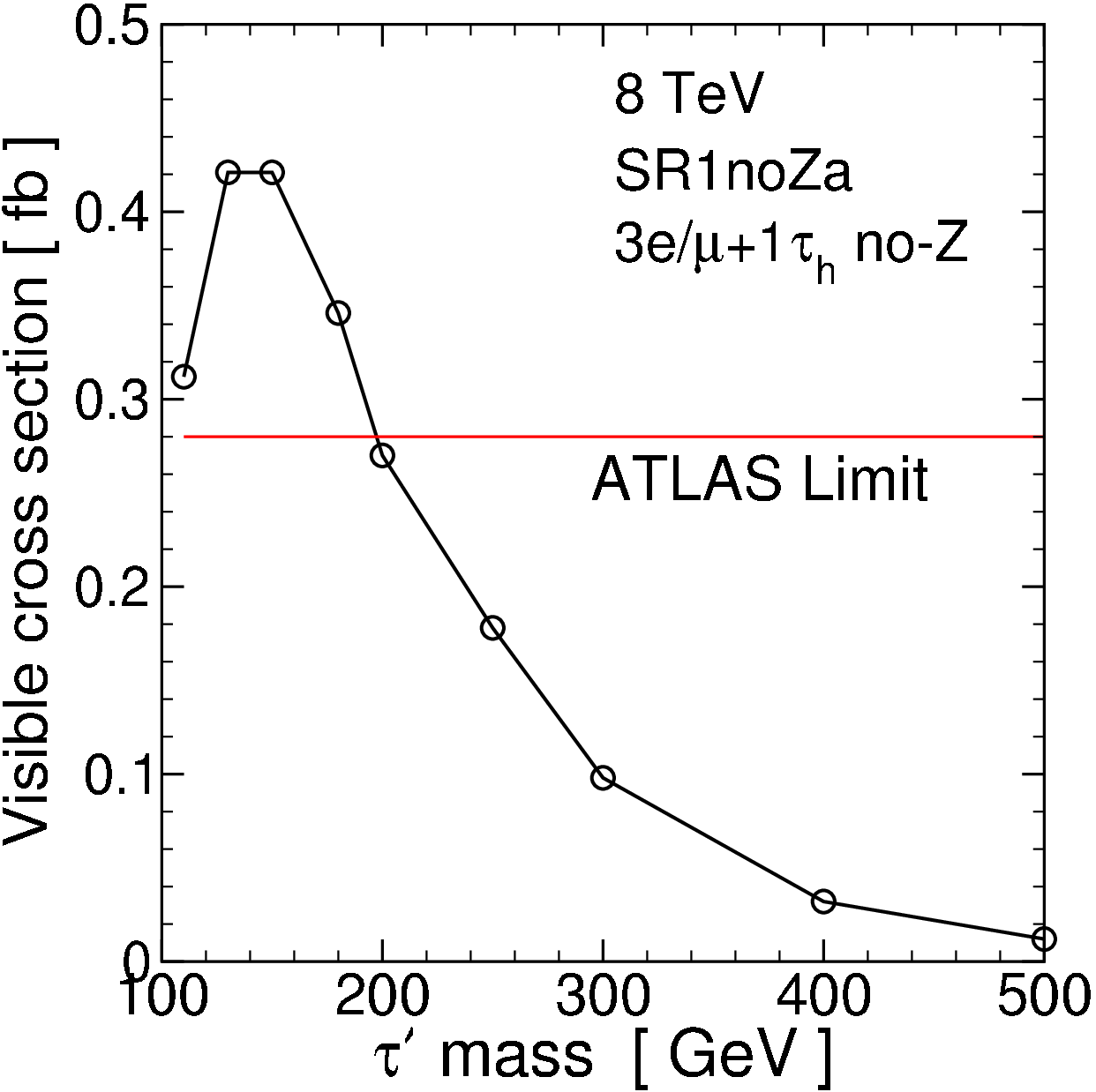}
\end{flushright}
\end{minipage}
\caption{\label{fig:4l}
Total visible cross sections $\sigma_v$, for the Doublet VLL model,
that pass selections
$\geq 4e/\mu$ on-$Z$ or SR0Z (left), and 
$3e/\mu+ \geq 1 \tau_h$ no-$Z$ or SR1noZa (right).
The results are shown
as a function of $M_{\tau^{\prime}} = M_{\nu'}$, at $\sqrt{s} = 8$ TeV,
and correspond to entries in Tables \ref{tab:table3} and \ref{tab:table4}. 
Also, the ATLAS limits at $\sqrt{s} = 8 $ TeV and $\int L$ dt = 20.3 fb$^{-1}$ 
from ref.~\cite{ATLAS4l} are represented by the horizontal red lines.
}
\end{figure}

It can be concluded from Tables \ref{tab:table3}, 
\ref{tab:table4}, and \ref{tab:table5} that the two signal regions
with the best reach for the Doublet VLL model  
are SR0Z (with at least 4 $e/\mu$ and a $Z$ candidate) and 
SR1noZa (with exactly 3 $e/\mu$, at least one $\tau_h$, and no $Z$ candidate). 
These results are shown in Figure \ref {fig:4l} as a function of $M_{\tau'}= M_{\nu'}$. 
As in the 3-lepton signal, we find that the predicted Doublet VLL model 
visible cross-section after cuts can exceed the ATLAS limit for masses 
below about 200 GeV. The same caveats apply as in the previous subsection, 
so we cannot claim an exclusion, but we simply 
note that these results are suggestive
that such an exclusion may be possible with existing LHC data at $\sqrt{s}=8$ TeV
using these signal regions.
\clearpage

\section{More inclusive four-lepton searches at $\sqrt{s} = 8$ TeV\label{sec:8TeV}}
\setcounter{equation}{0}
\setcounter{figure}{0}
\setcounter{table}{0}
\setcounter{footnote}{1}

The ATLAS searches of ref.~\cite{ATLAS4l} were aimed at supersymmetric models,
and therefore included cuts on $\meff$ and $\ETmiss$. These cuts are not 
necessarily particularly appropriate for vectorlike lepton searches. Therefore,
in this section we look at a different, simpler and more inclusive, 
strategy for 4-lepton searches 
to see if a better reach for the Doublet VLL Model can be achieved. 

\begin{figure}[!t]
\begin{minipage}[]{0.50\linewidth}
\includegraphics[width=8.2cm,angle=0]{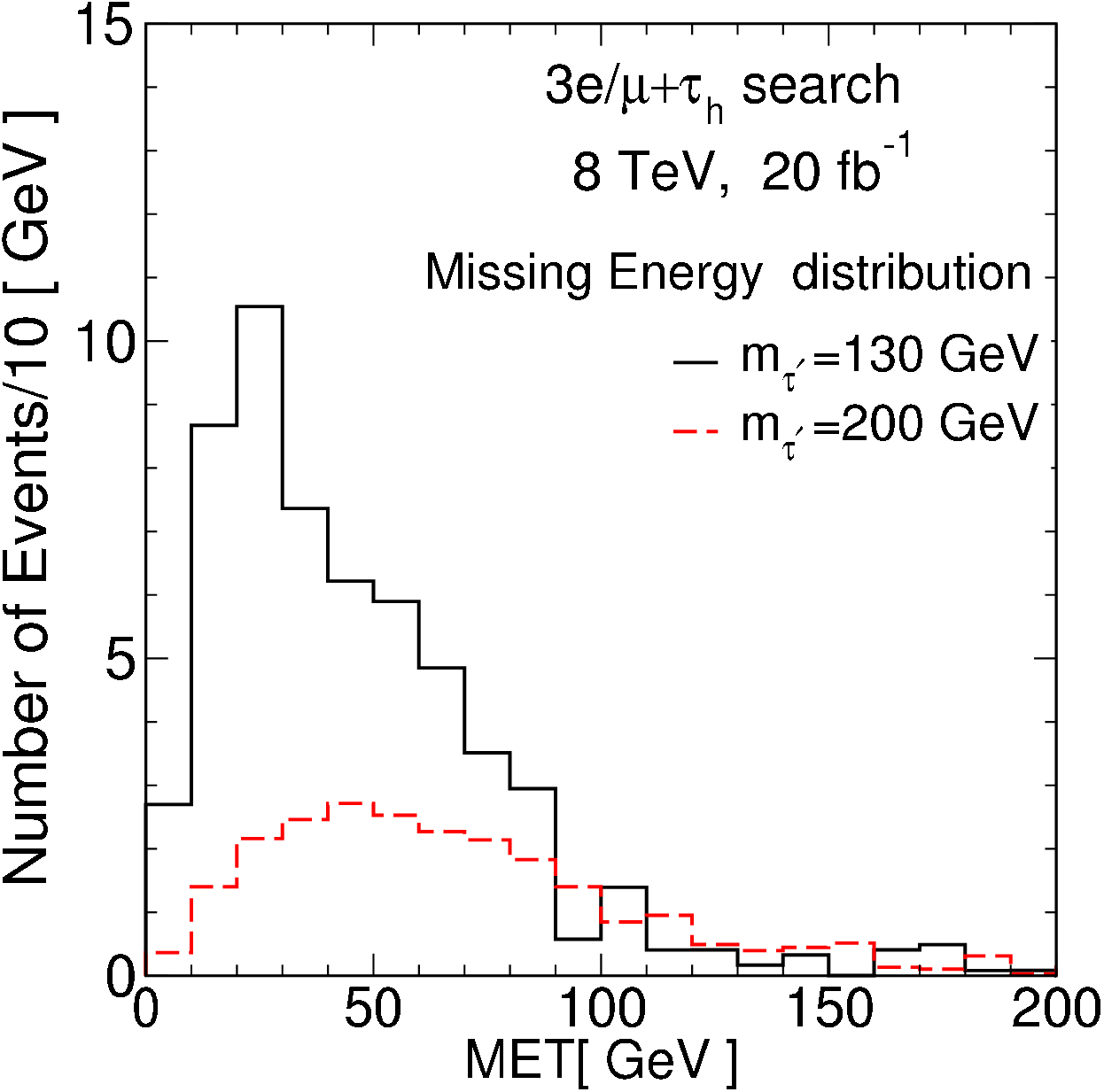}
\end{minipage}
\begin{minipage}[]{0.4\linewidth}
\phantom{x}
\end{minipage}
\begin{minipage}[]{0.45\linewidth}
\caption{\label{fig:ET}The expected $\ETmiss$ distribution
at $\sqrt{s} = 8$ TeV in the $\geq 3e/\mu+1\tau_h$ 
channel after inclusive event selection 
for the Doublet VLL model with
$M_{\tau^{\prime}}$ = 130 and 200 GeV. The distributions are normalized 
according to 20 fb$^{-1}$ of integrated luminosity.
} 
\end{minipage}
\end{figure}

In the following, lepton candidates must satisfy
\beq
p_T (e,\mu,\tau_h) &>& \mbox{15 GeV},\\
|\eta (e,\mu,\tau_{h})| &<& 2.5,
\\
\Delta R_{\ell,\ell'} &>& 0.1\>\> (\mbox{for each $\ell,\ell' = e,\mu,\tau_{h}$}). 
\\
\Delta R_{\ell j} &>& 0.3\>\> (\mbox{for each jet and $\ell = e,\mu,\tau_{h}$}). 
\eeq
We then require events to have at least 4 leptons, at least 2 of which must be $e/\mu$, to satisfy a trigger requirement on the leading $e/\mu$, and impose a veto of $b$-jets:
\beq
N(e,\mu,\tau_h) &\geq& 4,
\\
N(e,\mu) &\geq& 2,
\\
p_{T} (e_1/\mu_1) &>& \mbox{26 GeV},
\label{eq:pT26}
\\
N_{b{\rm-tag}} &=& 0.
\eeq
The last requirement is to help suppress $t\bar t+X$ backgrounds. We 
then consider 3 channels. The first one is $\geq 3e/\mu+1\tau_h$, which 
requires at least three $e/\mu$ and at least one $\tau_h$. Similarly we 
define a channel $\geq 2e/\mu+2\tau_h$ with at least two $e/\mu$ and at 
least two $\tau_h$, and a channel $\geq 4e/\mu$ by requiring at least 
four $e/\mu$. (For simplicity, we avoid using $\geq$ sign in front of 
the number of $\tau_h$ requirement here.) Events that pass each of the 
selections just mentioned form categories that we call inclusive. Events 
which pass a further cut that there is no pair of OSSF leptons (e,$\mu$) 
with $M_Z \pm 20$ are called no-$Z$. We do not include a separate 
on-$Z$ category, because we found that the reach is typically very 
similar to the inclusive category. We also do not impose a cut on 
$\ETmiss$, unlike the ATLAS 4-lepton signal region cuts, which were 
aimed at supersymmetry. The reason for this is illustrated in Figure 
\ref{fig:ET}, which shows the $\ETmiss$ distributions for the inclusive 
$\geq 3e/\mu+1\tau_h$ channel, for two different mass values $M_{\tau'} 
= 130$ and 200 GeV. This distribution shows that $\ETmiss<100$ for most 
of the signal events if vectorlike leptons are pair produced according 
to the Doublet VLL Model and events are selected for four lepton 
channels.

The breakdown of background contributions and the total background cross-section
after the cuts for the 6 signal regions above, 
obtained using simulations as described in section \ref{sec:simulation} 
with $\sqrt{s}=8$ TeV, 
are given in Table \ref{tab:table7}. 
The largest backgrounds, even in the no-$Z$ channels, come from $ZZ$, 
with sub-dominant contributions from $hZ$ and $WWZ$, and $WZ$ in the cases 
that use $\tau_h$. The backgrounds from $t\bar t W$ and $t \bar t Z$, and $t \bar t h$
are significantly reduced by our use of a $b$-tag veto.
The Doublet VLL model signal cross-sections for these 4-lepton search channels
are given in Table \ref{tab:table6} for several different values of $M_{\tau'}$, 
along with the total background results from the previous table.
\begin{table}
\caption{Background cross-sections $\sigma_b$  
for four-lepton channels at $\sqrt{s} = 8$ TeV, after inclusive and no-$Z$ selections
as described in section \ref{sec:8TeV}.} 
\label{tab:table7}
\begin{center}
\begin{tabular}{|c|c|c|c|c|c|c|}
\hline
~~~SM Backgrounds~~~ & \multicolumn{2}{c|}{$\sigma_b$ (fb) in $\geq 3e/\mu+1\tau_h$} & \multicolumn{2}{c|}{$\sigma_b$ (fb) in $\geq 2e/\mu+2\tau_h$} & \multicolumn{2}{c|}{$\sigma_b$ (fb) in $\geq 4e/\mu$}\\
\hline
  &  \phantom{xx}incl.\phantom{xx} & no-$Z$ &  \phantom{xx}incl.\phantom{xx} & no-$Z$ &  \phantom{xx}incl.\phantom{xx} & no-$Z$ \\
\hline
$pp\rightarrow W Z $ & 0.0398& 0.0000 &0.0066 &0.0066 &0.0000 &0.0000 \\
$pp\rightarrow Z Z $ & 0.3753& 0.0117 &0.1909 &0.0073 &6.4511 &0.0073\\
$pp\rightarrow t\overline t W$ & 0.0046	&0.0034 & 0.0018 &0.0017 &0.0000	&0.0000	\\
$pp\rightarrow t\overline t Z$ & 0.0087	&0.0016	 & 0.0018 &0.0010 & 0.0196 &0.0005\\
$pp\rightarrow t\overline{t} h$  &0.0038&0.0024 & 0.0027 &0.0024 &0.0019 &0.0009	\\
$pp\rightarrow hh $ & 0.0000& 0.0000 &0.0000 &0.0000 &0.0000 &0.0000\\
$pp\rightarrow hZ $ & 0.0465 &0.0017 & 0.0179 &0.0017 &0.0640 &0.0012	\\
$pp\rightarrow WWZ $ & 0.0094&0.0015 &0.0015 &0.0010 &0.0503 &0.0013	\\
$pp\rightarrow WZZ $ &0.0023&0.0002	&0.0005	&0.0001	 &0.0119 &0.0002	\\
$pp\rightarrow ZZZ $ &0.0005&0.0000 &0.0001	&0.0000 &0.0028	&0.0000\\
\hline
Total Background & 0.4909	&0.0225	 & 0.2237	&0.0218 & 6.6015	&0.0115\\
\hline
\end{tabular}
\end{center}
\caption{Signal and total background cross sections in the
$\geq 3e/\mu+1\tau_h$, $\geq 2e/\mu+2\tau_h$ and 
$\geq 4e/\mu$ channels after selection through inclusive and no-$Z$ 
requirements, for the Doublet VLL model, at $\sqrt{s} = 8$ TeV. 
}
\label{tab:table6}
\centering
\begin{center}
\begin{tabular}{|c|c|c|c|c|c|c|}
\hline
$M_{\tau^\prime}$ (GeV)	& \multicolumn{2}{c|}{$\sigma_{s}$ (fb) in $\geq 3e/\mu+1\tau_h$} & 
\multicolumn{2}{c|}{$\sigma_{s}$ (fb) in $\geq 2e/\mu+2\tau_h$} & \multicolumn{2}{c|}{$\sigma_{s}$ (fb) in $\geq 4e/\mu$} \\
\hline
& inclusive & no-$Z$ 
& inclusive & no-$Z$ 
& inclusive & no-$Z$ \\
\hline
110&	2.539	& 0.319 &	0.876 	&0.280&	1.548	&	0.087 \\
130&	2.869	& 0.508 &	1.396	&0.429&	1.941	&	0.126\\
150&	2.325	& 0.360 &	1.087	&0.347&	1.737   &	0.113\\
180&	1.634	& 0.322 &	0.649	&0.259&	1.144	&	0.121\\
200&	1.179	& 0.244 &	0.551	&0.237&	0.879	&	0.094\\
250&	0.528	& 0.147 &	0.252	&0.134&    0.402   &       0.060\\
300&	0.260	& 0.082 &	0.119	&0.067&    0.188   &       0.030\\
400&	0.075   & 0.027 &	0.033   &0.019&    0.064   &       0.014\\
500&    0.025   & 0.010 &	0.010   &0.007&    0.021   &       0.005\\
\hline
Total Background &0.491 & 0.023 & 0.224 & 0.0218  & 6.602 & 0.012\\
\hline
\end{tabular}
\end{center}
\end{table}

\begin{figure}[!tb]
\begin{minipage}[]{0.495\linewidth}
\includegraphics[width=7.5cm,angle=0]{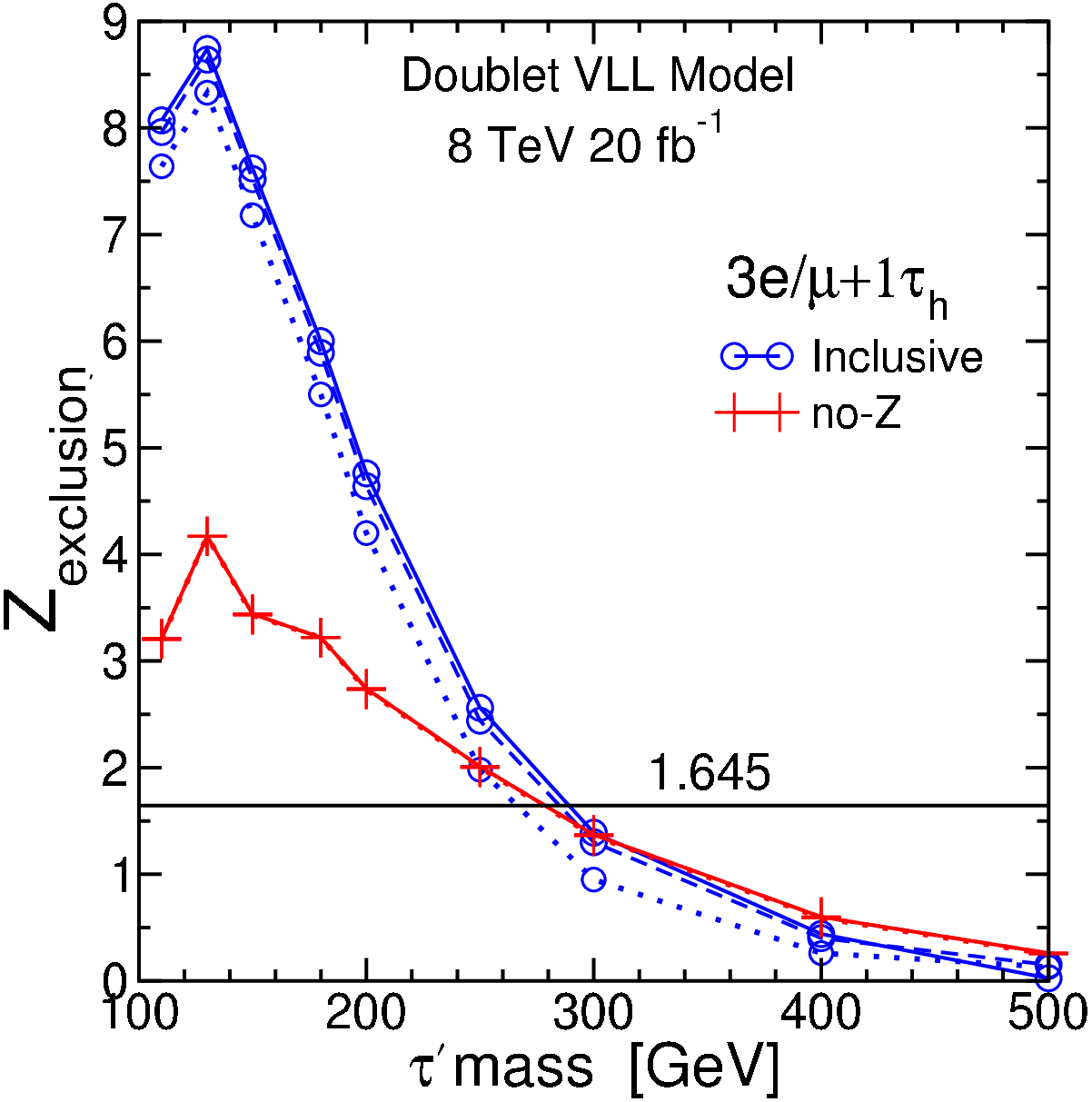}
\end{minipage}
\begin{minipage}[]{0.495\linewidth}
\begin{flushright}
\includegraphics[width=7.5cm,angle=0]{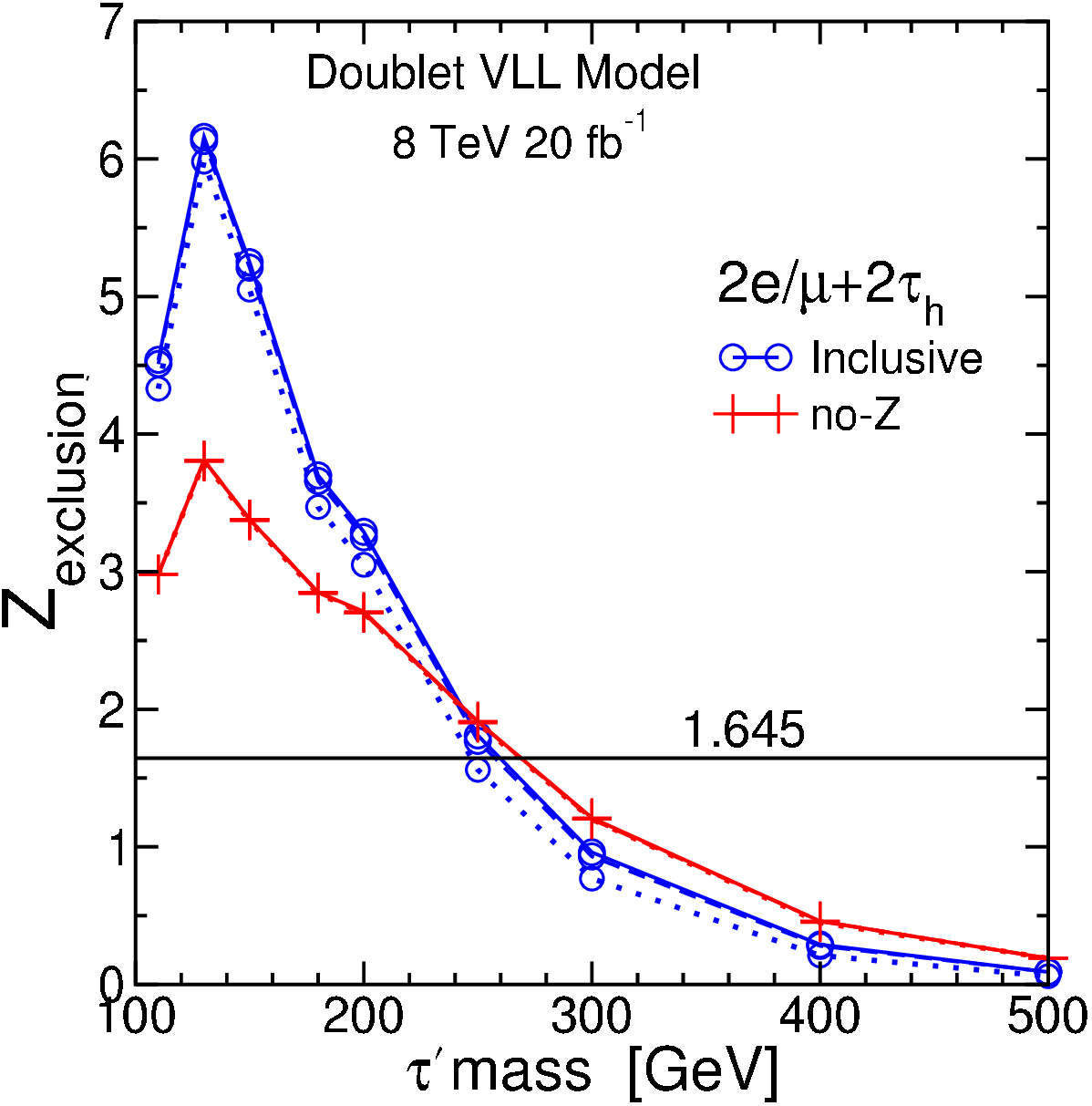}
\end{flushright}
\end{minipage}
\vspace{0.4cm}\\
\begin{minipage}[]{0.50\linewidth}
\includegraphics[width=7.5cm,angle=0]{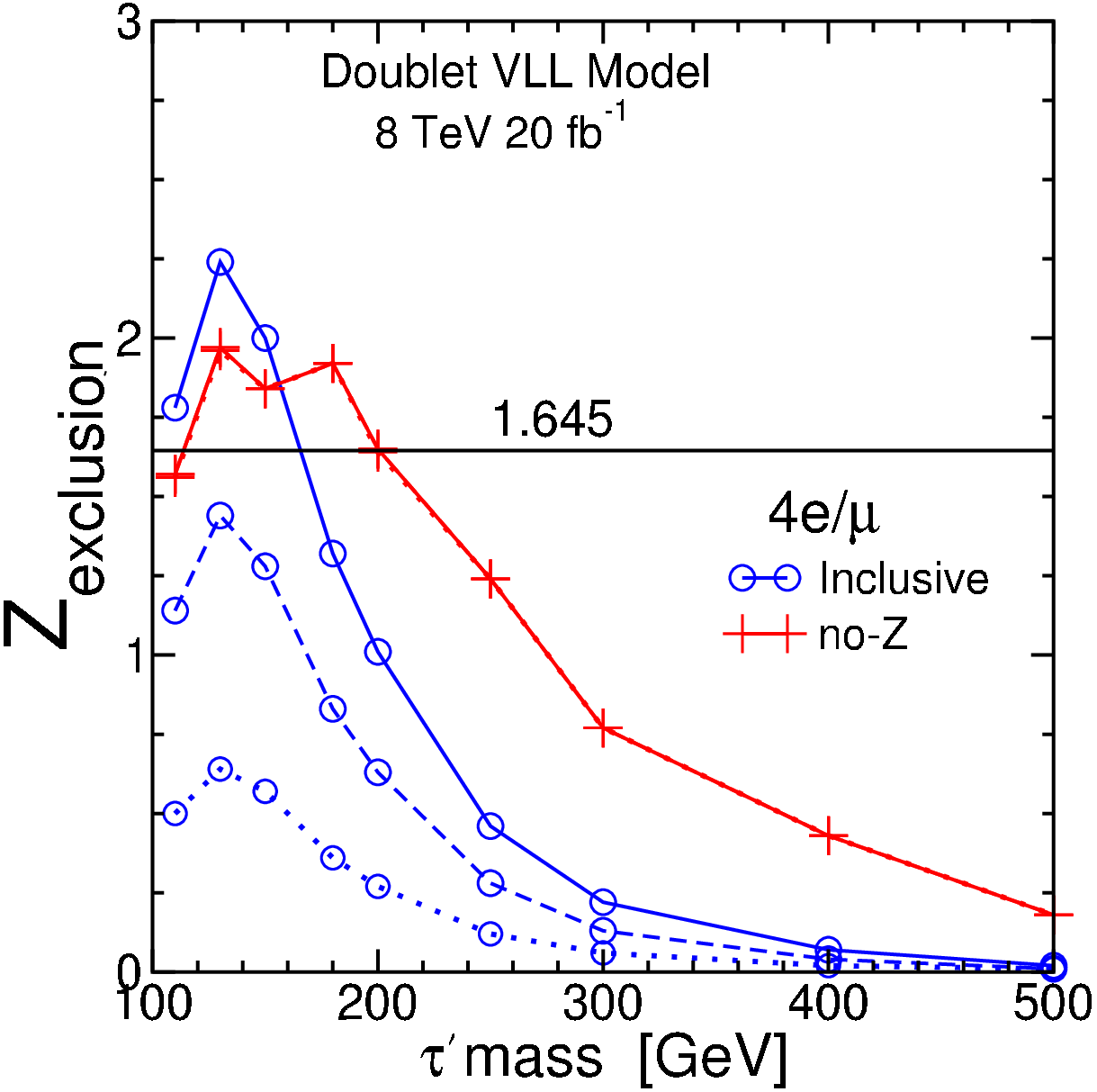}
\end{minipage}
\begin{minipage}[]{0.4\linewidth}
\phantom{x}
\end{minipage}
\begin{minipage}[]{0.45\linewidth}
\caption{\label{fig:Sig_4lept_d}
Median expected significances for exclusion 
at $\int L$ dt = 20 fb$^{-1}$  in 
$\geq 3e/\mu+1\tau_h$ (top left), $\geq 2e/\mu+2\tau_h$ (top right), and 
$\geq 4e/\mu$ (bottom left) channels, 
for the Doublet VLL model,
as a function of $M_{\tau^{\prime}} = M_{\nu'}$, 
at $\sqrt{s} = 8 $ TeV, based on the results of Table \ref{tab:table6}. 
The different lines in each plot
correspond to difference background variances,
$\sigmab= 0.1b$ (solid) and $\sigmab= 0.2b$ (dashed)
and $\sigmab= 0.5b$ (dotted), with blue lines (circle marks) for inclusive and 
red (plus marks) for no-$Z$ on each figure. 
The horizontal black lines are $\Zexc$=1.645 for 95$\%$ CL.}
\end{minipage}
\end{figure}

Using these results, we then calculate the median expected 
exclusion significance 
$\Zexc$ for 20 fb$^{-1}$ at $\sqrt{s}=8$ TeV, assuming $10\%$, $20\%$ and $50\%$ 
fractional uncertainty in the number of background events using 
eq.~(\ref{eq:Zexc}). Hence $\sigmab = 0.1b$, $0.2b$, and $0.5b$, 
where $b$ is the mean total number of background events to pass 
any selection, and $s$ is the corresponding number of signal events. We 
preferred to use this equation because in the real world it is impossible to 
know the backgrounds without uncertainty. This equation also allows us 
to calculate the significance even when $b$ and $s$ are small. 
Figure \ref {fig:Sig_4lept_d} shows the median expected exclusion 
significance predicted by 
eq.~(\ref{eq:Zexc}) for the 6 different 4-lepton channels.

By looking at Figure \ref {fig:Sig_4lept_d} we can say that the
$\geq 3e/\mu+1\tau_h$ channels predict the highest exclusion significance 
($\Zexc$) among the 4-lepton channels. 
We found $\Zexc \geq 1.645$ corresponding to an expected 95\% CL exclusion in the 
inclusive signal region when $M_{\tau'}\leq 265$ GeV, 
even with a fractional 
uncertainty in the background of up to 50\%. 
When the background uncertainty is lower, 
the exclusion reach goes up to about 285 GeV.
The no-$Z$ channel has both 
smaller signals and smaller backgrounds,
and also has exclusion power up to about $M_{\tau'} = 275$ GeV. 
For lower $M_{\tau'}$,
the expected exclusion significance is much higher in the inclusive case 
than in the no-$Z$ channel. Comparable, but slightly weaker, results are 
also found to hold for the $\geq 2e/\mu+2\tau_h$
inclusive and no-$Z$ channels. 
The 4$e/\mu$ channels are seen to be considerably weaker. 
In particular, the inclusive region suffers from a very high background 
and the majority of that is 
from $ZZ$, while the no-$Z$ 4e/$\mu$ channel has a low signal cross-section.
As we expected from the nature of eq.~(\ref{eq:Zexc}), lower 
exclusion significance decreases with increasing
uncertainty in background events for a particular value of 
$M_{\tau'}$, in the case of the inclusive channels. For the low-background
no-$Z$ channels, the dependence on background uncertainty is very mild. 
We have not attempted a combination of the different signals, but this would 
clearly increase the exclusion power. 

In this analysis, our expected exclusions are higher than what we got in 
the previous section where we considered analysis of the ATLAS 4-lepton 
signal regions for the VLL doublet model. There are several reasons for 
that. ATLAS considered $=3e/\mu$ and $=2e/\mu$ but we considered $\geq 
3e/\mu$ and $\geq 2e/\mu$ events and that gave us more events in $\geq 
3e/\mu+1\tau_h$ and $\geq 2e/\mu+2\tau_h$ channels. Our larger $Z$-mass 
window and $b$-jet veto tends to exclude more backgrounds. More 
importantly, the ATLAS 4-lepton signal regions used cuts on $\meff$ and 
on $\ETmiss$, which we did not find very useful, as illustrated above in 
Figure \ref {fig:ET}.  Since we do not have access to the data and our 
signal regions are quite different than those used by ATLAS and CMS 
multilepton searches, it is obvious that our results in this section 
should be considered only as indications of what might be excludable 
using existing $\sqrt{s}=8$ TeV data, rather than as actual exclusions.

\section{Multilepton searches at $\sqrt s =$ 13 TeV\label{sec:13TeV}}
\setcounter{equation}{0}
\setcounter{figure}{0}
\setcounter{table}{0}
\setcounter{footnote}{1}
We saw in the previous section that with the existing LHC data
at $\sqrt{s}=8$ TeV, it should be possible 
to set limits on the production of vectorlike leptons. In this section 
we perform a study of future prospects at $\sqrt{s}=13$ TeV, estimating the 
integrated luminosity required to make a 95$\%$ CL 
exclusion, or an expected $\Zdisc\geq 5$ discovery, 
as a function of $M_{\tau^\prime}$ in both the Doublet and Singlet VLL models. 
To do that, we define 4-lepton and 5-lepton signal regions, 
use simulations to find the visible cross-sections after cuts, 
and then solve $\Zexc=1.645$ using eq.~(\ref{eq:Zexc}) or 
$\Zdisc=5$ using eq.~(\ref{eq:Zdisc}) for the integrated luminosity. 
Our 4-lepton signal regions are the same as in the previous section, and are referred 
to as $\geq 3e/\mu+1\tau_h$, $\geq 2e/\mu+2\tau_h$ and $\geq 4e/\mu$.
We also consider 5-lepton signal regions, which essentially require one extra
$e/\mu$, and will be called 
$\geq 4e/\mu+1\tau_h$, $\geq 3e/\mu+2\tau_h$ and $\geq 5e/\mu$. 
In each case, we consider inclusive and no-$Z$ channels. 
For our study, we generated
100,000 signal events with $\sqrt{s}=13$ TeV, for each of $M_{\tau'} =$ 
110, 130, 150, 180, 200, 250, 300, 400, and 500 GeV.
We have also generated the
backgrounds at $\sqrt{s} = 13$ TeV 
and studied their contributions in each of 
these channels. We consider the Doublet VLL model
first, and then study the prospects for the Singlet VLL model 
as well as a more optimistic variant of it.

\subsection{Four-lepton searches for the Doublet VLL 
model\label{subsec:Doublet4l13TeV}}
\setcounter{equation}{0}
\setcounter{footnote}{1}

\begin{figure}[!tb]
\includegraphics[width=7.5cm,angle=0]{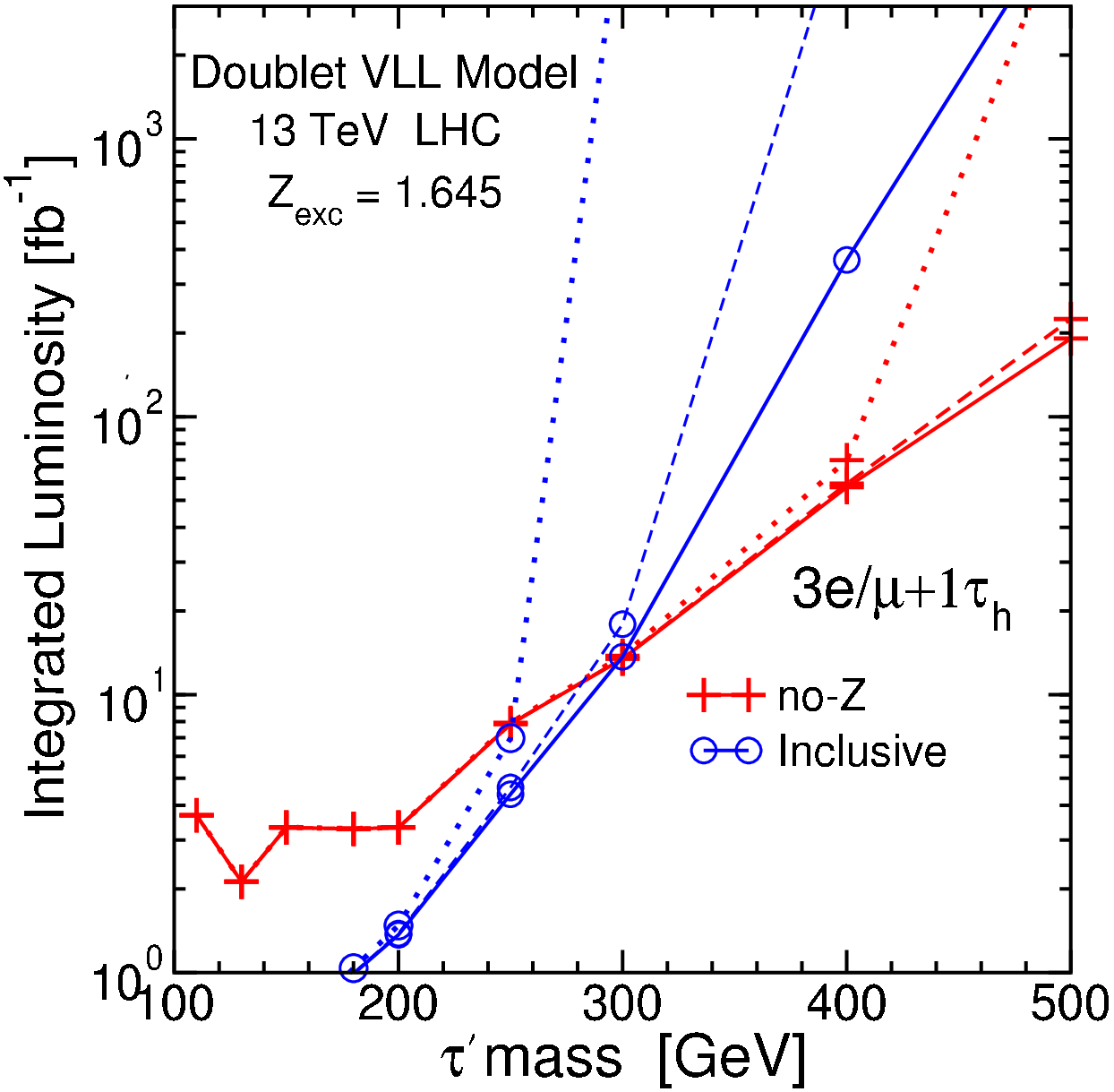}
\includegraphics[width=7.5cm,angle=0]{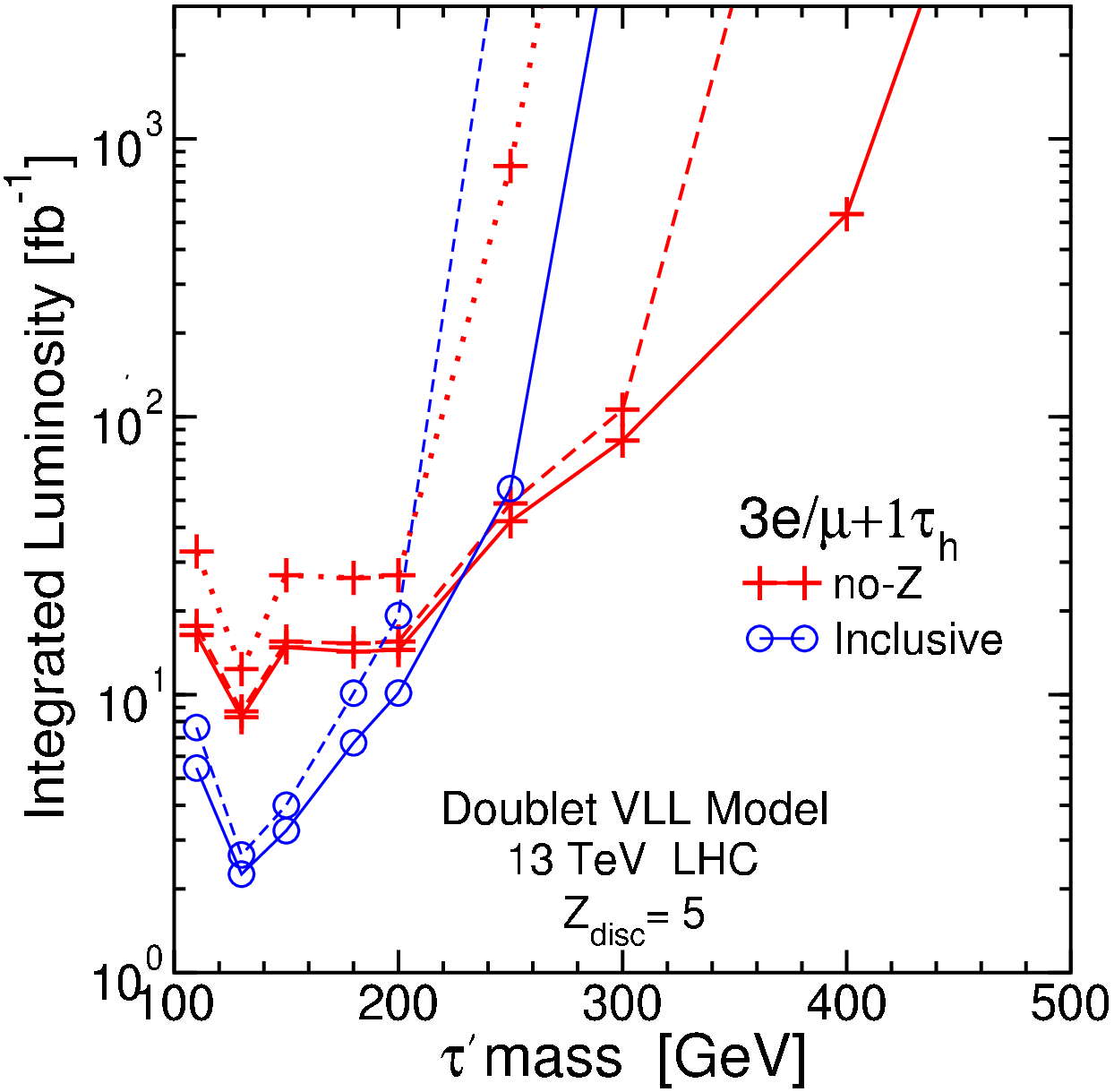}
\\
\includegraphics[width=7.5cm,angle=0]{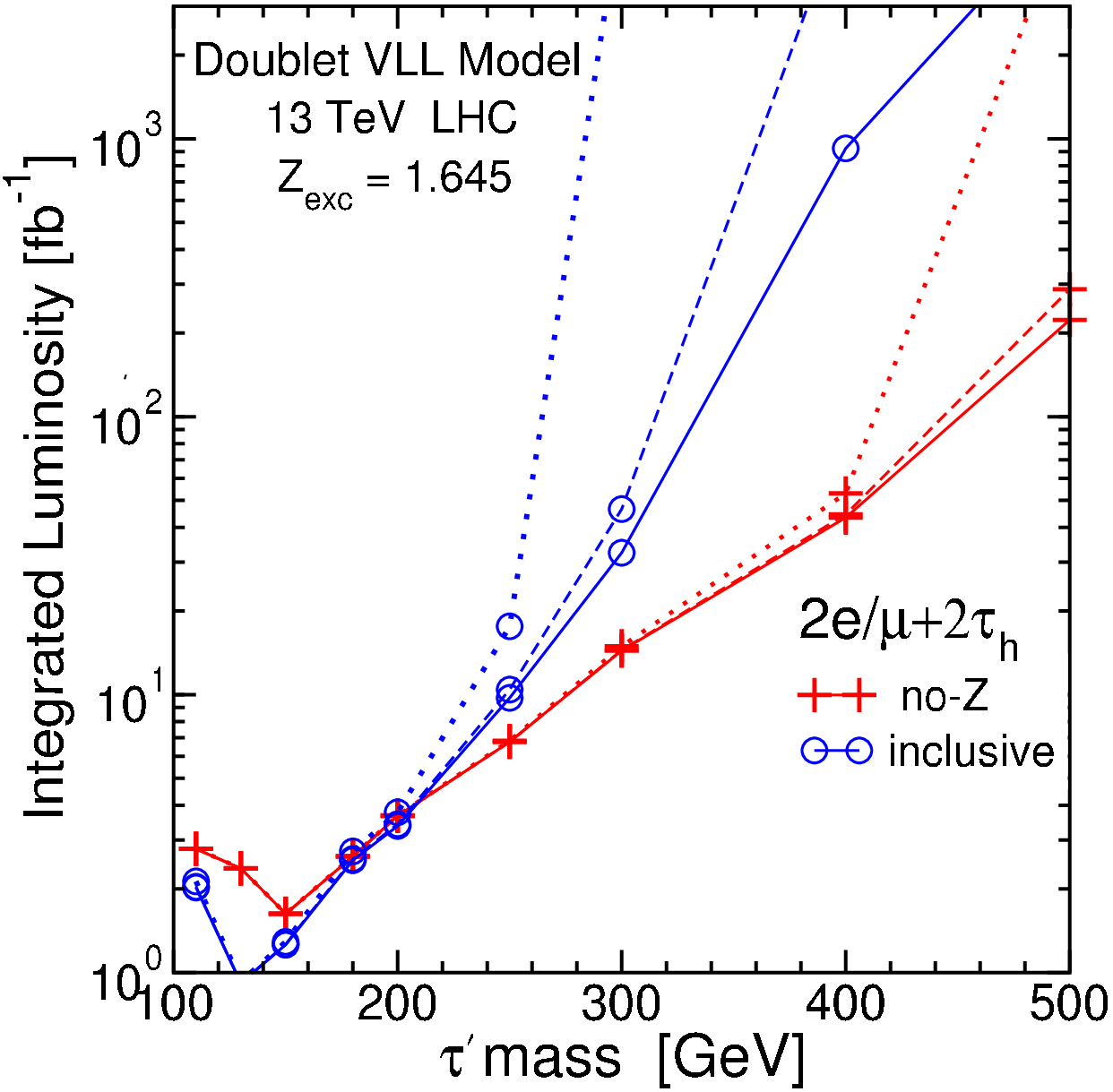}
\includegraphics[width=7.5cm,angle=0]{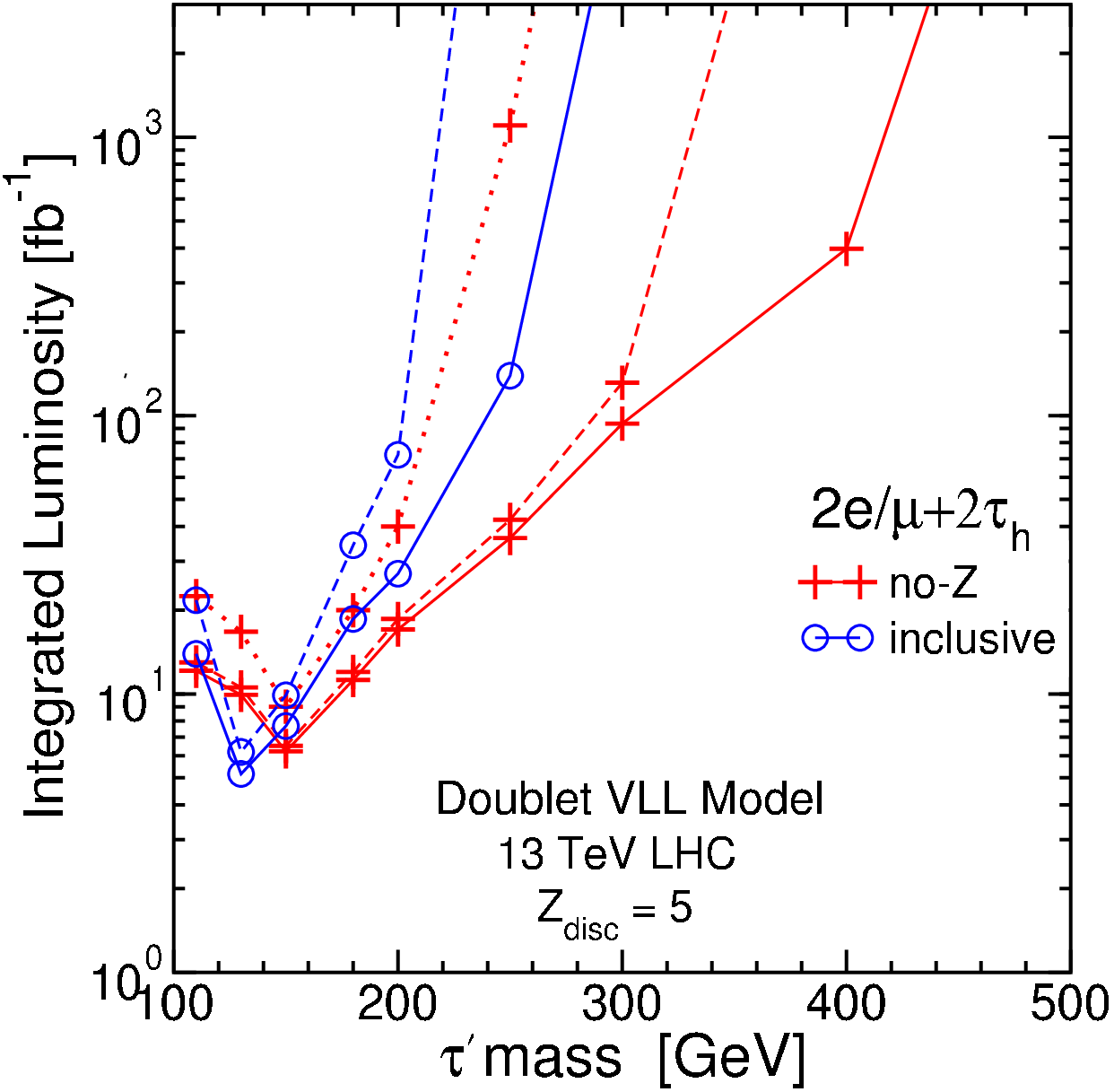}
\\
\includegraphics[width=7.5cm,angle=0]{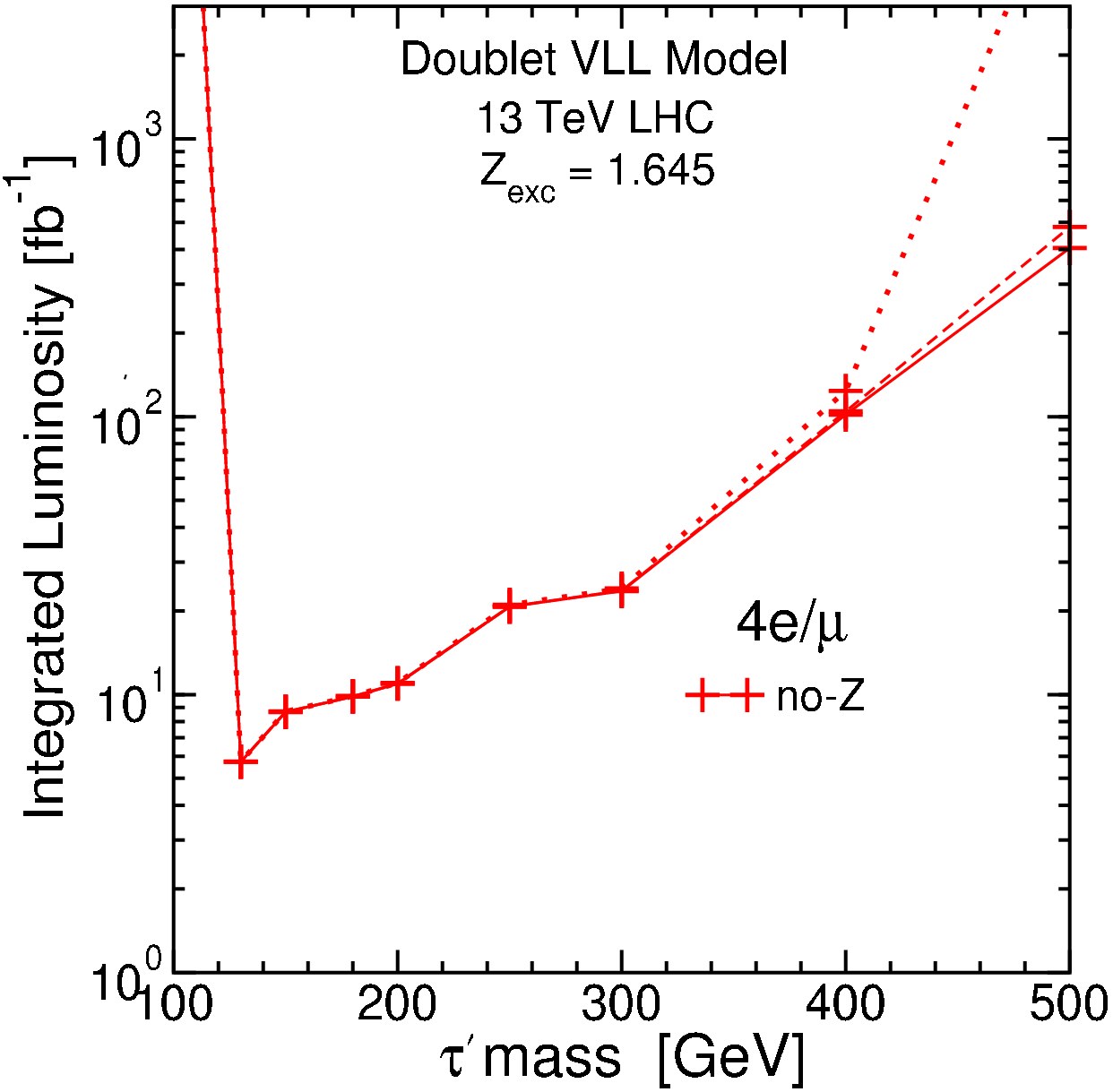}
\includegraphics[width=7.5cm,angle=0]{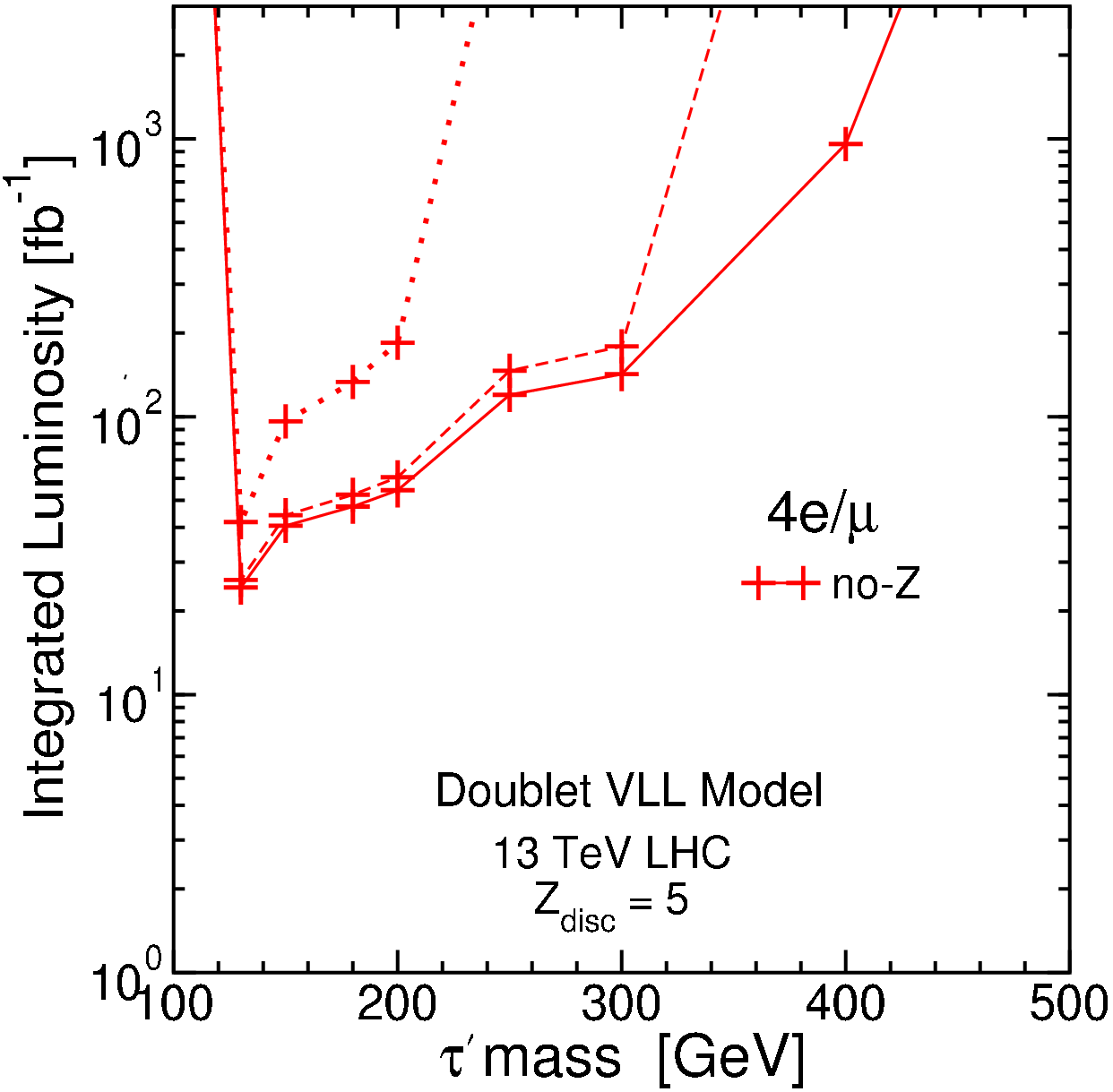}
\\
\caption{\label{fig:L_4lept_d}
Integrated luminosity needed for a median expected significance 
$\Zexc\geq 1.645$ for exclusion (left) and 
$\Zdisc\geq 5$ for discovery (right) in 
$\geq 3e/\mu+1\tau_h$, 
$\geq 2e/\mu+2\tau_h$, 
and 
$\geq 4e/\mu$ channels, for the Doublet VLL model,
as a function of $M_{\tau^{\prime}} = M_{\nu'}$, at $\sqrt{s} = 13 $ TeV,
based on the results of Table \ref{tab:table8}.
The different lines
correspond to assumed background uncertainties 
$\sigmab= 0.1$b (solid) and $\sigmab= 0.2$b (dashed)
and $\sigmab= 0.5$b (dotted), with blue lines (circle marks) for inclusive and 
red (plus marks) for no-$Z$ on each figure.
}
\end{figure}

At $\sqrt{s}=13$ TeV, we selected events using the same requirements 
as described in the previous section. The individual 
background cross-sections after cuts are listed in 
Table \ref{tab:table9} for each of the 6 signal regions.
The Doublet VLL model signal cross sections are given in 
Table \ref{tab:table8} for various masses $M_{\tau'}$.
\begin{table}
\caption{Background cross-sections $\sigma_b$ 
for 4-lepton channels at $\sqrt{s} = 13$ TeV,
after inclusive and no-$Z$ selections as described in section \ref{sec:8TeV}.} 
\label{tab:table9}
\begin{center}
\begin{tabular}{|c|c|c|c|c|c|c|}
\hline
~~~SM Backgrounds~~~ & \multicolumn{2}{c|}{$\sigma_b$ (fb) in $\geq 3e/\mu+1\tau_h$} & \multicolumn{2}{c|}{$\sigma_b$ (fb) in $\geq 2e/\mu+2\tau_h$} & \multicolumn{2}{c|}{$\sigma_b$ (fb) in $\geq 4e/\mu$}\\
\hline
  & \phantom{xx} incl.\phantom{xx} & no-$Z$ & \phantom{xx} incl. \phantom{xx}& no-$Z$ &  \phantom{x}incl.\phantom{x} & no-$Z$ \\
\hline
$pp\rightarrow W Z $ &            0.0637&0.0000&0.0127	&0.0127 &0.0000	&0.0000\\
$pp\rightarrow Z Z $  &          0.7840&0.0242 &0.4555	&0.0302 &14.7263 &0.0121\\
$pp\rightarrow t\overline t W$ & 	0.0080 &0.0057 &0.0028	&0.0025 &0.0000	&0.0000\\
$pp\rightarrow t\overline t Z$ & 0.0249&0.0045 &0.0059	&0.0033 &0.0508	&0.0014\\
$pp\rightarrow t\overline{t} h$  & 0.0071 &0.0049 &0.0071	& 0.0056 &0.0052  &0.0026\\
$pp\rightarrow hh $             & 0.0012&0.0004 &0.0008	&0.0004 &0.0016	&0.0004\\
$pp\rightarrow hZ $            & 0.1377&0.0084 &0.0588	&0.0051  &0.2418	& 0.0067\\
$pp\rightarrow WWZ $            & 0.0193&0.0034 &0.0025	&0.0017 &0.0986	&0.0026\\
$pp\rightarrow WZZ $             &0.0062&0.0004 &0.0015	&0.0004  &0.0423	&0.0005\\
$pp\rightarrow ZZZ $              &0.0030&0.0001 &0.0013	&0.0002 &0.0282	&0.0002\\
\hline
Total Background &  1.055	& 0.0520 &0.549	&0.0619 & 15.1950&	0.0265\\
\hline
\end{tabular}
\end{center}
\caption{Signal and total background cross sections in the
$\geq 3e/\mu+1\tau_h$, $\geq 2e/\mu+2\tau_h$ and $\geq 4e/\mu$ channels after selection through inclusive and no-$Z$ 
requirements, for the Doublet VLL model, at $\sqrt{s} = 13$ TeV.}
\label{tab:table8}
\centering
\begin{center}
\begin{tabular}{|c|c|c|c|c|c|c|}
\hline
$M_{\tau^\prime}$ (GeV)	& \multicolumn{2}{c|}{$\sigma_{s}$ (fb) in $\geq 3e/\mu+1\tau_h$} & 
\multicolumn{2}{c|}{$\sigma_{s}$ (fb) in $\geq 2e/\mu+2\tau_h$} & \multicolumn{2}{c|}{$\sigma_{s}$ (fb) in$\geq 4e/\mu$} \\
\hline
& inclusive & no-$Z$ 
& inclusive & no-$Z$ 
& inclusive & no-$Z$ \\
\hline
110&	3.048&  0.490& 1.353 & 0.636& 2.632& 0.000\\
130&	5.126&  0.782& 2.402 & 0.729 &2.877& 0.302\\
150&	4.123&  0.533& 1.896 & 1.008 &3.050& 0.214\\
180&	2.712&  0.538& 1.158& 0.670 & 1.913&0.193\\
200&	2.173&   0.533&  0.958& 0.507 &1.558&0.177\\ 
250&	1.037&  0.267& 0.495& 0.312 &0.740& 0.108\\
300&	0.549&   0.178&0.257& 0.178 &0.456& 0.098\\
400&	0.188&   0.068& 0.095& 0.085 &0.170&0.036\\
500&	0.081&   0.033&  0.035& 0.033 & 0.066&0.016\\
\hline
Total Background & 1.055 & 0.052& 0.549 & 0.062 & 15.195 & 0.027\\
\hline
\end{tabular}
\end{center}
\end{table}
By solving eqs.~(\ref{eq:Zexc}) and (\ref{eq:Zdisc}) we get 
the median expected integrated luminosities needed for 95\% CL exclusion
and $\Zdisc>5$ discovery, as a function of $M_{\tau^\prime}$. These results
are shown in Figure \ref{fig:L_4lept_d}, for the cases of assumed
10\%, 20\%, and 50\% fractional uncertainties in the background.
Just as in the $\sqrt{s}=8$ TeV case, we find that the
reach is best in the channels that include at least one $\tau_h$ candidates.

Assuming the signal is absent, then with 10 fb$^{-1}$ one expects to 
be able to make
a 95\% CL exclusion for $M_{\tau^\prime}$ up to about 270 GeV, using the
no-$Z$ version of either of the $\geq 3e/\mu+1\tau_h$ or 
$\geq 2e/\mu+2\tau_h$ channels.
This is true even if the assumed fractional uncertainty in the background 
is as large as 50\%,
simply because the no-$Z$ background levels are small. 
The inclusive $\geq 3e/\mu+1\tau_h$
channel seems to do slightly better for exclusion with 10 fb$^{-1}$,
but only if the fractional uncertainty in the background is less than 20\%.
With 100 fb$^{-1}$, the no-$Z$ channels are clearly better, and can exclude 
up to about $M_{\tau^\prime}$ = 440 GeV (or 400 GeV), provided that the 
fractional uncertainty in the background is not more than  20\% (or 50\%).

Under the same circumstances, a discovery of the Doublet VLL model 
could be possible
up to about $M_{\tau^\prime}$ = 300 GeV with 100 fb$^{-1}$, using either
the no-$Z$ selection for either of the $\geq 3e/\mu+1\tau_h$ or 
$\geq 2e/\mu+2\tau_h$ channels. The discovery reach in these channels degrades
to about 210 GeV if $\Delta_b = 0.5 b$.
In general, 
the inclusive search is seen to be
better at low masses where the signal cross-section is large enough to overcome 
the significant backgrounds, while the no-$Z$
channel performs much better at high masses.
 
Also, because of the higher background, the
inclusive channels tend to be more sensitive to a given assumed level of
fractional background uncertainties than the no-$Z$ channels. 
With an assumption of a 50\% fractional uncertainty in the background, 
the exclusion reach is completely eliminated for $M_{\tau'}$ above 400 GeV. 
The discovery reach similarly is absent for $M_{\tau'}$ above 300 GeV if the
fractional uncertainty in the background is larger than 20\%.
The real-world background uncertainties will likely be 
larger for the $2\tau_h$ cases than the $1\tau_h$ cases.
It may well also be possible to combine these channels to 
enhance the significance 
of an exclusion or discovery, but we do not attempt this here.
The $\geq 4e/\mu$ search with a no-$Z$
requirement is seen to be much less powerful, 
and the inclusive channel (not shown) 
is quite weak as it suffers from a comparatively very large $ZZ$ background. 
Similarly to the $\sqrt{s}=8$ TeV inclusive search of the previous section,
we found that an on-$Z$ selection (also not shown) 
would not do any better than the inclusive selection.


\subsection{Five-lepton searches for the Doublet VLL model
\label{subsec:Doublet5l13TeV}}
\setcounter{equation}{0}
\setcounter{footnote}{1}

In this section, we consider 5-lepton search channels 
for the Doublet VLL model. 
These have the advantage
that the backgrounds tend to be extremely small. We use the same criteria
for lepton identification and isolation as in section \ref{sec:8TeV}, 
and again require as a trigger at least one high-$p_T$ electron or muon, 
as in eq.~(\ref{eq:pT26}). We then define three channels
$\geq 4 e/\mu + 1\tau_h$ and $\geq 3e/\mu + 2 \tau_h$ and $\geq 5e/\mu$.
For each of these, we further consider the inclusive and no-$Z$ categories,
as before. 

The individual backgrounds are given in Table \ref{tab:table11}.
\begin{table}
\caption{Background cross-sections $\sigma_b$ after inclusive and no-$Z$ 
selections, for five lepton channels at $\sqrt{s} = 13$ TeV. 
\label{tab:table11}}
\begin{center}
\begin{tabular}{|c|c|c|c|c|c|c|}
\hline
~~~SM Backgrounds~~~ & \multicolumn{2}{c|}{$\sigma_b$ (fb) in $\geq 4e/\mu+1\tau_h$} & \multicolumn{2}{c|}{$\sigma_b$ (fb) in $\geq 5e/\mu$} & \multicolumn{2}{c|}{$\sigma_b$ (fb) in $\geq 3e/\mu+2\tau_h$}\\
\hline
  &  incl. & no-$Z$ &  incl. & no-$Z$ &  incl. & no-$Z$ \\
\hline
$pp\rightarrow Z Z $  &   		0.00202&  0&       0&       0&       0.00202     	& 0 \\
$pp\rightarrow t\overline t W$&         0&        0&       0&       0&       0.00026    	& 0.00013\\ 
$pp\rightarrow t\overline t Z$ &	0.00215&  0.00013& 0&       0&       0.00124       	& 0.00007\\
$pp\rightarrow t\overline{t} h$&        0&        0&       0&       0&       0.00150      	& 0.00075\\
$pp\rightarrow WWZ $ & 			0.00012&  0&       0&       0&       0.00003      	& 0\\
$pp\rightarrow WZZ $ &			0.00076&  0.000004& 0.00498&  0.000015&  0.00019      	& 0.0000073\\
$pp\rightarrow ZZZ $ & 			0.00044& 0.000004& 0.00150& 0.000002&   0.00008      	& 0.0000031\\
\hline
Total Background & 0.00548  &	0.00014  & 0.00648  & 0.000017  & 0.00533  & 0.00096    \\
\hline
\end{tabular}
\end{center}
\end{table}
We note that all of the backgrounds are quite small, 
amounting to only a few events expected
even in 1000 fb$^{-1}$ in the inclusive cases,
and fewer than 1 event in 1000 fb$^{-1}$ in the no-$Z$ 5-lepton cases.
In some background channels, 
our simulations did not find any events that passed all of 
the selection criteria. The $pp \rightarrow ZZ$ and 
$pp \rightarrow t \bar t Z$ backgrounds are the largest in the inclusive
cases for $\geq 4 e/\mu + 1\tau_h$ and $\geq 3e/\mu + 2 \tau_h$, but these
are effectively eliminated if the no-$Z$ requirement is included. 
We note that these backgrounds rely on the rate for jets to fake $\tau_h$,
which we took to be 0.001 as noted above. In the real world, these backgrounds
will have to be determined for the relevant topologies by using 
control regions.

For the signal, the contributions come mostly from 
$h\tau Z\tau$ and $Z\tau Z\tau$ and $W\tau Z\tau$ events resulting from 
vectorlike lepton production. Results are shown in Table \ref{tab:table10}
as a function of $M_{\tau'}$ for the Doublet VLL model. 
In these simulations, we
forced the $Z$ and $W$ to decay leptonically (including to $\tau$ leptons)
for better statistics. 
\begin{table}
\caption{Signal and background cross sections for the Doublet VLL model, 
at $\sqrt{s} = 13$ TeV in the
$\geq 4e/\mu+1\tau_h$, $\geq 3e/\mu+2\tau_h$ and $\geq 5e/\mu$ channels 
after selection through inclusive and no-$Z$ 
requirements.
\label{tab:table10}
}
\centering
\begin{center}
\begin{tabular}{|c|c|c|c|c|c|c|}
\hline
$M_{\tau^\prime}$ (GeV)	& \multicolumn{2}{c|}{$\sigma_{s}$ (fb) in $\geq 4e/\mu+1\tau_h$} & 
\multicolumn{2}{c|}{$\sigma_{s}$ (fb) in $\geq 3e/\mu+2\tau_h$} & \multicolumn{2}{c|}{$\sigma_{s}$ (fb) in $\geq 5e/\mu$} \\
\hline
& inclusive & no-$Z$ 
& inclusive & no-$Z$ 
& inclusive & no-$Z$ \\
\hline
110&	0.156&	0.01104 & 0.091&	0.01060 &0.081	&0.00029\\
130&	0.382&	0.01682 & 0.204&	0.00950 &0.204  &0.00569\\
150&	0.315&	0.00799 & 0.176&	0.01275 &0.176	&0.00345\\
180&	0.217&	0.00842 & 0.112&	0.00882 &0.138	&0.00188\\
200&	0.169&	0.00507 & 0.081&	0.00749 &0.106	&0.00146\\
250&	0.088&	0.00314 & 0.042&	0.00446 &0.057	&0.00085\\
300&	0.046&	0.00257 & 0.021&	0.00326 &0.032	&0.00058\\
400&	0.015&	0.00086 & 0.007&	0.00120 &0.011	&0.00032\\
500&	0.006&	0.00037 & 0.002&	0.00044 &0.005	&0.00016\\
\hline
Total Background & 0.00548 & 0.00014  & 0.00533  & 0.00096 & 0.00648 & 0.000017\\
\hline
\end{tabular}
\end{center}
\end{table}
Inclusive and on-$Z$ requirements give essentially the 
same results, so we do not consider separately an on-$Z$ search. 

\begin{figure}[!tb]
\includegraphics[width=7.5cm,angle=0]{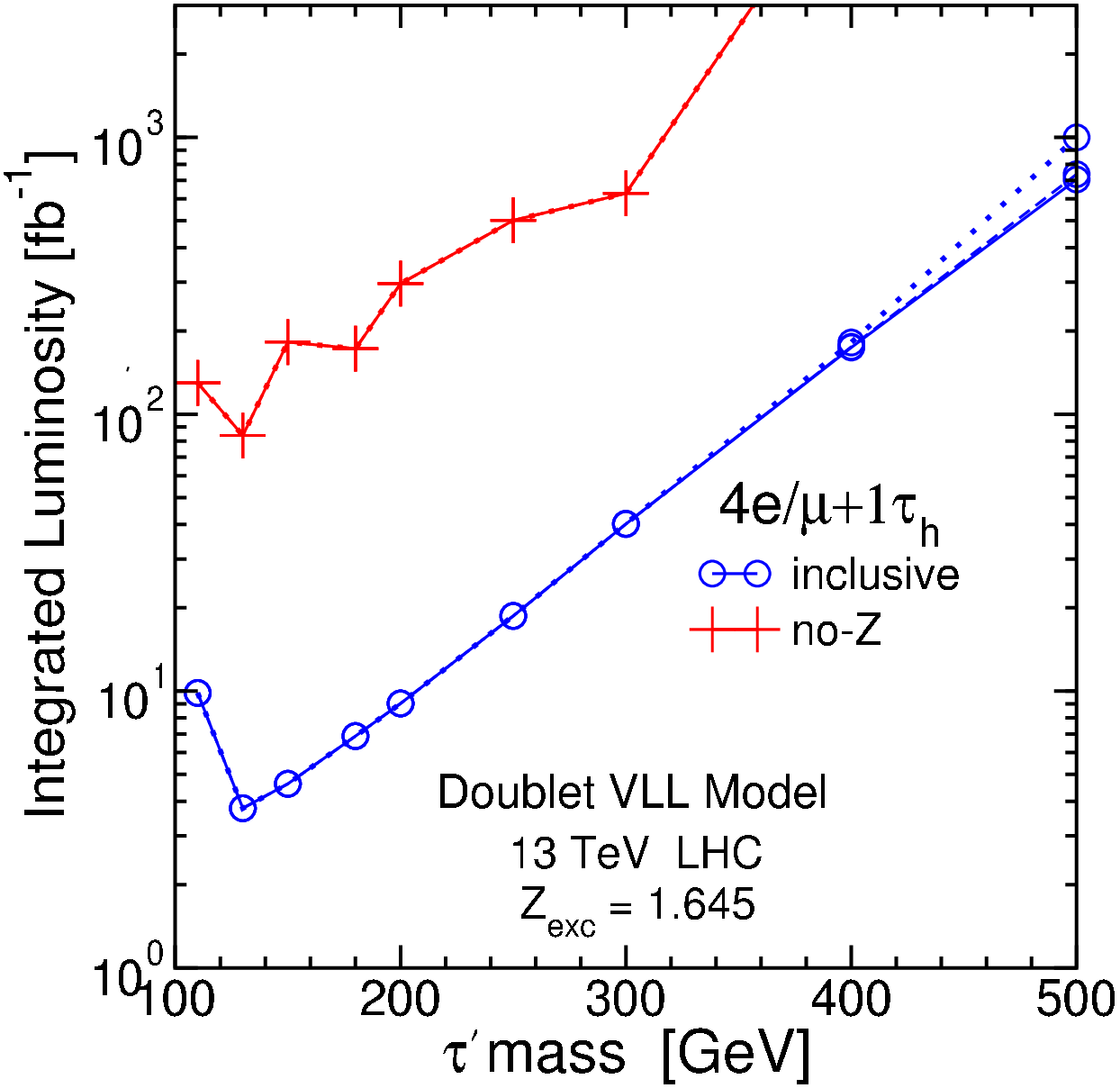}
\includegraphics[width=7.5cm,angle=0]{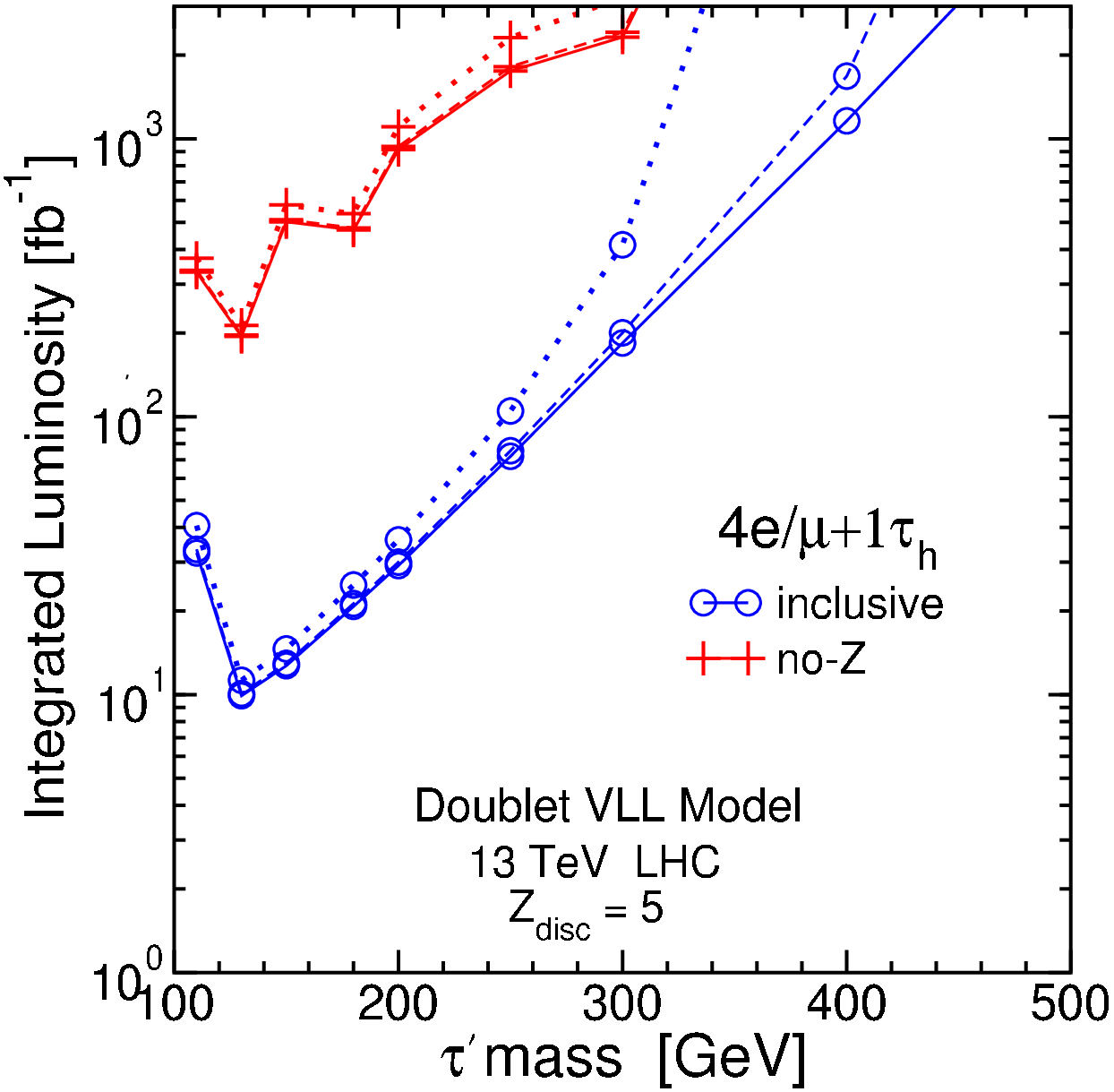}
\\
\includegraphics[width=7.5cm,angle=0]{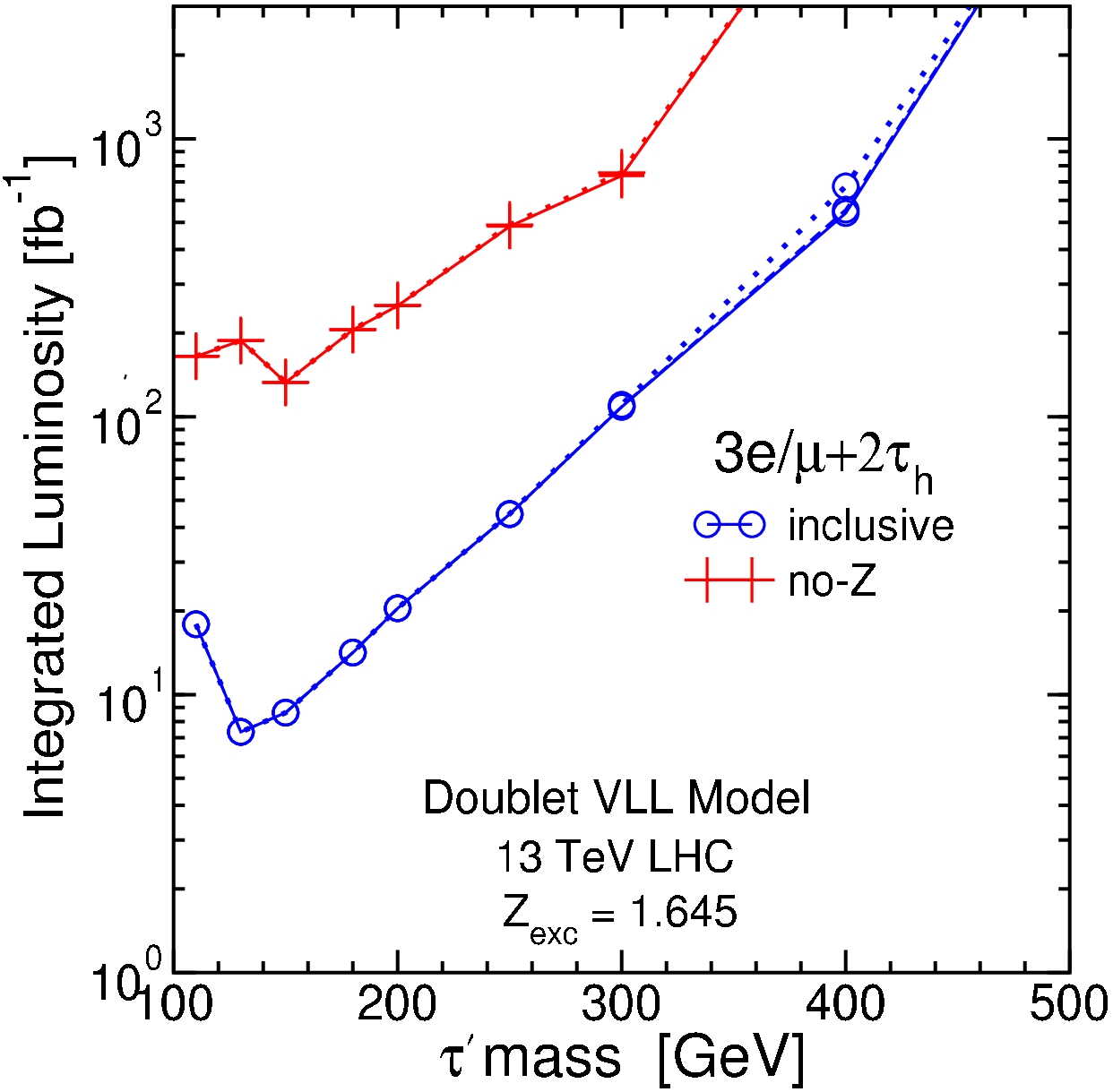}
\includegraphics[width=7.5cm,angle=0]{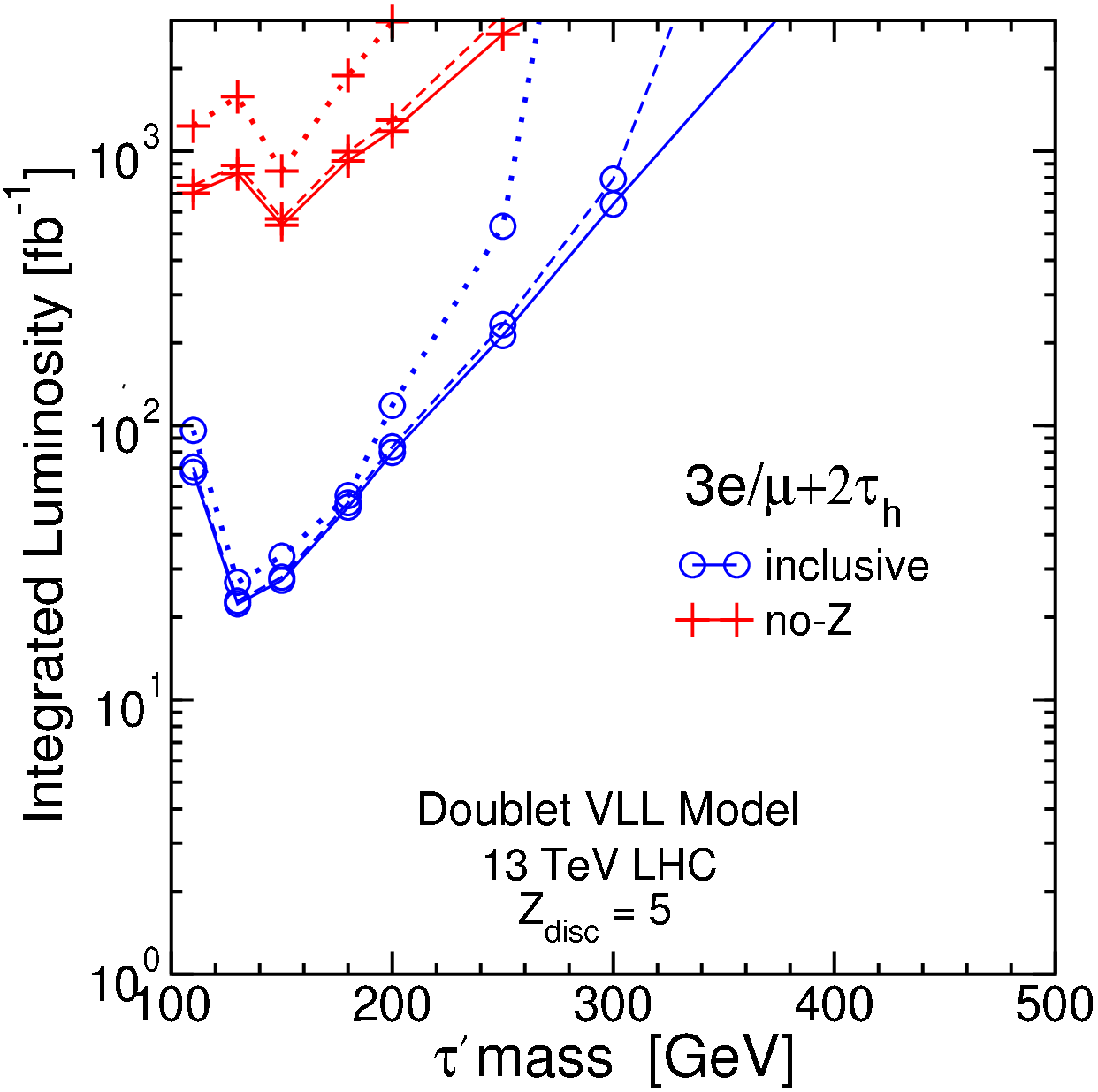}
\\
\includegraphics[width=7.5cm,angle=0]{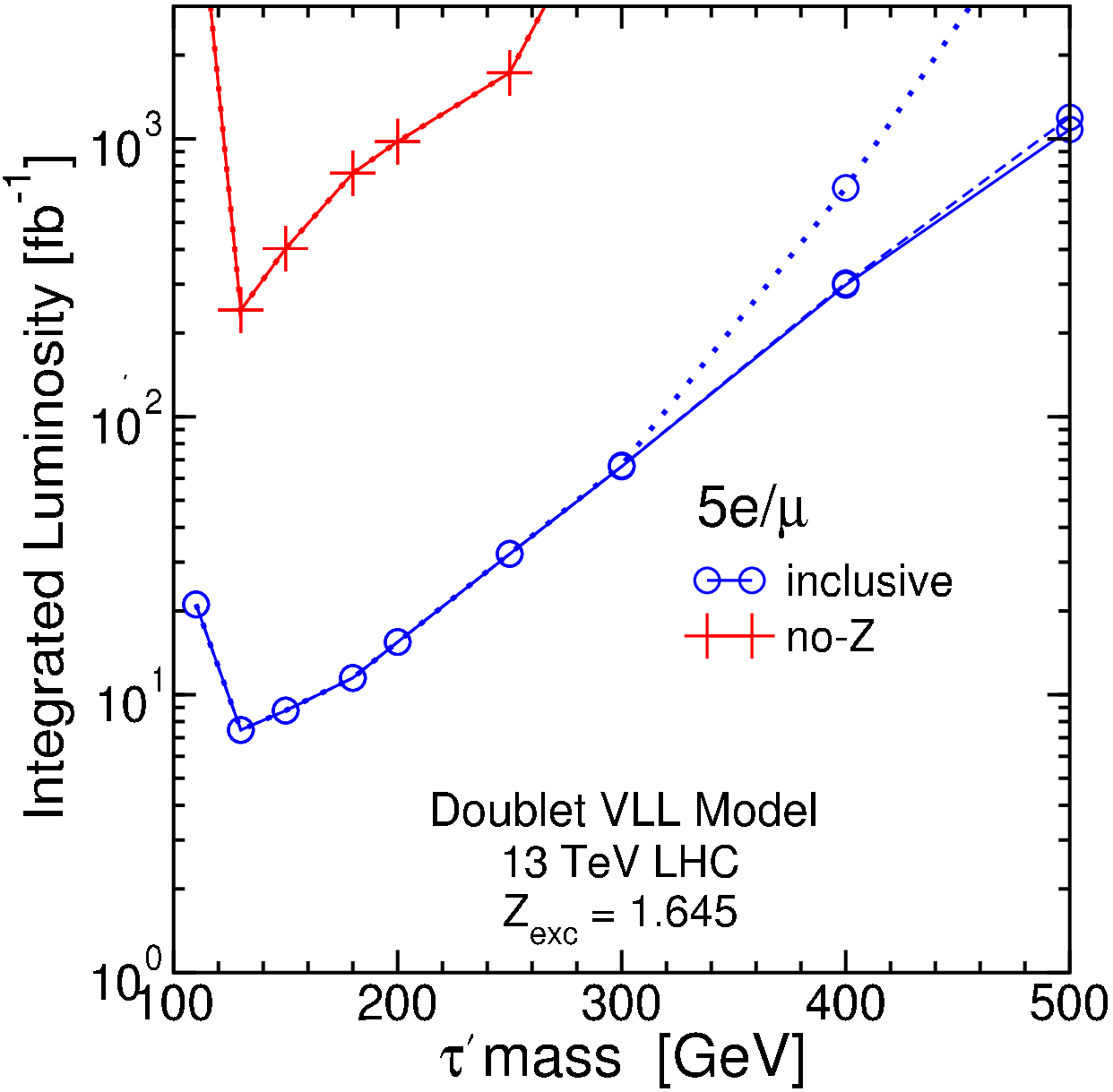}
\includegraphics[width=7.5cm,angle=0]{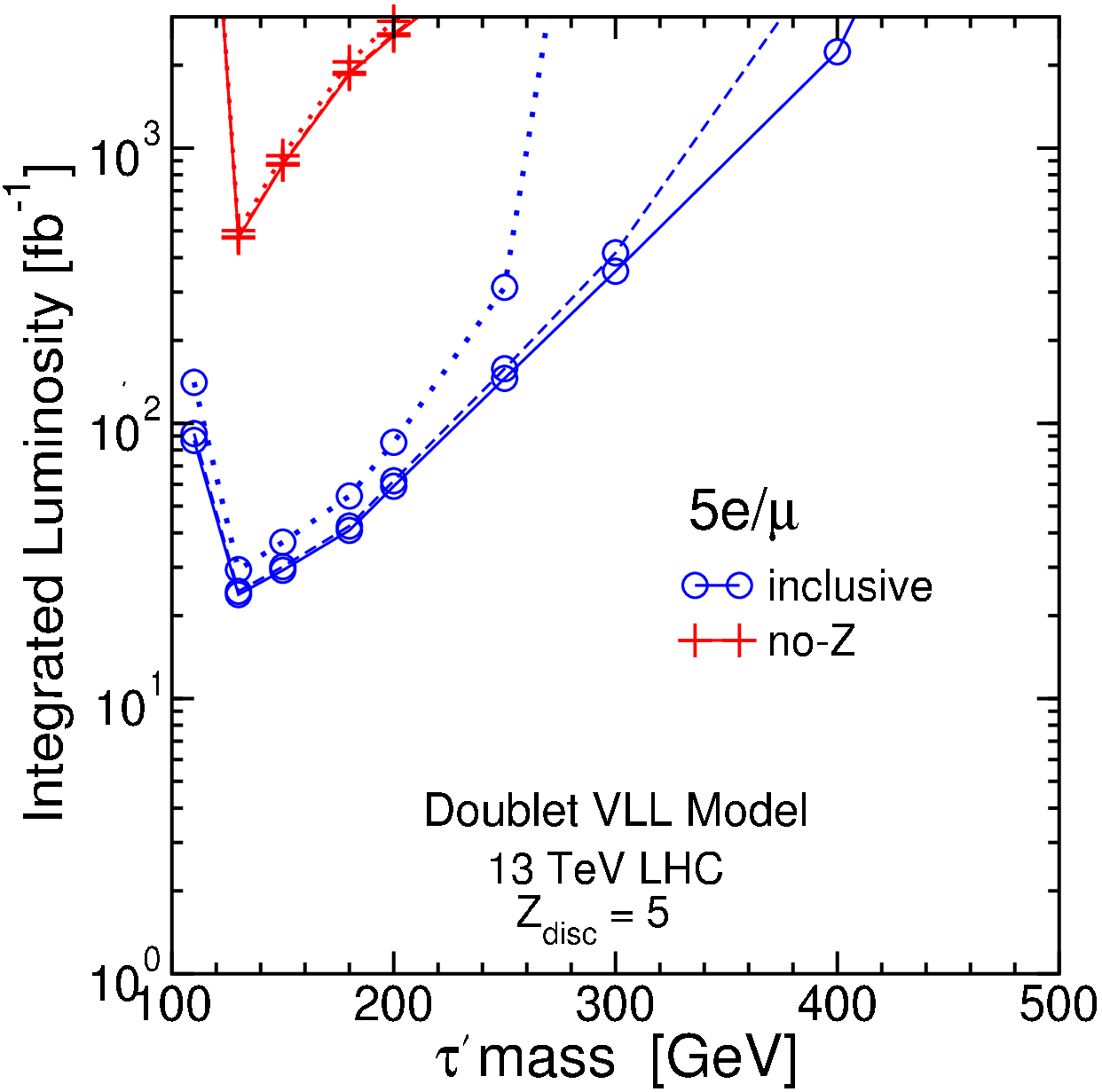}
\\
\caption{\label{fig:L_5lept_d}
Integrated luminosity needed for a median expected significance
$\Zexc\geq 1.645$ for 95\% CL exclusion (left) and 
$\Zdisc\geq 5$ for discovery (right) in 
$\geq 4e/\mu+1\tau_h$, $\geq 3e/\mu+2\tau_h$, and 
$\geq 5e/\mu$ channels,
for the Doublet VLL model,
as a function of $M_{\tau^{\prime}} = M_{\nu'}$, at $\sqrt{s} = 13 $ TeV,
based on the results of Table \ref{tab:table10}.
In the figures different lines
correspond to $\sigmab= 0.1$b (solid) and $\sigmab= 0.2$b (dashed)
and $\sigmab= 0.5$b (dotted), with blue lines (circle marks) for inclusive,  
red (plus marks) for no-$Z$ on each figure.}
\end{figure}

Results for the luminosities needed to achieve a median expected
$\Zexc\geq 1.645$ (95\% CL exclusion) 
or $\Zexc\geq 5$ (discovery) are presented in Figure \ref{fig:L_5lept_d}.
The 5-lepton search strategy is statistics-limited, 
rather than background limited. Therefore, for 10 fb$^{-1}$ of integrated
luminosity, these channels are not
competitive with the 4-lepton searches of the previous section.
Clearly, very high integrated luminosities are required 
if a no-$Z$ search is performed, because of the very low signal yields. 
The best search strategy for achieving a 95\% CL exclusion seems to be
the $\geq 4e/\mu+1\tau_h$ inclusive search, which with 100 fb$^{-1}$
can exclude up to $M_{\tau'} = 340$ GeV even if the fractional background 
uncertainty is taken to be 50\%.
However, for smaller background uncertainties this search 
is less effective than the 4-lepton no-$Z$ 
searches described in the previous section
(compare Figure \ref{fig:Sig_4lept_d}). 

Similarly, the potential for Doublet VLL discovery using the 
5-lepton searches is
somewhat worse than in the 4-lepton search of the previous section
if one
assumes that the background uncertainties in both cases
are taken to be 10\% or lower,
but the situation is reversed if the assumed fractional
background uncertainties are higher.
With 100 fb$^{-1}$, the $\geq 4e/\mu+1\tau_h$ inclusive search could discover
the Doublet VLL model for masses up to about $M_{\tau'} =$ 250 GeV, 
even if the fractional
uncertainty in the background is 50\%. With an integrated luminosity of
1000 fb$^{-1}$, there is a possibility to discover the Doublet VLL model
up to $M_{\tau'} =$ 400 GeV in the same channel, provided that the background
uncertainty is 10\% or lower. 
The other 5-lepton channels provide somewhat less discovery reach.

\subsection{Four-lepton searches for the Singlet VLL model}
\setcounter{equation}{0}
\setcounter{footnote}{1}

\begin{figure}[!tb]
\begin{minipage}[]{0.495\linewidth}
\includegraphics[width=8.0cm,angle=0]{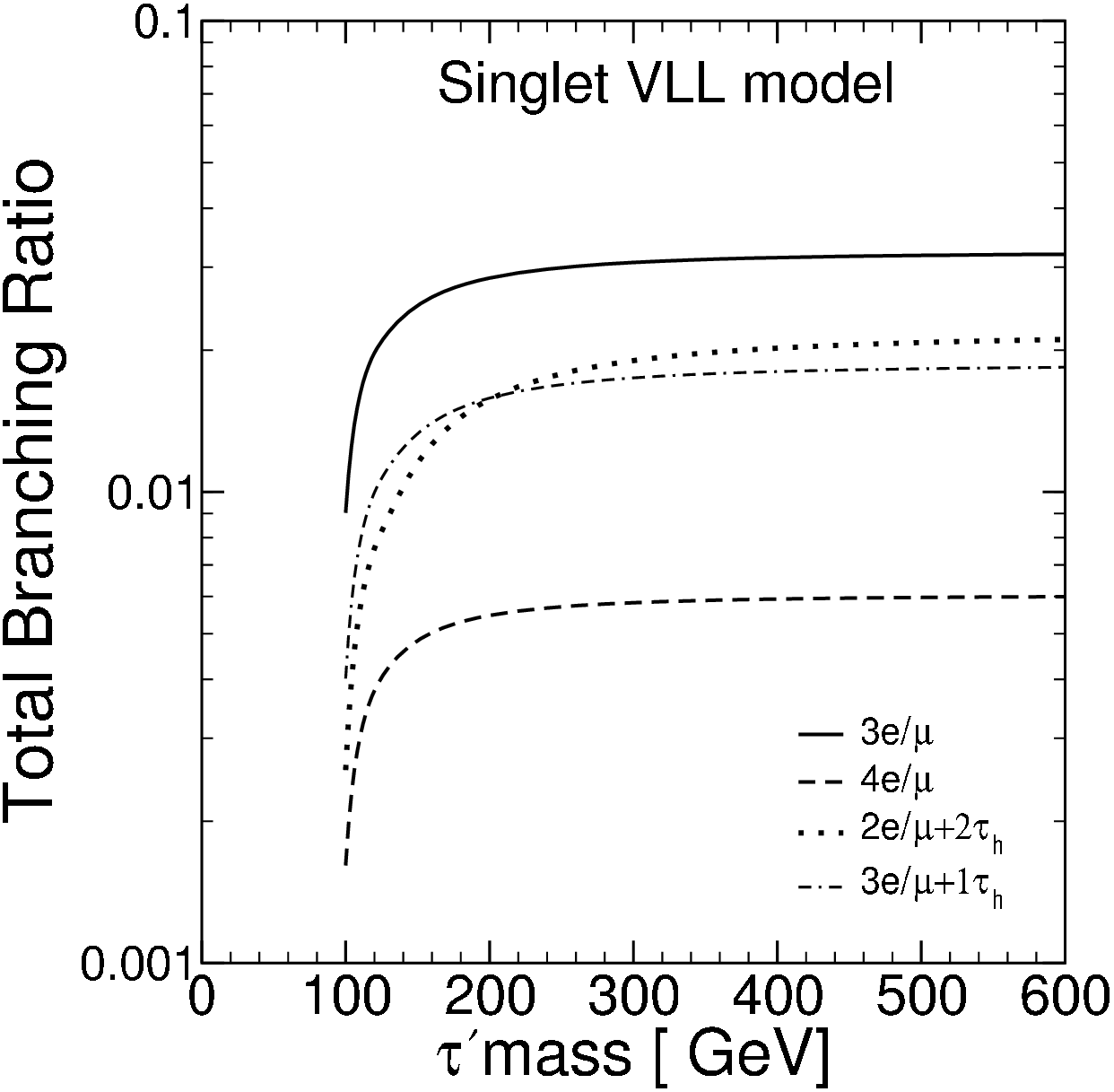}
\end{minipage}
\begin{minipage}[]{0.495\linewidth}
\begin{flushright}
\includegraphics[width=8.0cm,angle=0]{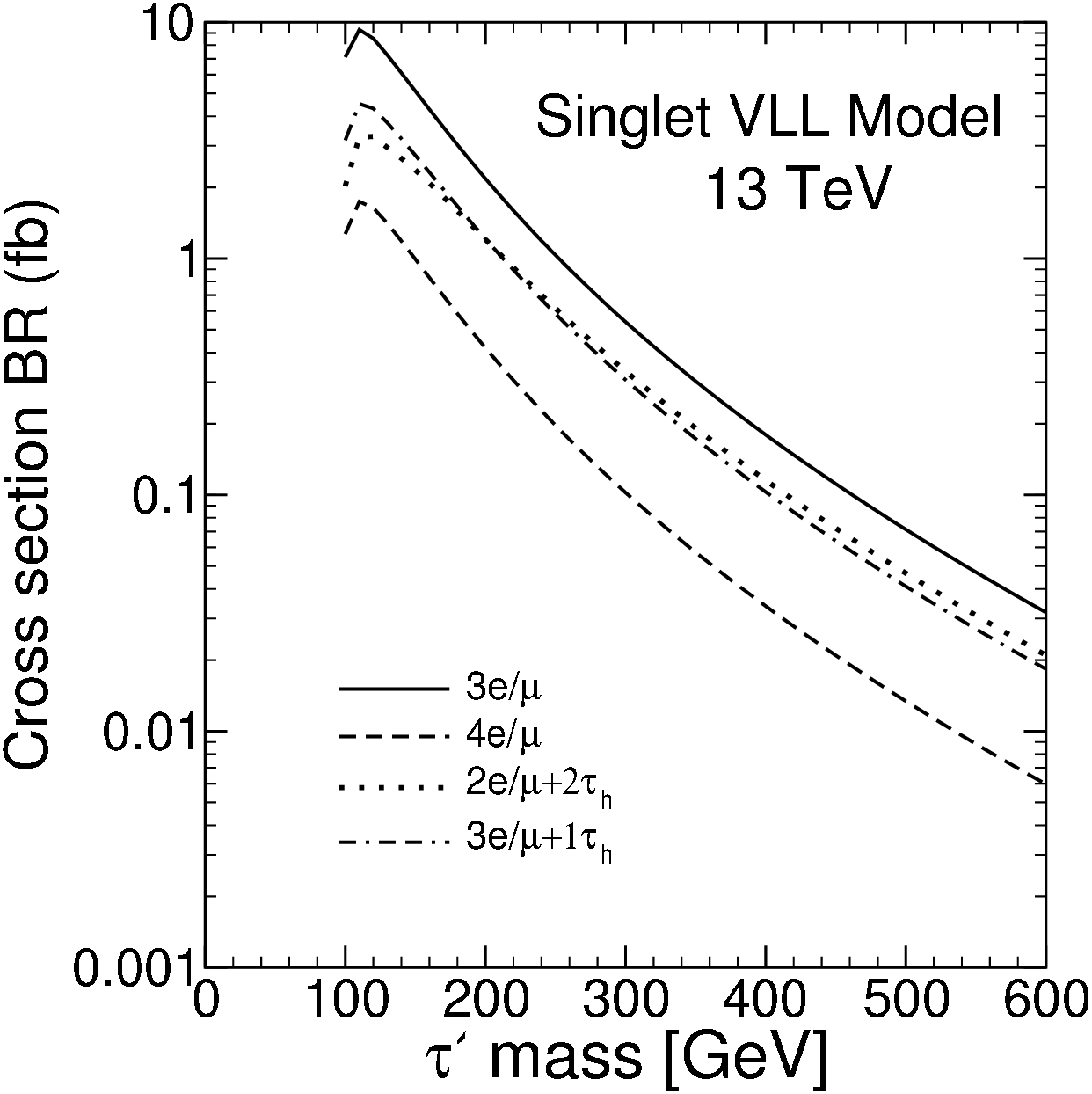}
\end{flushright}
\end{minipage}
\caption{\label{fig:BR}The total Branching ratio (left) and cross section $\times$
Branching Ratio (right) that contribute to different channels 
are shown as a function of $\tau^{\prime}$ mass for singlet VLL model. 
In the plot at right, the cross sections 
predicted by the Singlet VLL model at $\sqrt{s} = 13$ TeV are used.}
\end{figure}

In the Singlet VLL model we only have the production 
of $\tau^{\prime+}\tau^{\prime-}$. Hence the signal cross section that contributes 
to the visible final states is much smaller than for the Doublet 
VLL model. The challenge is illustrated by Figure \ref {fig:BR}, 
in which the plot on left shows the dependence on $M_{\tau^\prime}$ of the total 
branching fraction of $\tau^{\prime +} \tau^{\prime -}$
into different individual multilepton channels 
$\geq 3e/\mu$, $\geq 4e/\mu$, $\geq 3e/\mu+1\tau_h$ 
and $\geq 2e/\mu + 2\tau_h$, and
the plot on right shows the cross section$\times$BR, 
before putting in any cuts or detection efficiencies.
From Figure \ref {fig:BR}, it is 
evident that a 3-lepton search gives the biggest contribution 
in signal cross section, but that suffers from a large background. We therefore will
concentrate on 4-lepton and 5-lepton searches.

\begin{figure}[!tb]
\begin{minipage}[]{0.495\linewidth}
\includegraphics[width=7.5cm,angle=0]{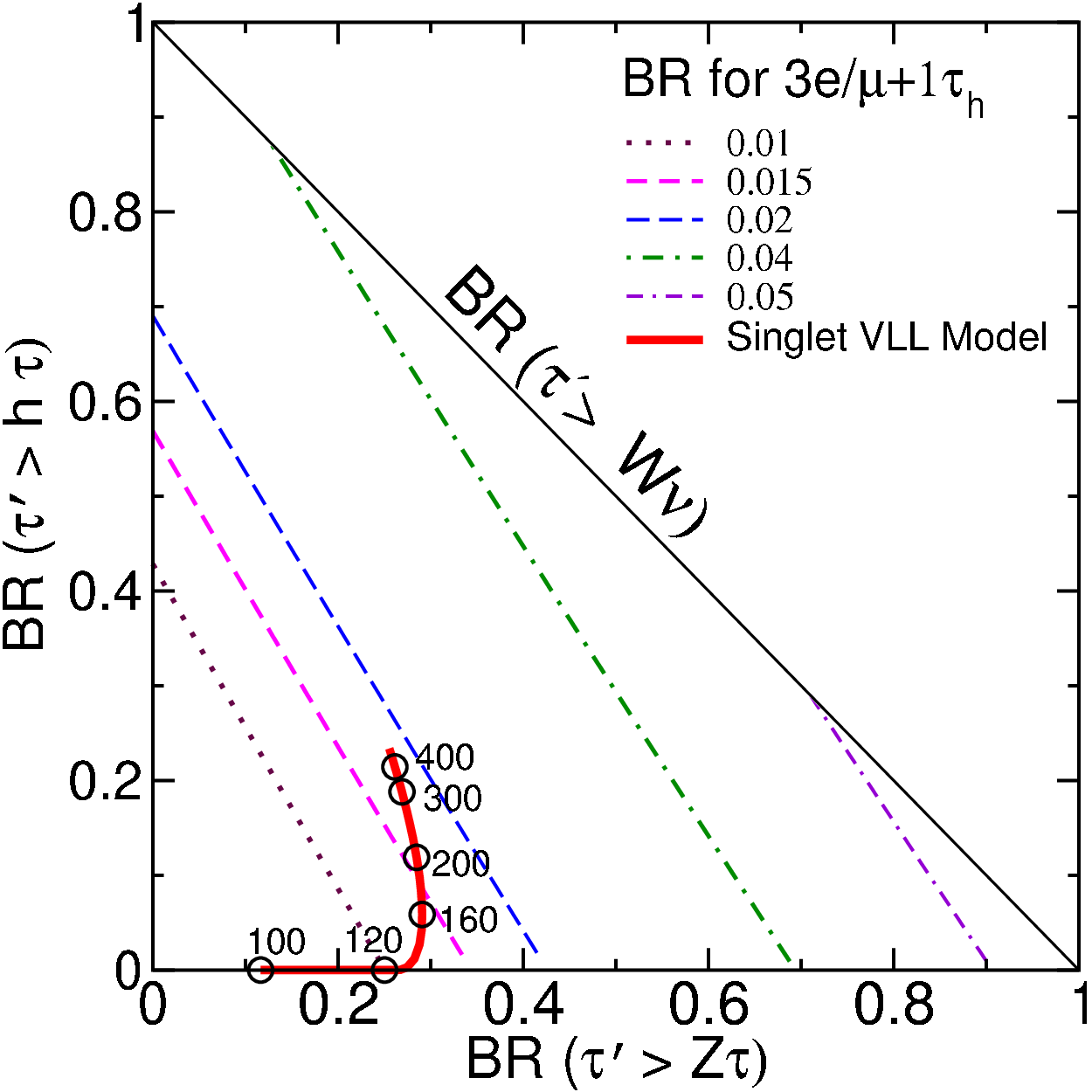}
\end{minipage}
\begin{minipage}[]{0.495\linewidth}
\begin{flushright}
\includegraphics[width=7.5cm,angle=0]{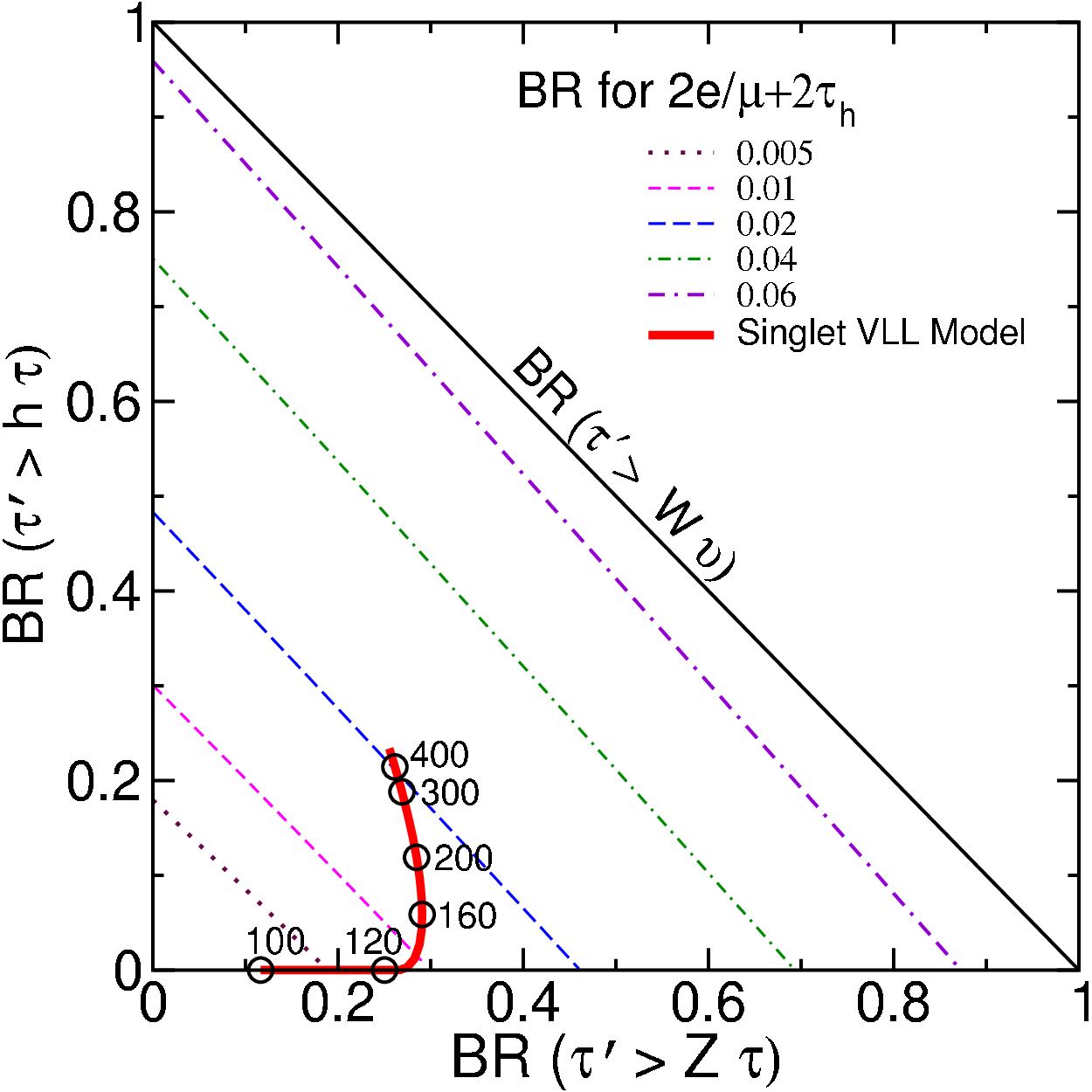}
\end{flushright}
\end{minipage}
\caption{\label{fig:triangular}Contour lines for total branching ratio of
$\tau^{\prime+}\tau^{\prime-}$ into $\geq 3e/\mu+1\tau_h$ (left) and
$\geq 2e/\mu+2\tau_h$ (right) channels 
are shown in the plane 
of BR$(\tau' \rightarrow h \tau)$ and 
BR$(\tau' \rightarrow Z \tau)$, assuming BR$(\tau' \rightarrow W\nu)$ = 
1 - BR$(\tau' \rightarrow h \tau)$ - BR$(\tau' \rightarrow Z \tau)$.
The prediction of the Singlet VLL model is shown
by the thick red curve, with circles corresponding to
$M_{\tau^{\prime}}$ = 100, 120, 160, 200, 300, 400 GeV.
}
\end{figure}

In Figure \ref{fig:triangular}, we show results for the 4-lepton
channels $\geq 3e/\mu+1\tau_h$ and $\geq 2e/\mu+2\tau_h$, again before any cuts
or detection efficiencies,
and this time allowing the branching ratios of $\tau'$ into the three possible
final states $Z\tau$ and $h\tau$ and $W\nu$ to float,
subject to the constraint
BR$(\tau' \rightarrow Z \tau)$ + BR$(\tau' \rightarrow h \tau)$ + 
BR$(\tau' \rightarrow W\nu)$ = 1. The effects of leptonic $\tau$ decays
have been included.
Within this plane of more general possibilities, the thick red curve shows the
prediction of the Singlet VLL model (following from
the results shown in Figure \ref{fig:BRs}), and the other
contour lines have constant branching ratios of 
$\tau^{\prime +} \tau^{\prime -}$ into 
$\geq 3e/\mu+1\tau_h$ and $\geq 2e/\mu+2\tau_h$. The predicted large 
BR$(\tau' \rightarrow W\nu)$ in the Single VLL Model, especially for
low $M_{\tau'}$ is seen to be the reason for low signal yields for 4-lepton
and 5-lepton channels.

\begin{figure}[!tb]
\begin{minipage}[]{0.495\linewidth}
\includegraphics[width=8.0cm,angle=0]{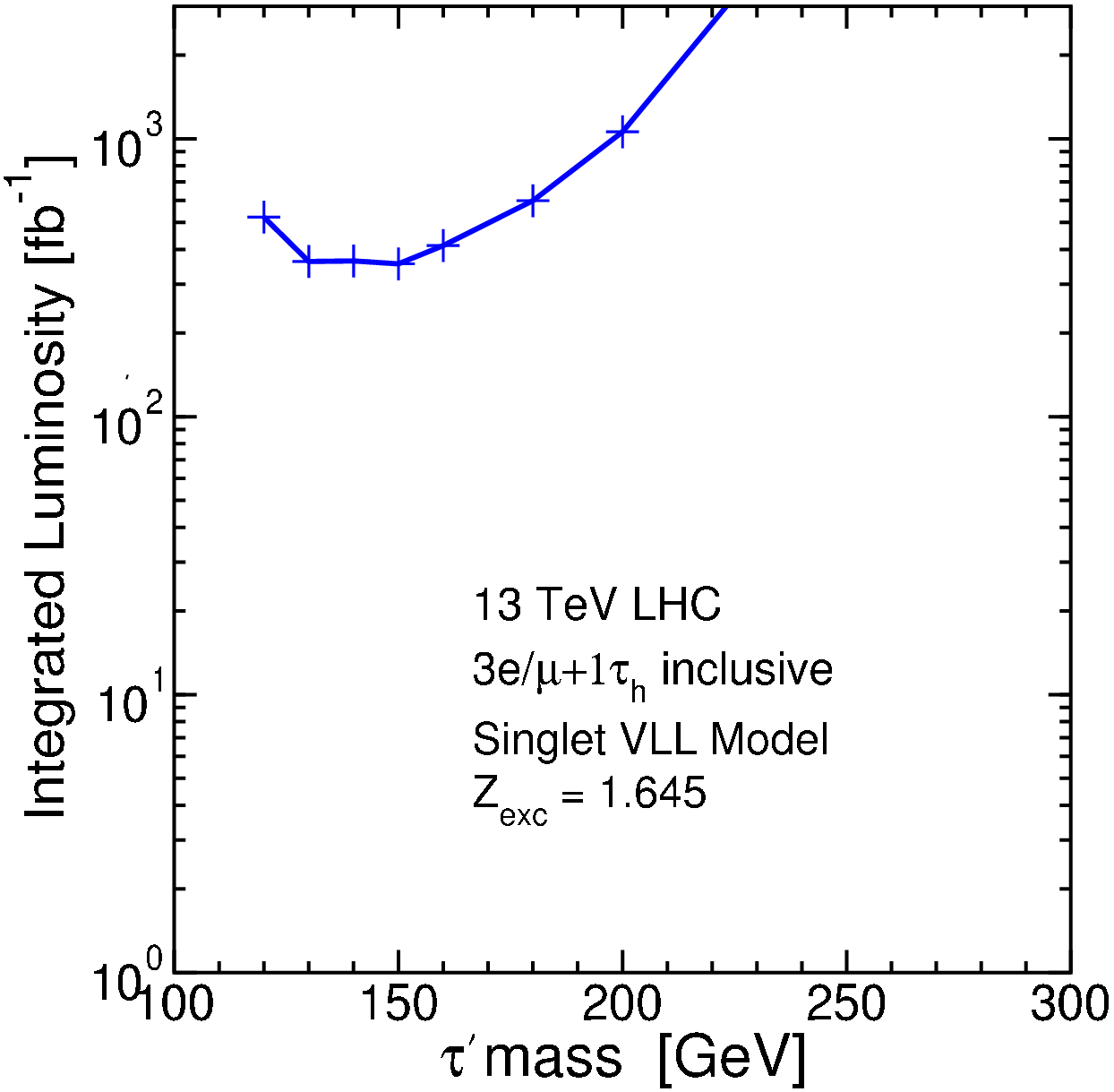}
\end{minipage}
\begin{minipage}[]{0.495\linewidth}
\caption{\label{fig:4lsinglet_theory}
Integrated luminosity needed for a median expected
$\Zexc\geq 1.645$ (95\% CL) exclusion 
in the inclusive $\geq 3e/\mu+1\tau_h$ channel,
for the Singlet VLL model, 
as a function of $M_{\tau^{\prime}}$, at $\sqrt{s} = 13 $ TeV,
based on the results of Table \ref{tab:table12}. 
The line with plus 
symbols represents the required luminosity in the most optimistic case of
no uncertainty in the background cross-section ($\sigmab = 0$).
We do not find any prospects for exclusion if the fractional uncertainty
in the background is 10\% or higher.
}
\end{minipage}
\end{figure}

In the following, we performed 4-lepton searches by generating events 
using the same cuts and selections 
as we did for the Doublet VLL model at $\sqrt{s}=13$ TeV 
in subsection \ref{subsec:Doublet4l13TeV}. The results for individual
backgrounds were already listed above in Table \ref{tab:table9}.
The signal and total background cross sections to pass the inclusive and no-$Z$ 
selections are listed in Table \ref{tab:table12}, for various $M_{\tau'}$. 
We find that the no-$Z$ selection 
is not effective for the purposes of setting a 95$\%$ CL exclusion or claim a  
discovery in 4-lepton searches, for the Singlet VLL model. 
Even in the case of the inclusive selections, we found
that in Singlet VLL model no reasonable integrated luminosity would be able to 
set a 95$\%$ CL exclusion or claim a  
discovery in $\geq 4e/\mu$, $\geq 2e/\mu + 2\tau_h$ and $\geq 3e/\mu+1\tau_h$ 
channels at 13 TeV, if the uncertainty in the background is 10$\%$. This is because
in this case the  
background cross section is always greater than the signal cross section by 
more than a factor of 10. However, in the most optimistic possible case that 
there is no uncertainty at all in the background cross-section, 
then it would be possible to set 
a 95$\%$ CL exclusion in $\geq 3e/\mu+1\tau_h$ channel with 
350 fb$^{-1}$ luminosity for 130 GeV $< M_{\tau'} <$ 150 GeV, 
as shown in Figure \ref {fig:4lsinglet_theory}. With 1000 fb$^{-1}$, the
exclusion reach in this case would extend up to about $M_{\tau'} =200$ GeV.
\begin{table}
\caption{Signal and background cross sections in the
$\geq 3e/\mu+1\tau_h$, $\geq 2e/\mu+2\tau_h$ and  $\geq 4e/\mu$ channels 
after selecting events through inclusive and no-$Z$ 
requirements, for the Singlet VLL model, at $\sqrt{s} = 13$ TeV.}
\label{tab:table12}
\centering
\begin{center}
\begin{tabular}{|c|c|c|c|c|c|c|}
\hline
$M_{\tau^\prime}$ (GeV)	& \multicolumn{2}{c|}{$\sigma_{s}$ (fb) in $\geq 3e/\mu+1\tau_h$} & 
\multicolumn{2}{c|}{$\sigma_{s}$ (fb) in $\geq 2e/\mu+2\tau_h$} & \multicolumn{2}{c|}{$\sigma_{s}$ (fb) in $\geq 4e/\mu$} \\
\hline
& inclusive & no-Z 
& inclusive & no-Z 
& inclusive & no-Z \\
\hline
120&	0.0672&	0.0027&	0.0143&	0.0023&	0.0528&	0.0007\\
130&	0.0811&	0.0035&	0.0208&	0.0027&	0.0644&	0.0011\\
140&	0.0809&	0.0037&	0.0225&	0.0050&	0.0679&	0.0012\\
150&	0.0819&	0.0060&	0.0215&	0.0043&	0.0651&	0.0016\\
160&	0.0758&	0.0047&	0.0210&	0.0057&	0.0565&	0.0016\\
180&	0.0627&	0.0057&	0.0208&	0.0052&	0.0491&	0.0021\\
200&	0.0469&	0.0047&	0.0161&	0.0055&	0.0404&	0.0021\\
250&	0.0259&	0.0038&	0.0094&	0.0038&	0.0230&	0.0017\\
300&	0.0150&	0.0026&	0.0056&	0.0025&	0.0141&	0.0010\\
\hline
Total Background & 1.055 & 0.052& 0.549 & 0.062 & 15.195 & 0.026\\
\hline
\end{tabular}
\end{center}
\end{table}

\begin{figure}[!tb]
\begin{minipage}[]{0.495\linewidth}
\includegraphics[width=8.0cm,angle=0]{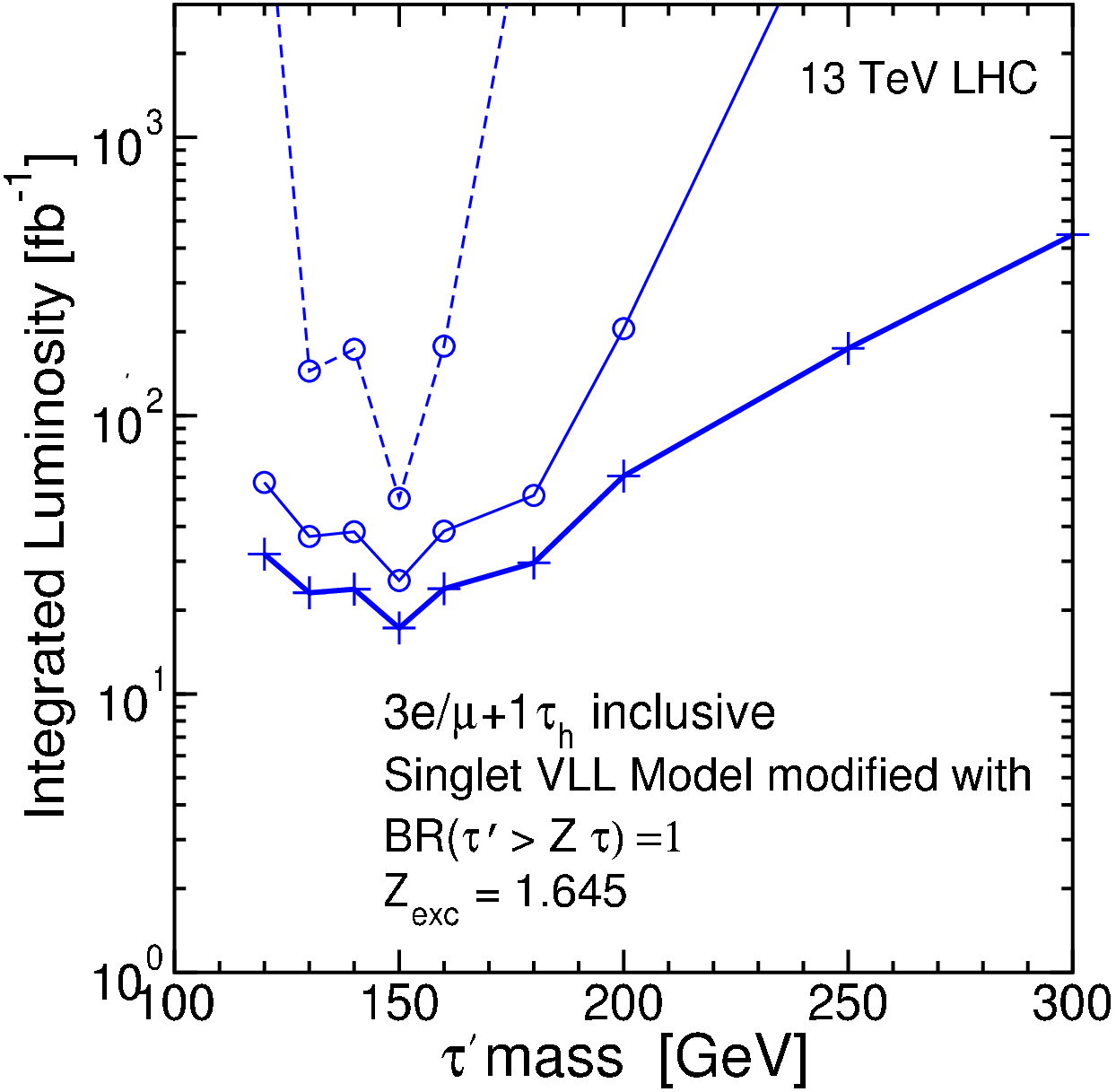}
\end{minipage}
\begin{minipage}[]{0.495\linewidth}
\begin{flushright}
\includegraphics[width=8.0cm,angle=0]{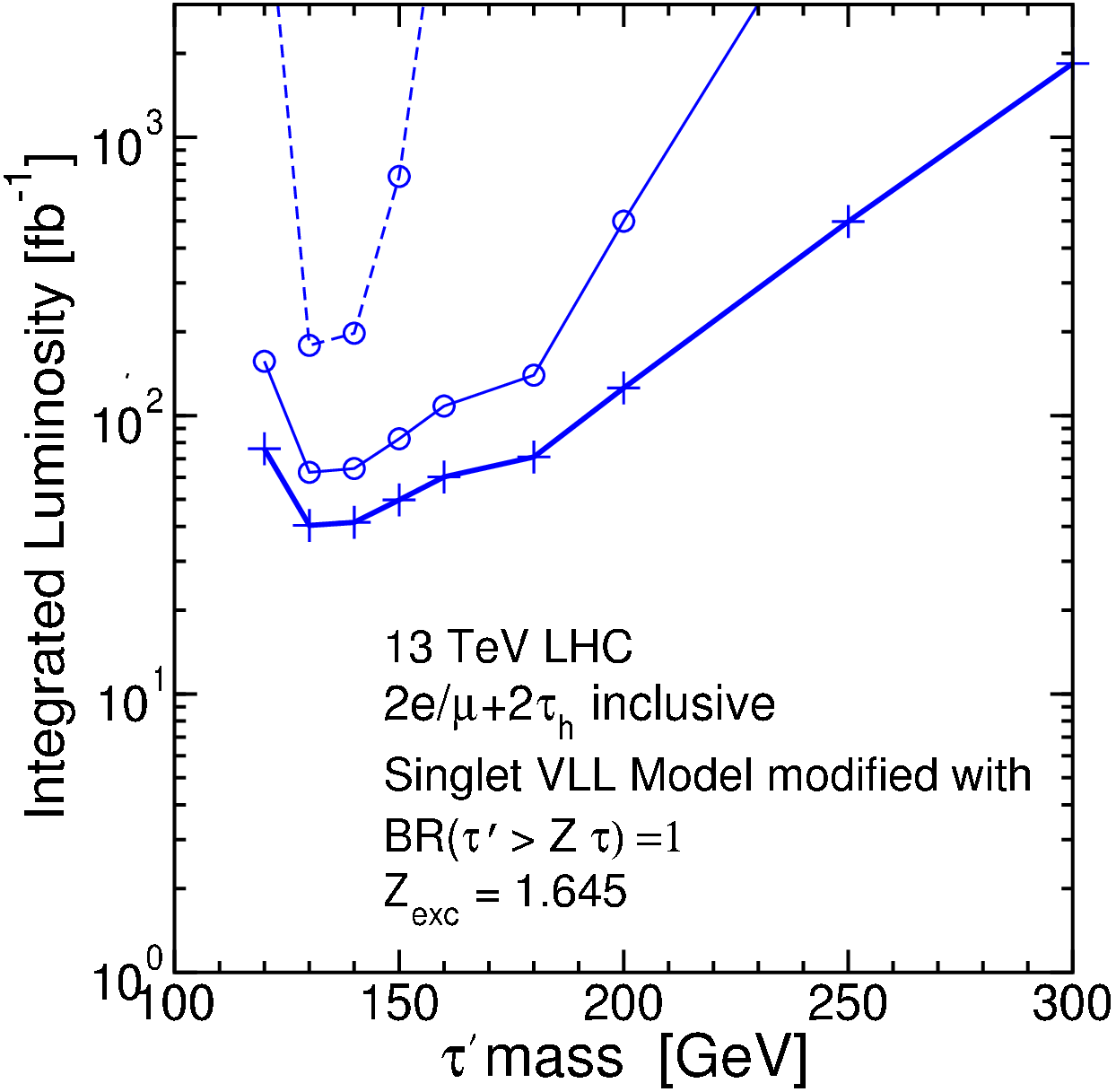}
\end{flushright}
\end{minipage}
\begin{minipage}[]{0.495\linewidth}
\includegraphics[width=8.0cm,angle=0]{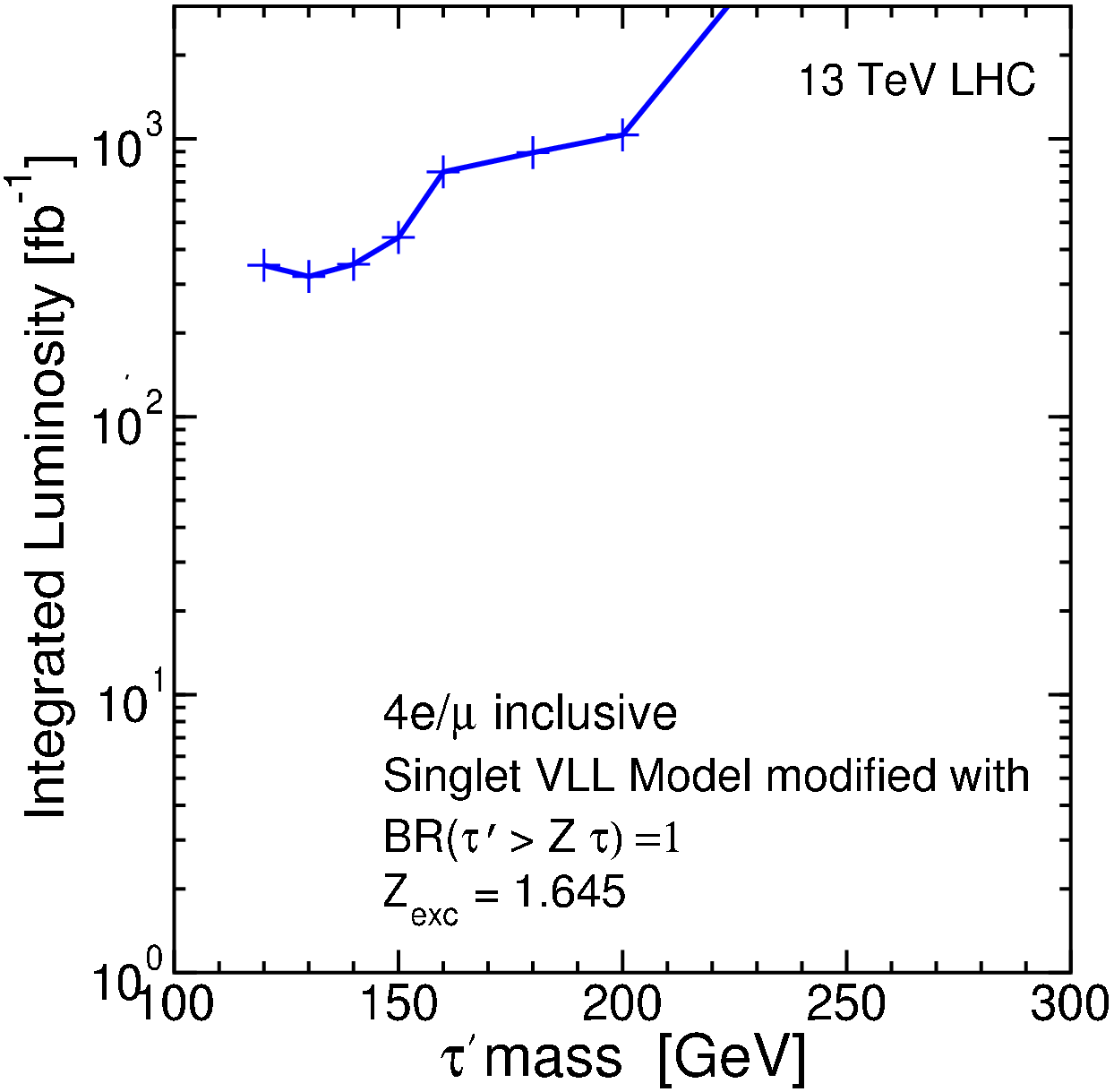}
\end{minipage}
\begin{minipage}[]{0.495\linewidth}
\caption{\label{fig:4lsinglet}
Integrated luminosities needed for $\Zexc= 1.645$ (95\% CL) exclusion, 
in the inclusive $\geq 3e/\mu+1\tau_h$ and $\geq 2e/\mu+2\tau_h$ and 
$\geq 4e/\mu$ channels, for the modified 
Singlet VLL model with BR($\tau^{\prime}\rightarrow Z\tau$)=1,
as a function of $M_{\tau^{\prime}}$, at $\sqrt{s} = 13 $ TeV,
based on the results from Table \ref{tab:table13}.
The different lines in each plot correspond to assumed background 
uncertainties $\sigmab = 0$ 
(thicker solid, with $+$ signs), $\sigmab= 0.1b$ (solid, with circles) 
and $\sigmab= 0.2b$ (dashed, with circles).}
\end{minipage}
\end{figure}

In view of the rather pessimistic nature of these results, we now study 
a more optimistic variant of the Singlet VLL model in which 
BR($\tau^{\prime}\rightarrow Z\tau$)=1 is forced, but the production 
cross-section is not changed. We emphasize that this is somewhat 
arbitrary, as we do not have in mind a specific model that actually makes this 
prediction, although this scenario is at least consistent in the sense
that the couplings involved in the production 
are different from those involved in the decays.
The signal cross-sections to pass the inclusive and no-$Z$ 
selections in this case are given in Table \ref{tab:table13}. We find 
that in this case, it is possible to set 95$\%$ exclusions in the $\geq 
3e/\mu+1\tau_h$ and $\geq 2e/\mu+2\tau_h$ channels with the inclusive 
search (and also with the no-$Z$ search, although that will require a 
much higher luminosity).
\begin{table}
\caption{Signal and background cross sections in the
$\geq 3e/\mu+1\tau_h$, $\geq 2e/\mu+2\tau_h$ and 
$\geq 4e/\mu$ channels after selecting events 
through inclusive and no-$Z$  
requirements, for the 
modified Singlet VLL model with BR($\tau^{\prime}\rightarrow Z\tau$)=1 at 
$\sqrt{s} = 13$ TeV.}
\label{tab:table13}
\centering
\begin{center}
\begin{tabular}{|c|c|c|c|c|c|c|}
\hline
$M_{\tau^\prime}$ (GeV)	& \multicolumn{2}{c|}{$\sigma_{s}$ (fb) in $\geq 3e/\mu+1\tau_h$} & 
\multicolumn{2}{c|}{$\sigma_{s}$ (fb) in $\geq 2e/\mu+2\tau_h$} & \multicolumn{2}{c|}{$\sigma_{s}$ (fb) in $\geq 4e/\mu$} \\
\hline
& inclusive & no-Z 
& inclusive & no-Z 
& inclusive & no-Z \\
\hline
120&	0.2898&	0.0043&	0.1341&	0.0043&	0.3071&	0.0000\\
130&	0.3449&	0.0133&	0.1891&	0.0100&	0.3217&	0.0000\\
140&	0.3393&	0.0130&	0.1865&	0.0285&	0.3057&	0.0000\\
150&	0.4048&	0.0288&	0.1685&	0.0164&	0.2733&	0.0021\\
160&	0.3386&	0.0099&	0.1520&	0.0215&	0.2081&	0.0017\\
180&	0.3014&	0.0188&	0.1391&	0.0144&	0.1921&	0.0022\\
200&	0.2052&	0.0084&	0.1026&	0.0130&	0.1784&	0.0038\\
250&	0.1181&	0.0059&	0.0501&	0.0079&	0.1019&	0.0024\\
300&	0.0728&	0.0044&	0.0257&	0.0047&	0.0642&	0.0009\\
\hline
Total Background & 1.055 & 0.052& 0.549 & 0.062 & 15.195 & 0.026\\
\hline
\end{tabular}
\end{center}
\end{table}
Figure \ref {fig:4lsinglet} shows the integrated luminosity needed to 
set $\Zexc\geq 1.645$ exclusion in in these two channels. We found that 
it is possible to obtain a 95$\%$ CL exclusion for $M_{\tau^\prime}\leq$ 
190 GeV with 100 fb$^{-1}$ in the $\geq 3e/\mu+1\tau_h$ inclusive search 
with 10$\%$ uncertainty in the background events. However, we found that 
it is not possible to satisfy the $\Zdisc\geq 5$ discovery criteria in 
the 4-lepton channels at $\sqrt{s} = 13$ TeV, for any reasonable 
integrated luminosity and any value of $M_{\tau'}$.

\subsection{Five-lepton searches for the Singlet VLL model} 
\setcounter{equation}{0}
\setcounter{footnote}{1}

\begin{figure}[!tb]
\begin{minipage}[]{0.495\linewidth}
\includegraphics[width=8.0cm,angle=0]{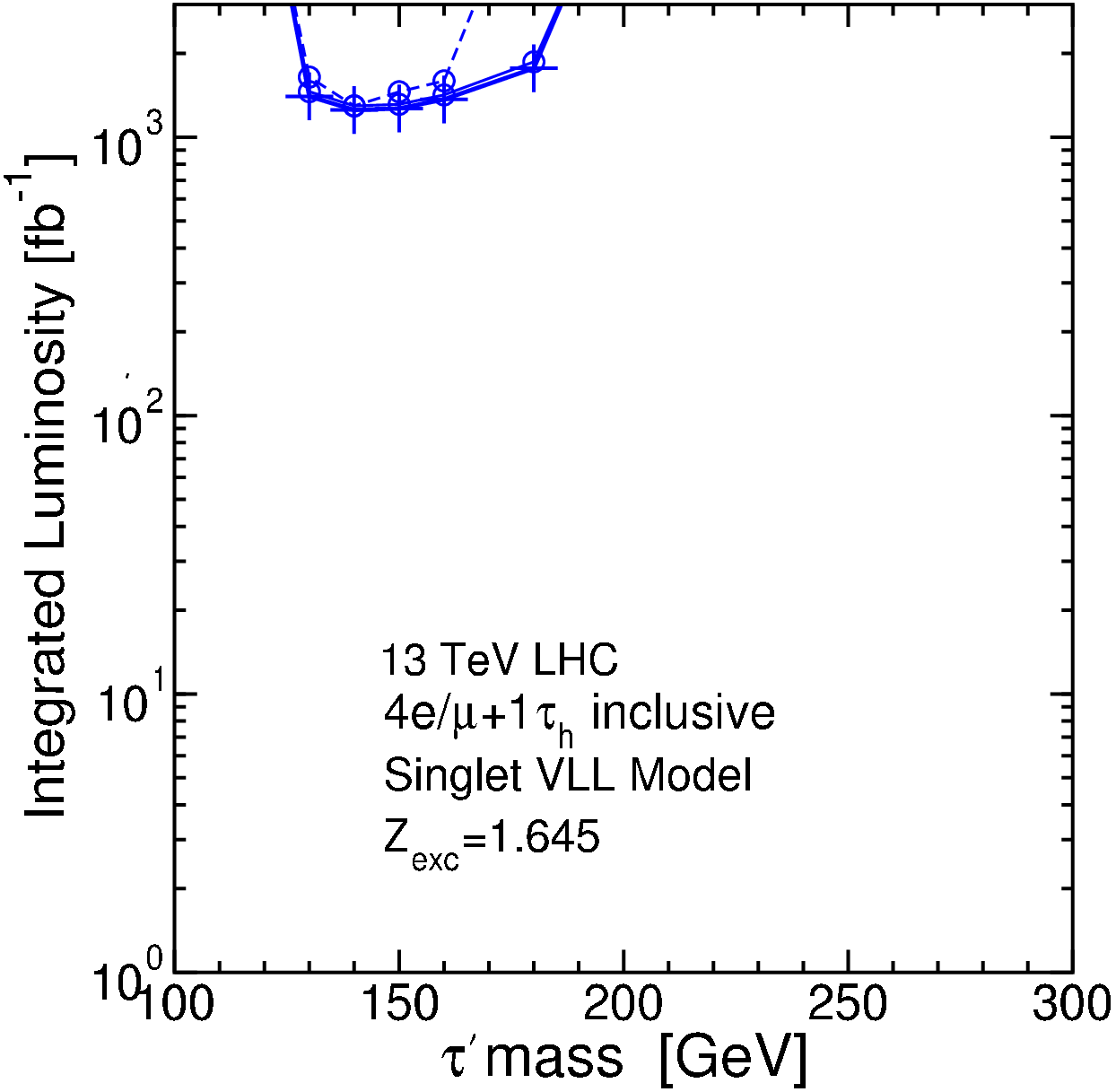}
\end{minipage}
\begin{minipage}[]{0.495\linewidth}
\begin{flushright}
\includegraphics[width=8.0cm,angle=0]{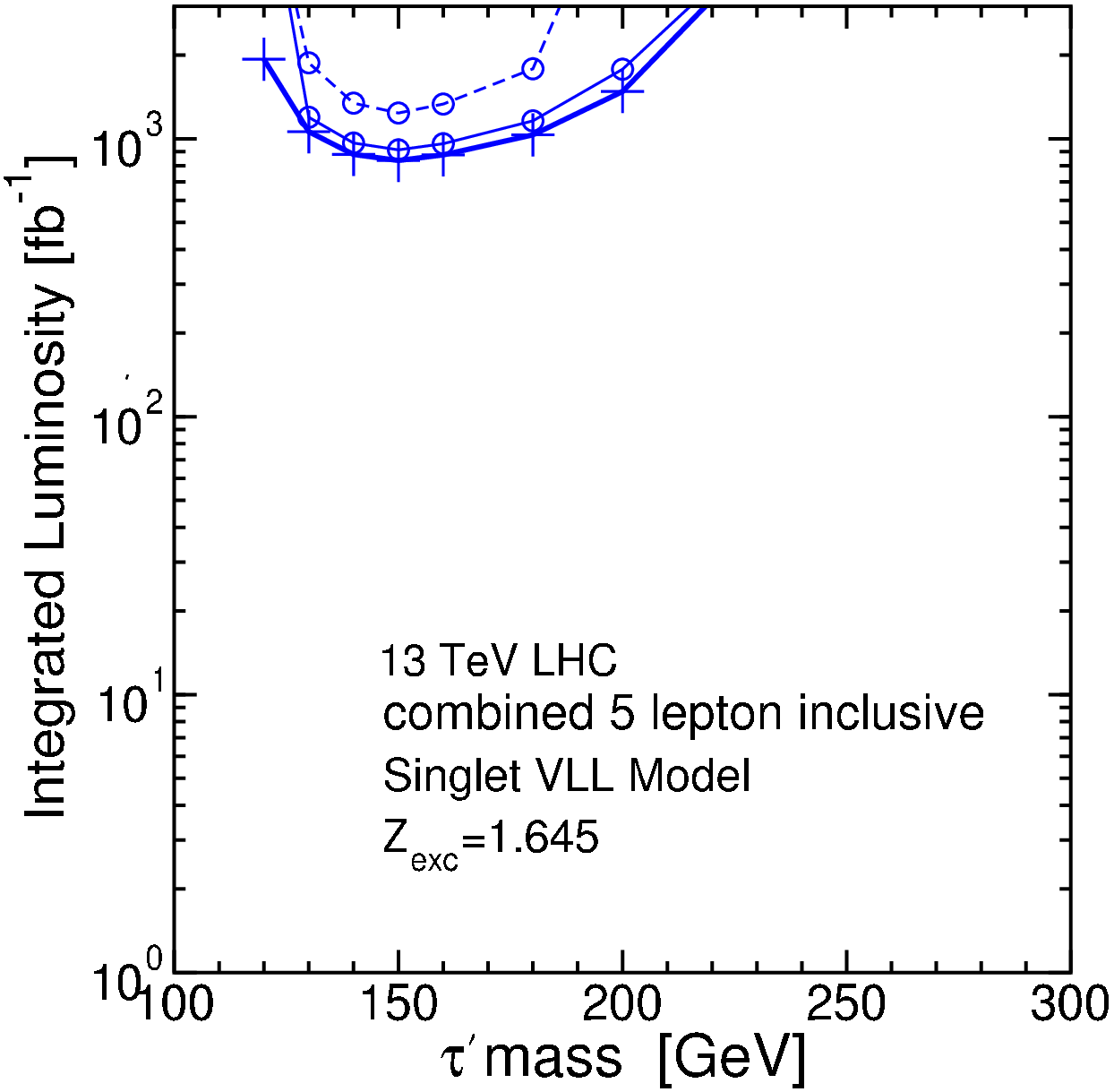}
\end{flushright}
\end{minipage}
\caption{\label{fig:5lsinglet}
Integrated luminosities needed for median expected 
$\Zexc = 1.645$ (95\% CL) exclusion in the $\geq 4e/\mu+1\tau_h$ and
combined 5-lepton inclusive searches, for the Singlet VLL model,
as a function of $M_{\tau^{\prime}}$, at $\sqrt{s} = 13 $ TeV,
based on the results from Table \ref{tab:table14}.
The different lines in each plot correspond to assumed background 
uncertainties $\sigmab = 0$ (thicker solid, with $+$ signs),
$\sigmab= 0.1b$ (solid, with circles) and 
$\sigmab= 0.2b$ (dashed, with circles). 
}
\end{figure}

In this section, we consider the 5-lepton channels at $\sqrt{s}=13$ TeV
for the Singlet 
VLL model 
to study the prospectives for exclusion or discovery,
in the same manner as we did 
in subsection \ref{subsec:Doublet5l13TeV} for the Doublet VLL model. 
The individual 
backgrounds were already given above in Table \ref{tab:table11}.
The visible signal cross sections after cuts are quite small,
even to pass the inclusive selections, 
as can be seen from Table \ref{tab:table14}. 
\begin{table}
\caption{Signal and total background cross sections in the
$\geq 4e/\mu+1\tau_h$ and $\geq 3e/\mu+2\tau_h$ and 
$\geq 5e/\mu$ channels after inclusive selections, 
for the Singlet VLL model, at $\sqrt{s} = 13$ TeV.
}
\label{tab:table14}
\centering
\begin{center}
\begin{tabular}{|c|c|c|c|}
\hline
$M_{\tau^\prime}$ (GeV)	& $\sigma_{s}$ (fb) in $\geq 4e/\mu+1\tau_h$ &  
$\sigma_{s}$ (fb) in $\geq 3e/\mu+2\tau_h$ & $\sigma_{s}$ (fb) in $\geq 5e/\mu$ \\
\hline
120&	0.00285& 0.00071 & 0.00186 \\
130&	0.00393& 0.00113 & 0.00246 \\
140&	0.00420& 0.00123 & 0.00293 \\
150&	0.00417& 0.00134 & 0.00310 \\
160&	0.00399& 0.00138 & 0.00302 \\
180&	0.00343& 0.00142 & 0.00277 \\
200&	0.00294& 0.00120 & 0.00211 \\
250&	0.00176& 0.00082 & 0.00141 \\
300&	0.00094& 0.00049 & 0.00080 \\
\hline
Total Background & 0.0055 & 0.0065 & 0.0053 \\
\hline
\end{tabular}
\end{center}
\end{table}
Note that the three individual 5-lepton channels have comparable signal and
background levels. Therefore,
because of the low cross-section yields, we consider 
not only the individual $\geq 4e/\mu+1\tau_h$ and $\geq 3e/\mu+2\tau_h$
and $\geq 5e/\mu$ inclusive cross-sections, but  
also their combination given by the sum of the three channels.
We then find that with a very high integrated luminosity,
of 1000 fb$^{-1}$, it may be possible to set a 95$\%$ CL exclusion
for a narrow range of 140 GeV $< M_{\tau'}<$ 165 GeV, using the
$\geq 4e/\mu+1\tau_h$ channel or the combined 5-lepton channel. 
These results
are shown in Figure \ref{fig:5lsinglet} as a function of $M_{\tau^\prime}$.   

\begin{figure}[!tb]
\includegraphics[width=7.5cm,angle=0]{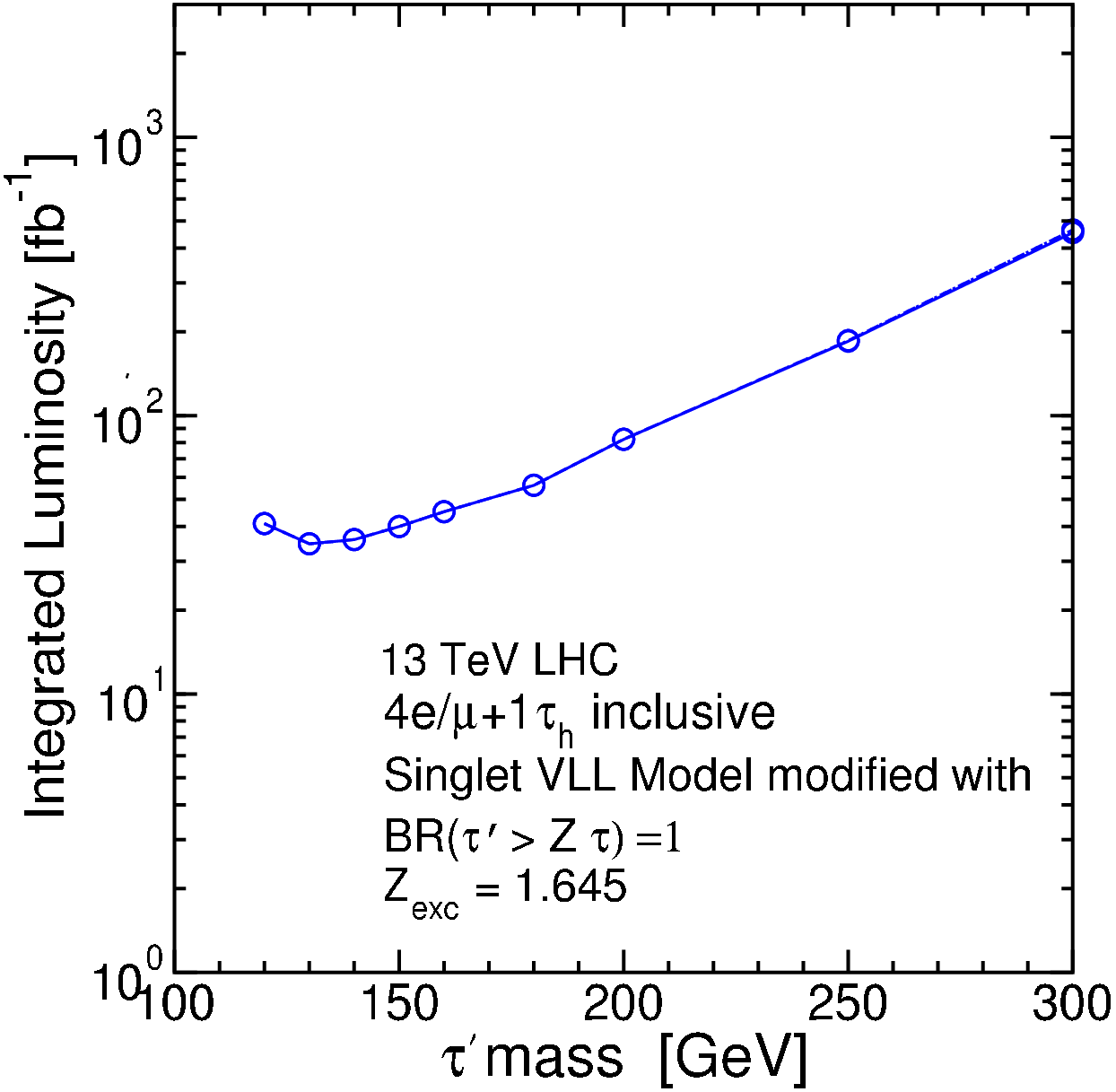}
\includegraphics[width=7.5cm,angle=0]{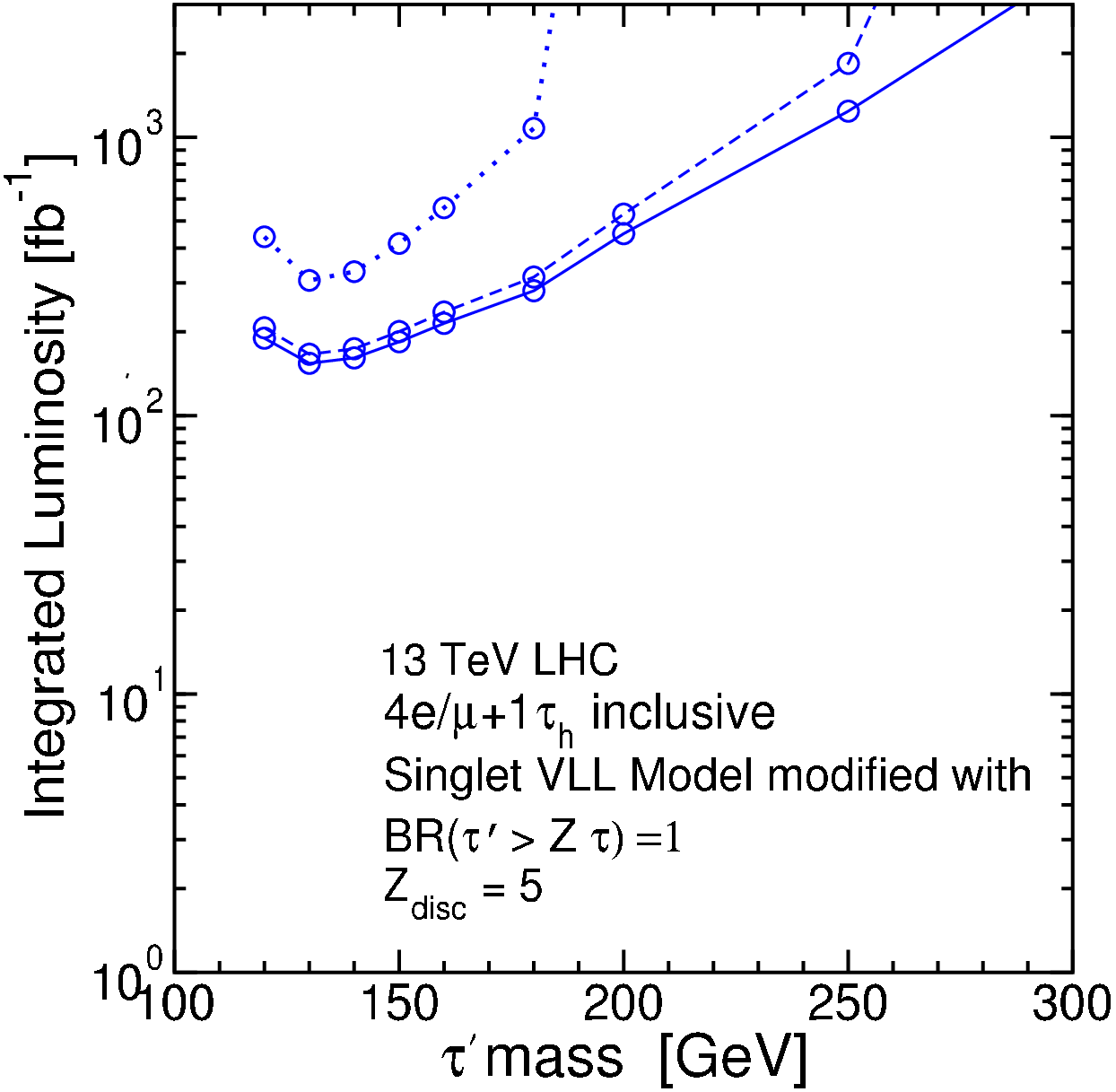}
\\
\includegraphics[width=7.5cm,angle=0]{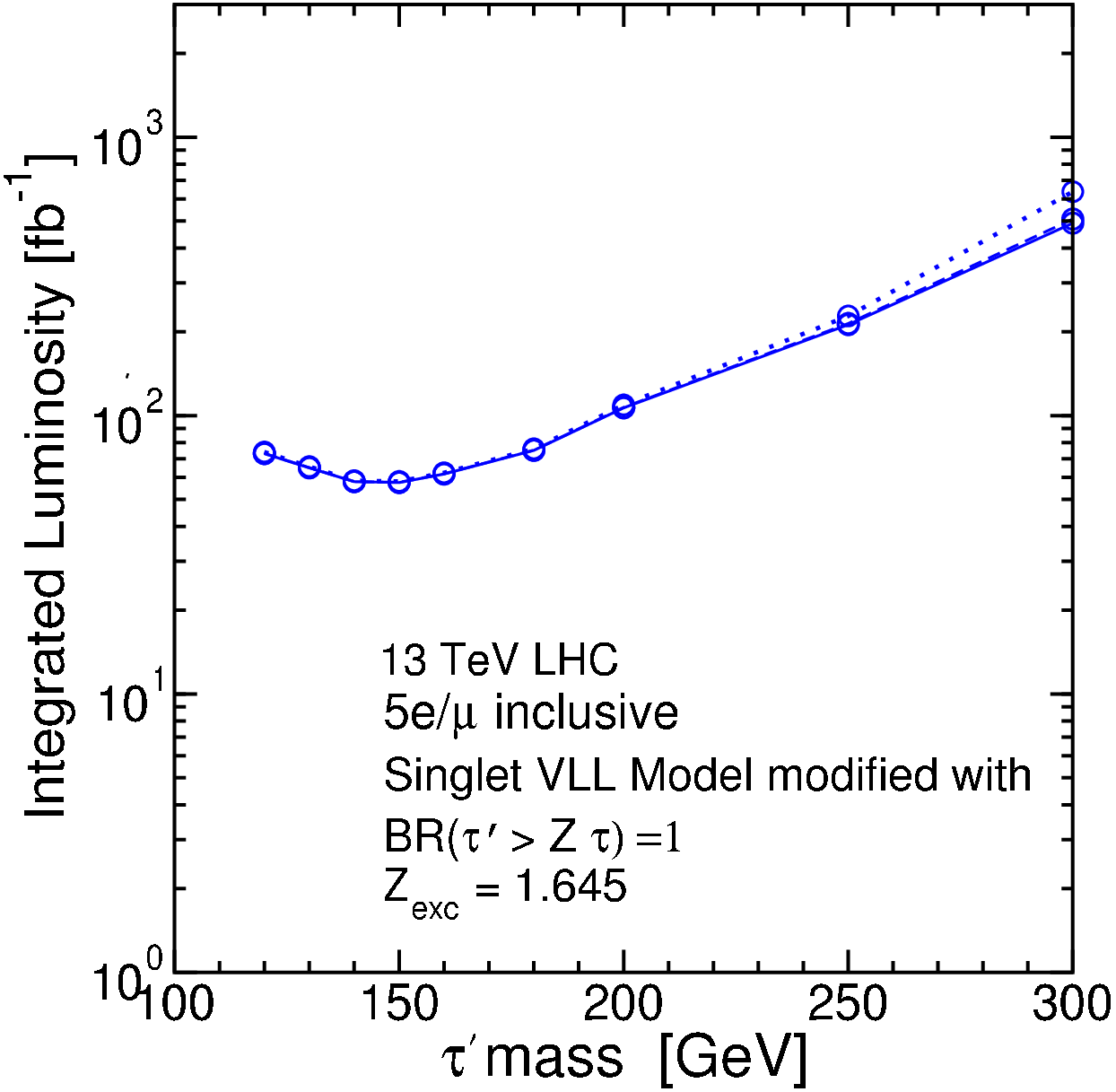}
\includegraphics[width=7.5cm,angle=0]{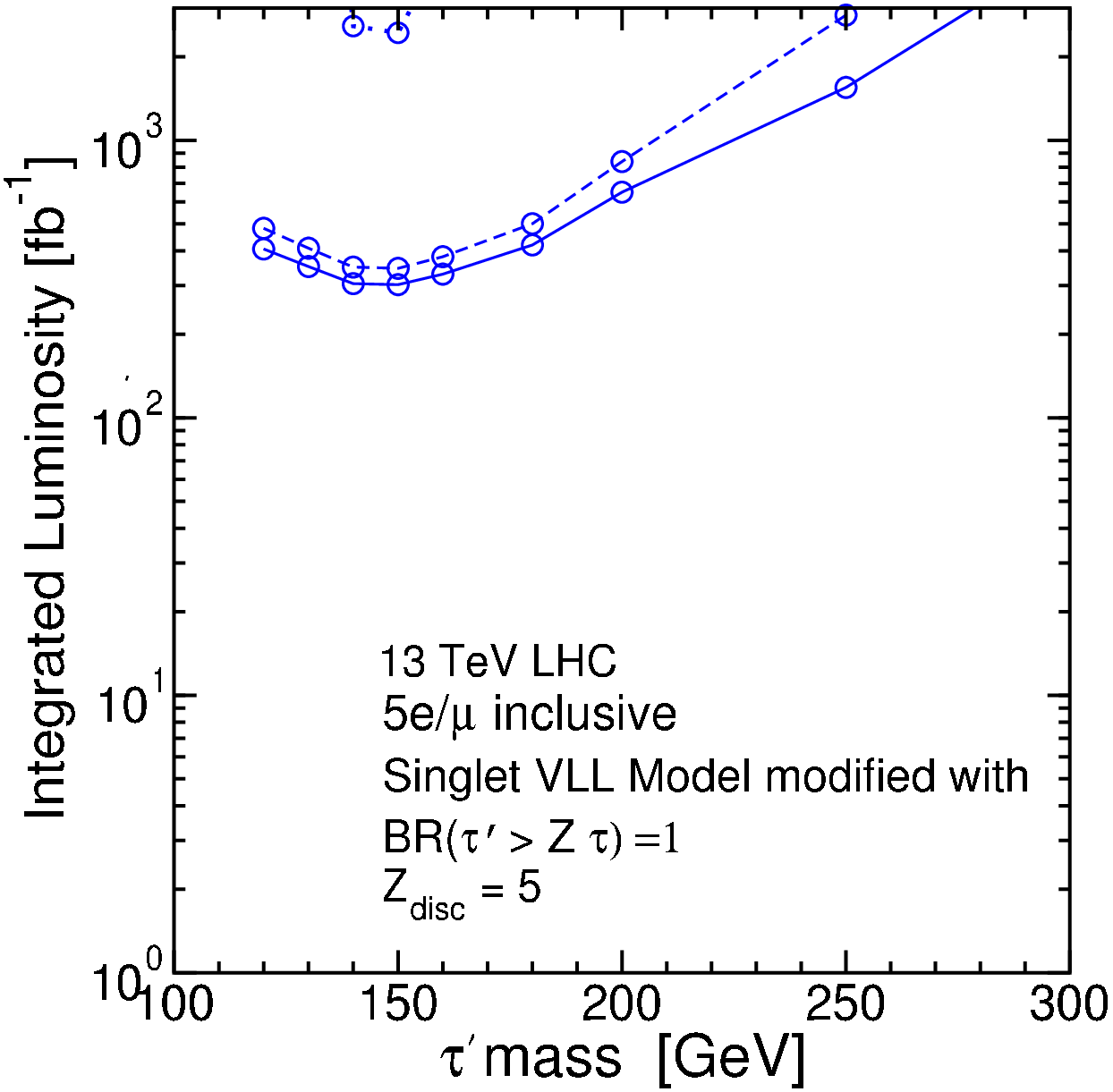}
\\
\includegraphics[width=7.5cm,angle=0]{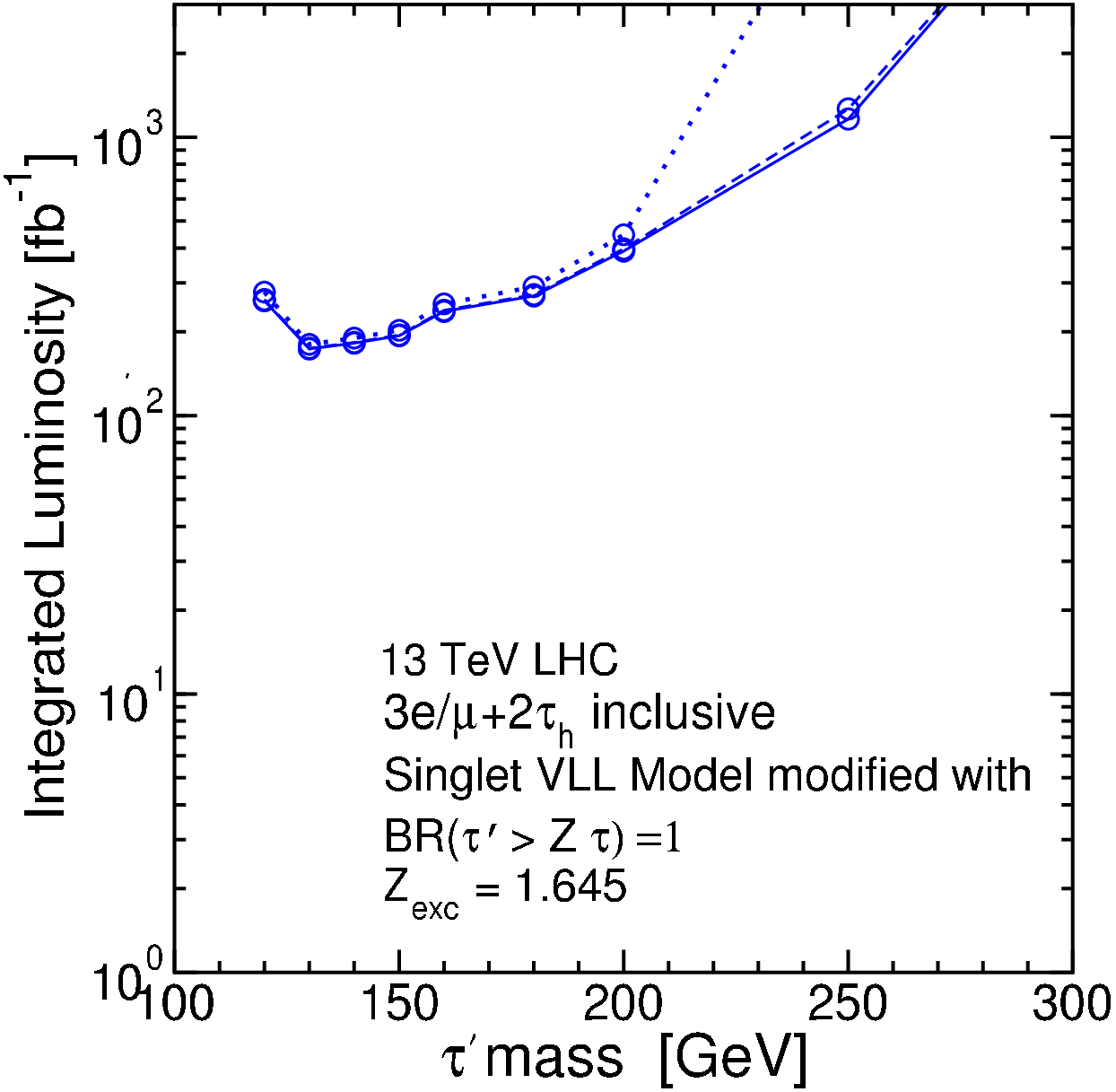}
\includegraphics[width=7.5cm,angle=0]{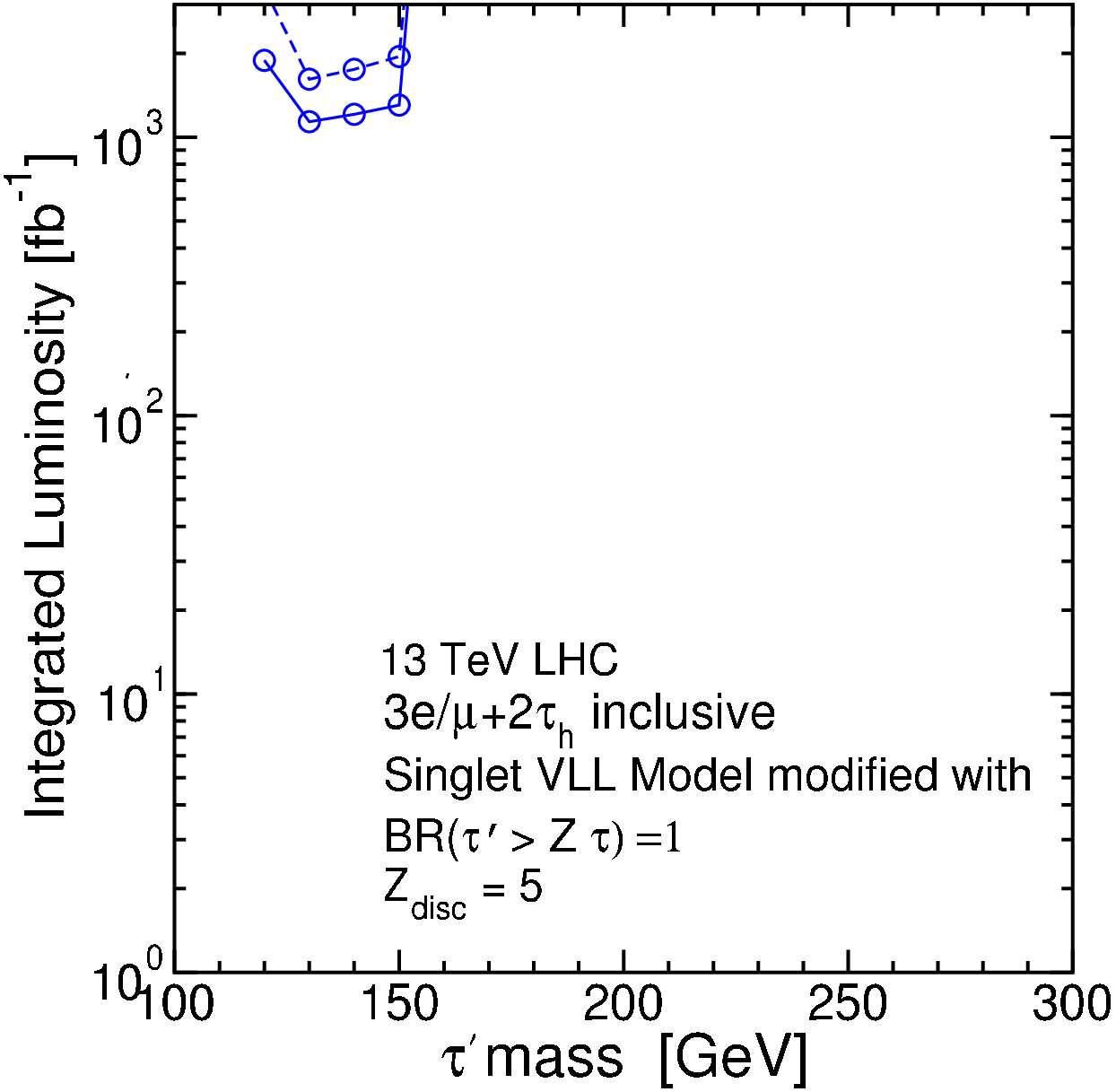}
\\
\caption{\label{fig:L_5lept_singlet}
Integrated luminosities needed for $\Zexc= 1.645$ (95\% CL) 
exclusion (left) or $\Zdisc= 5$ discovery (right), 
in the $\geq 4e/\mu+1\tau_h$ and $\geq 5e/\mu$ and 
$\geq 3e/\mu+2\tau_h$ inclusive channels (from top to bottom), 
for the Singlet VLL model with BR($\tau^{\prime}\rightarrow Z\tau$)=1,
as a function of $M_{\tau^{\prime}}$, at $\sqrt{s} = 13$ TeV,
based on the results from Table \ref{tab:table15}.
The different lines in each plot correspond to assumed 
background uncertainties $\sigmab= 0.1b$ (solid), 
$\sigmab= 0.2b$ (dashed), and $\sigmab= 0.5b$ (dotted).
}
\end{figure}

\begin{figure}
\includegraphics[width=7.5cm,angle=0]{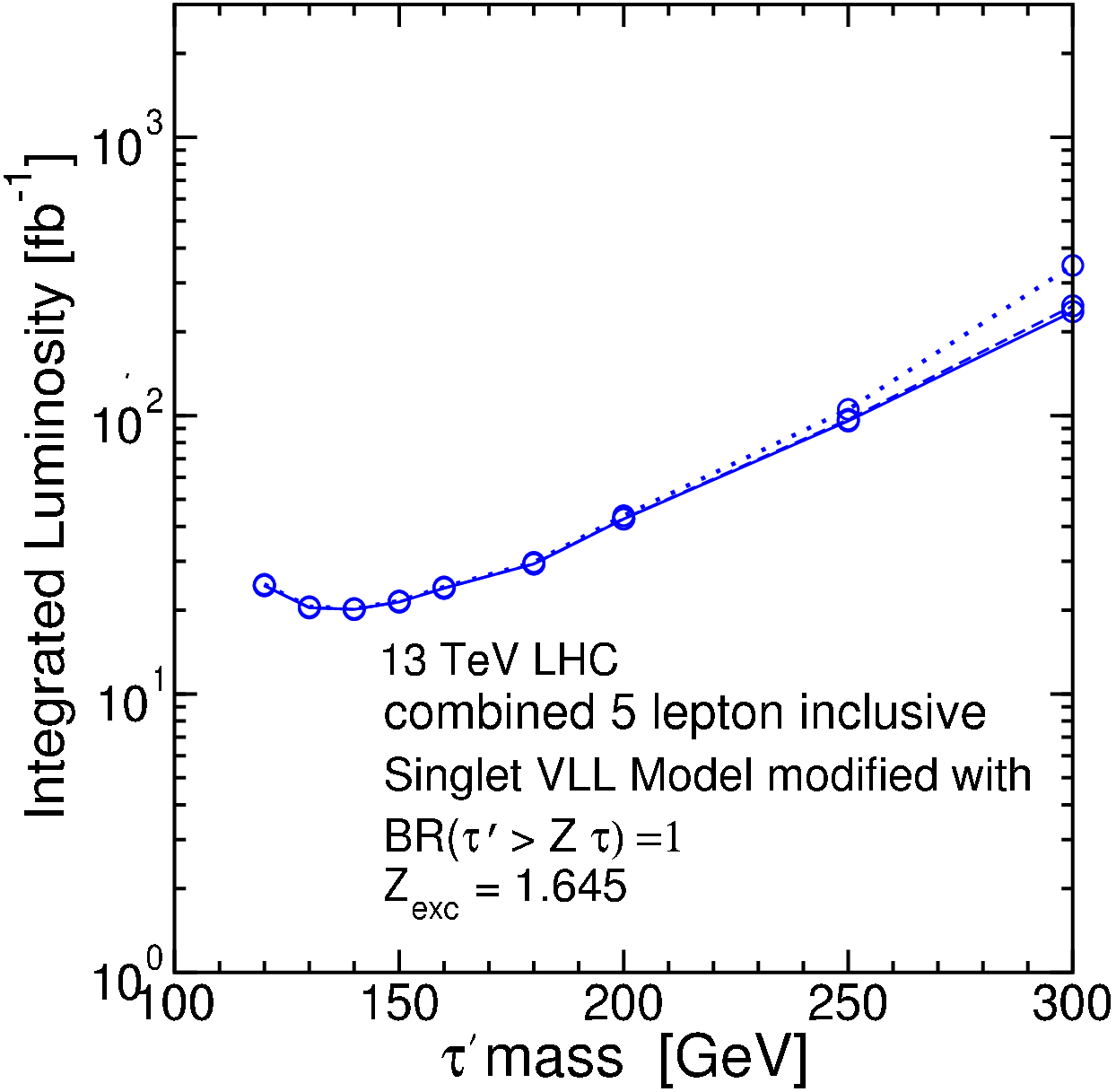}
\includegraphics[width=7.5cm,angle=0]{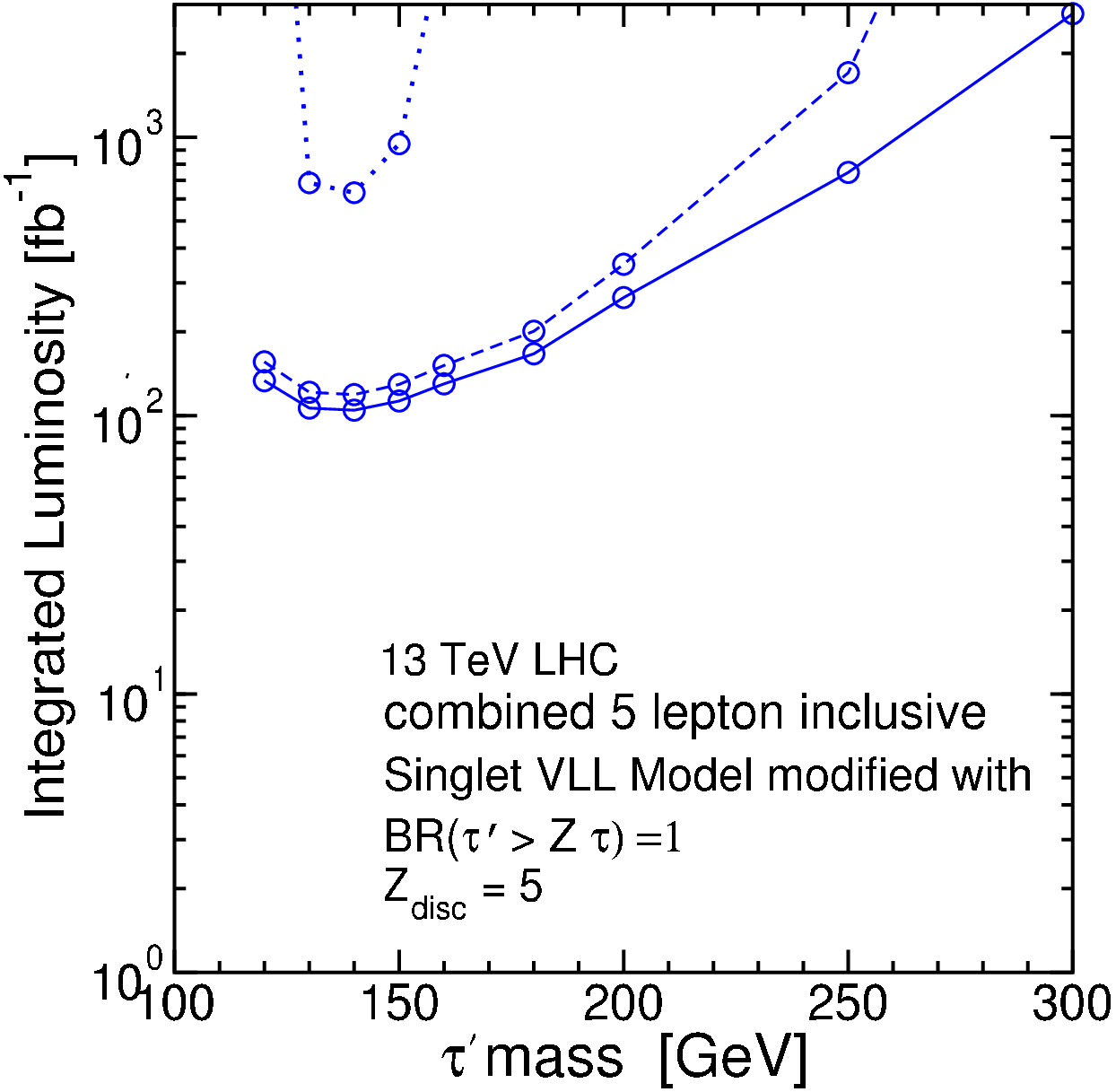}
\caption{\label{fig:L_5lept_tot_singlet}
Integrated luminosities needed for $\Zexc = 1.645$ (95\% CL) exclusion (left) 
or $\Zdisc = 5$ discovery (right) in the combined 5-lepton inclusive 
channel, for the modified Singlet VLL model with 
BR($\tau^{\prime}\rightarrow Z\tau$)=1, as a function of 
$M_{\tau^{\prime}}$, at $\sqrt{s} = 13$ TeV,
based on the results from Table \ref{tab:table15}. 
The different lines in each plot correspond to assumed
background uncertainties 
$\sigmab= 0.1b$ (solid),  
$\sigmab= 0.2b$ (dashed), and $\sigmab= 0.5b$ (dotted).
}
\end{figure}

For the modified Singlet VLL model with BR($\tau^{\prime}\rightarrow Z\tau$)=1,
the more optimistic results for visible cross-sections after cuts for
the $\geq 3e/\mu+2\tau_h$, $\geq 4e/\mu+1\tau_h$ and $\geq 5e/\mu$ inclusive
signal regions are presented in Table \ref{tab:table15}.
\begin{table}
\caption{Signal and total background cross sections in the
$\geq 4e/\mu+1\tau_h$ and $\geq 3e/\mu+2\tau_h$ and $\geq 5e/\mu$  
channels after inclusive selections, 
for the modified Singlet VLL model with 
BR($\tau^{\prime}\rightarrow Z\tau$)=1, at $\sqrt{s} = 13$ TeV.
}
\label{tab:table15}
\centering
\begin{center}
\begin{tabular}{|c|c|c|c|}
\hline
$M_{\tau^\prime}$ (GeV)	& $\sigma_{s}$ (fb) in $\geq 4e/\mu+1\tau_h$ &  
$\sigma_{s}$ (fb) in $\geq 3e/\mu+2\tau_h$  & $\sigma_{s}$ (fb) in $\geq 5e/\mu$\\
\hline
120&	0.0453& 0.0113 &0.0297 \\
130&	0.0520& 0.0149 &0.0325 \\
140&	0.0505& 0.0144 & 0.0355 \\
150&	0.0462&  0.0138 &0.0357 \\
160&	0.0418& 0.0120 & 0.0338 \\
180&	0.0351& 0.0110 & 0.0291 \\
200&	0.0261& 0.0086 & 0.0224 \\
250&	0.0144&  0.0044 & 0.0138 \\
300&	0.0079& 0.0024 & 0.0080 \\
\hline
Total Background & 0.0055 & 0.0065 & 0.0053 \\
\hline
\end{tabular}
\end{center}
\end{table}
The predicted luminosities required for 95\% CL exclusion with $\Zexc\geq 1.645$ 
and $\Zdisc\geq 5$ discovery are shown in figure \ref{fig:L_5lept_singlet} 
for these three channels at $\sqrt{s}=13$ TeV.
In addition, Figure \ref {fig:L_5lept_tot_singlet} shows the results
for the combined 5-lepton signal region obtained by 
summing these three channels.
With an integrated luminosity of 100 fb$^{-1}$, a
95$\%$ CL exclusion can be expected in the combined 5-lepton search
up to about $M_{\tau'}=250$ GeV, even with a 50\% fractional uncertainty in
the background. The best of the individual channels for this search
is $\geq 4e/\mu+1\tau_h$. A potential $\Zdisc>5$ discovery 
would require more than 
100 fb$^{-1}$ in this most optimistic case of 
BR($\tau^{\prime}\rightarrow Z\tau$)=1,
even for $M_{\tau'}$ less than 150 GeV, and even after 
combining the three individual 5-lepton channels,
and discovering $M_{\tau'}=200$ GeV would require 350 fb$^{-1}$. 
The discovery potential degrades completely for the $\sigmab = 0.5b$ case.
\clearpage

\section{Outlook\label{sec:Outlook}}
\setcounter{equation}{0}
\setcounter{figure}{0}
\setcounter{table}{0}
\setcounter{footnote}{1}

In this paper we have studied possibilities for discovering or excluding 
vectorlike leptons at the LHC in different multilepton searches. 
We mainly looked at two different cases, 
the Singlet VLL model and Doublet VLL model, with small mixing
allowing decays of vectorlike leptons to tau leptons, 
as described in the Introduction. (A previous paper \cite{Dermisek:2014qca}
had already considered the more	optimistic case	of decays to muons.)
We pointed out that there is an opportunity to 
set limits on vectorlike lepton production in the Doublet VLL model using 
existing LHC data at $\sqrt{s}=8$ TeV. This could be done with
searches similar to the Run 1 ATLAS 
3-lepton and 4-lepton searches \cite{ATLASmultilepton,ATLAS4l}
based on $\int L$ dt = 20.3 fb$^{-1}$.
In the 3-lepton channels, we found that our estimates for the visible 
signal cross-section exceeds the ATLAS limit up to about 
$M_{\tau'}$ = 200 GeV for both $\geq 3e/\mu$ and $2e/\mu + \geq 1 
\tau_h$ searches with off-Z and no-OSSF selections. While looking at the 
4-lepton channels we again found that up to about $M_{\tau'} = 200$ GeV,
our estimates for the visible cross section exceeds 
the ATLAS limit in two searches, $\geq 4e/\mu$ on-$Z$ and 
$3e/\mu+ \geq 1 \tau_h$ no-$Z$. 

We then presented a simpler 4-lepton search strategy more appropriate 
for the the Doublet VLL model. We came up with a different set of 
selections that we followed throughout the rest of the paper. Imposing a 
$b$-jet veto became very useful to reduce some of the background cross 
sections involving top quarks. We also used a set of equations as 
mentioned in the Introduction to calculate median expected significances 
for exclusion ($\Zexc$). We found that the highest exclusion 
significance can be reached with the $\geq 3e/\mu+1\tau_h$ channel, and 
$M_{\tau'}$ masses up to about 275 GeV could be excluded with 95\% CL in both 
inclusive and no-$Z$ searches with 20 fb$^{-1}$ at $\sqrt{s}=8$ TeV, 
provided that there is indeed no signal present.

We also studied the future prospects for vectorlike leptons at $\sqrt{s} 
= 13$ TeV. In the Doublet VLL model, we estimated the integrated 
luminosities needed to set 95$\%$ CL exclusion and discovery with 
$\Zdisc\geq$5 in 4-lepton and 5-lepton channels as a function of 
$M_{\tau'} = M_{\nu'}$. We find that it should be possible to set an 
exclusion up to $M_{\tau'}$ = 440 GeV with 100 fb$^{-1}$ of integrated 
luminosity, using several different 4-lepton channels with no-$Z$ 
selections, even assuming 20$\%$ fractional uncertainty in the 
background. Discovery is possible with 100 fb$^{-1}$ for $M_{\tau'}$ up 
to about 300 GeV using the same channels. In 5-lepton searches 
we found the inclusive search to do much better than the no-$Z$ 
channels, because of the statistics-limited nature of the signal. 
The 5-lepton searches have the advantage of extremely small backgrounds. With 
inclusive 5-lepton searches we found that there is a chance to discover 
the vectorlike leptons in the Doublet VLL model up to $M_{\tau'}$=250 
GeV with integrated luminosity of 100 fb$^{-1}$, even with 50$\%$ 
fractional uncertainties assumed for the backgrounds.

The Singlet VLL model is much more difficult. We find that even setting
a 95\% CL exclusion is not possible with 4-lepton searches 
unless the background is known with less than 10\% uncertainty.
Even in the optimistic scenario that the background is known exactly,
we find no expected exclusions with less than 350 fb$^{-1}$, and
it would take 1000 fb$^{-1}$ to exclude up to $M_{\tau'}= 200$ GeV. 
Using 5-lepton searches, excluding any range
of masses for the Singlet VLL model requires on the order of
1000 fb$^{-1}$ of integrated luminosity. 

We also considered a modified Singlet VLL model, obtained 
by (arbitrarily) setting BR($\tau^{\prime}\rightarrow Z\tau$)=1,
while assuming the same production cross-section. While we did not specify
a model exhibiting these characteristics, it is at least consistent
in the sense that the Lagrangian terms governing production are distinct from
the mixing terms governing the decays.
Here, the 5-lepton signal seem to be the best, with 100 fb$^{-1}$ providing
an expected exclusion up to $M_{\tau'}=250$ GeV, while discovery up to 
$M_{\tau'}=200$ GeV would required 350 fb$^{-1}$.

In this paper, we have only looked at signals based on relatively clean 
multi-lepton final states, including up to 2 hadronic taus. 
There are other channels which can be 
looked at, including those with more than 2 $b$-jets from Higgs and $Z$ 
decays. For example, these could include the channel 4$b$+2$\tau_h$, for 
which physics backgrounds should be small, but detector backgrounds are 
harder to estimate.

We note that the projections made in this paper are heavily dependent on 
our simulation tools, and only the experimental collaborations can 
provide real exclusions (or discovery), based on background estimates 
driven and verified by data and knowledge of detector responses. 
However, we believe that it is clear that the opportunity to conduct
searches for vectorlike 
leptons that decay to taus should be pursued at the LHC.

{\it Acknowledgments:} We thank Jahred Adelman and Glen Cowan for helpful
discussions about the treatment of significances in the presence
of background uncertainties.
This work was supported in part by the National
Science Foundation grant number PHY-1417028.


\end{document}